\documentclass{jfm}
\usepackage{graphicx}
\usepackage{epstopdf, epsfig}

\shorttitle{Direct numerical simulations of the supersonic TGV via the Boltzmann equation}
\shortauthor{T. Dzanic, W. Trojak and L. Martinelli}

\title{Direct numerical simulations of the supersonic Taylor--Green vortex via the Boltzmann equation}

\author{
Tarik Dzanic\aff{1}\corresp{\email{dzanic1@llnl.gov}},
Will Trojak\aff{2}
\and
Luigi Martinelli\aff{3}
}

\affiliation{
\aff{1}Center for Applied Scientific Computing, Lawrence Livermore National Lab, Livermore, CA 94550, USA
\aff{2}NVIDIA, Bristol BS32 4TU, UK
\aff{3}Department of Mechanical and Aerospace Engineering, Princeton University, Princeton, NJ 08544, USA
}

\usepackage{amsmath}
\usepackage{upgreek}
\usepackage{pdflscape}
\usepackage{listings}
\usepackage{multirow}
\usepackage[percent]{overpic}
\usepackage{multicol}
\usepackage{color}
\usepackage{stmaryrd}
\usepackage{xfrac}
\usepackage{booktabs}
\usepackage[section]{placeins} 
\usepackage[section]{algorithm}
\usepackage{calc}
\usepackage[titletoc]{appendix}
\usepackage{siunitx}
\usepackage{mathtools}
\usepackage{tabularx}
\usepackage{amsmath}
\usepackage{diagbox}
\usepackage[capitalise]{cleveref}
\usepackage{xcolor}

\usepackage{tikz-dimline}
\usepackage{graphicx}
\usepackage{graphics}
\usepackage{wrapfig}
\usepackage{float}
\usepackage{subfloat}
\usepackage{caption}
\usepackage{subcaption}
\usepackage[percent]{overpic}
\usepackage{varwidth}
\usepackage{tikz}
\usepackage{pgfplots}
\usepackage{adjustbox}
\usetikzlibrary{arrows,matrix,positioning,fit}
\usetikzlibrary{shapes,positioning}
\usetikzlibrary{backgrounds}
\usepackage{tikz-layers}
\usepgfplotslibrary{fillbetween}
\usetikzlibrary{intersections}
\pgfplotsset{compat=1.14}
\usepgfplotslibrary{colorbrewer}
\usepgfplotslibrary{patchplots}
\usepgfplotslibrary[colorbrewer]
\usetikzlibrary{pgfplots.colorbrewer}
\usetikzlibrary[pgfplots.colorbrewer]
\usepgfplotslibrary{units}
\usetikzlibrary{spy}
\usepackage{pgfplotstable}
\usepackage{arrayjobx}
\graphicspath{ {./figs/} }
\usetikzlibrary{external}

\newlength\myheight
\newlength\mydepth
\settototalheight\myheight{Xygp}
\newcommand*\inlinegraphics[1]{%
  \settototalheight\myheight{Xygp}%
  \settodepth\mydepth{Xygp}%
  \raisebox{-\mydepth}{\includegraphics[height=\myheight]{#1}}%
}
\newcommand\orcid[1]{\href{https://orcid.org/#1}{\inlinegraphics{orcid_16x16.png}}}

\makeatletter
\def\BState{\State\hskip-\ALG@thistlm}
\makeatother

\newcommand{\half}{{\frac{1}{2}}}

\newcommand{\tref}{{\text{ref}}}

\colorlet{PlotColor1}{Spectral-A}
\colorlet{PlotColor2}{Spectral-L}
\colorlet{PlotColor3}{Spectral-O}
\colorlet{PlotColor4}{Spectral-D}

\begin{document}

\maketitle

\begin{abstract}
We explore the dynamics of the three-dimensional compressible Taylor--Green vortex from the perspective of kinetic theory by directly solving the six-dimensional Boltzmann equation. This work studies the connections between molecular-scale information encoded in the high-dimensional distribution function (e.g., molecular entropy measures) and macroscopic turbulent flow characteristics. We present high-order direct numerical simulations at Mach numbers of $0.5$ to $1.25$ and Reynolds numbers of $400$ to $1600$ performed using up to $137{\times}10^{9}$ degrees of freedom. The results indicate that the Kullback--Leibler divergence of the distribution function $f$ and its local equilibrium state $M[f]$ (i.e., the relative entropy functional $\langle f \log (f/M[f]) \rangle$) is strongly related to the macroscopic viscous dissipation rate, with the relative entropy value matching the sum of the solenoidal and dilatational dissipation rates very closely across a range of Mach and Reynolds numbers. Furthermore, we present the behavior of subgrid-scale quantities under spatial averaging/filtering operations performed directly on the particle distribution function, which shows notable similarities between the subgrid-scale dissipation and subgrid-scale relative entropy. These observations imply that information encoded in the distribution function, particularly certain measures of its deviation from equilibrium, may be useful for closure-modeling for compressible turbulence.

\end{abstract}


\section{Introduction}\label{sec:introduction}

The fundamental understanding of the mechanisms of transition to turbulence and turbulent flow behavior remains a significant challenge for both fundamental research and practical applications. The bulk of studies into fundamental turbulent flow dynamics have generally focused on incompressible (or weakly compressible) flow regimes. With increasing Mach number, the effects of compressibility and the coupling of thermodynamic and hydrodynamic behavior can cause significantly more complex vortex dynamics, including shock-vortex interactions and dilatational effects. The investigation of this behavior has shown increased interest over the years, partially driven by the difficulties of predicting aerothermodynamic behavior in hypersonic aeronautics and high-speed flight. 

However, as with the study of most turbulent flow behavior, the complexities of the Navier--Stokes equations governing fluid flow present a significant obstacle towards analysis of the underlying vortex dynamics, particularly so in regimes where compressibility effects and shocks are dominant. This analysis is largely hampered by the mathematical nature of the governing equations which consist of a coupled system of equations with nonlinear transport and diffusion. The inherently nonlinear nature of turbulence requires closure models for unresolved scales, and these models often rely on phenomenological assumptions or heuristic models. Furthermore, the Navier--Stokes framework is fundamentally limited to macroscopic fields, offering limited insight into the microscopic processes underpinning macroscopic fluid motion.

Kinetic theory, rooted in the Boltzmann equation, provides a complementary perspective to fluid dynamics that extends beyond the continuum framework. By describing the evolution of a particle distribution function in phase space, kinetic theory captures both macroscopic flow behavior and the underlying microscopic interactions. This multiscale perspective enables a more detailed description of turbulence as the solution of the Boltzmann equation provides significantly more information than the associated continuum equations (e.g., higher-order moments, molecular entropy measures, etc.), offering new avenues to study energy transfer, dissipation mechanisms, and fluctuations at scales inaccessible to traditional continuum approaches. For example, the chaotic and nonlinear behavior of compressible turbulence may be better understood when analyzed in terms of the evolution of a high-dimensional distribution function which may potentially represent such complex dynamics in a simpler manner~\citep{Chen2003}. Furthermore, the mathematical simplicity of the governing equation, consisting of the predominantly linear evolution of a scalar field, makes it more amenable to closure and subgrid-scale modeling~\citep{Chen1999}, which may offer insights into more appropriate models for the associated macroscopic governing equations (e.g., ~\citep{Girimaji2007, Ansumali2004}). 

As such, the primary objective of this work is to investigate the dynamics of compressible turbulence as encoded in the solution of the Boltzmann equation, with the aim of studying potential connections between the additional molecular-scale information it provides and insights relevant to understanding and analyzing compressible turbulence -- information that is inaccessible to traditional macroscopic models. While full kinetic simulations of high-speed three-dimensional turbulent flows have been performed (e.g., via stochastic~\citep{Gallis2017} or deterministic methods~\citep{Chen2020, Wilde2021}), limited analysis of the behavior of kinetic-scale quantities and their influence on macroscopic turbulent flow behavior has been conducted. In particular, this work focuses on studying two novel aspects of the kinetic representation of compressible turbulence. The first is the behavior of various molecular entropy measures (e.g., Boltzmann H-entropy), which act as Lyapunov functionals for the underlying dynamical system. This work aims to establish connections between these entropy measures and macroscopic turbulent flow characteristics such as viscous dissipation rates to provide a kinetic perspective to energy transfer mechanisms within turbulent flows. The second focus lies in analyzing subgrid-scale quantities (e.g., subgrid-scale dissipation) under spatial averaging/filtering operations \emph{performed directly on the particle distribution function}, with the goal of ultimately contributing to the future development of subgrid-scale closure models derived from kinetic theory. 

This objective is pursued through direct numerical simulations (DNS) of the three-dimensional compressible Taylor--Green vortex across a range of flow conditions, spanning from subsonic to supersonic and laminar to transitional/turbulent regimes. The Boltzmann equation is solved using a high-order flux reconstruction scheme paired with a discrete velocity model using a Bhatnagar--Gross--Krook (BGK)-type collision model~\citep{Bhatnagar1954}. Verification is first performed with respect to established Navier--Stokes results in the literature, followed by analysis of the aforementioned kinetic behavior of the flow. The remainder of this manuscript is organized as follows. An overview of the representative flow problem is given in \cref{sec:problem}, followed by details of the Boltzmann equation and its mathematical properties in \cref{sec:equations}. We then give a brief description of the numerical approach in \cref{sec:numerical}. The results of the numerical experiments and subsequent analysis are then shown in \cref{sec:results}, followed by conclusions in \cref{sec:conclusion}.


\section{Model problem}\label{sec:problem}
The model problem in this work is a compressible form of the three-dimensional Taylor--Green vortex~\citep{Taylor1937}. This flow problem, which consists of a periodic decaying vortex system, is a canonical configuration widely used to study the transition to turbulence, energy cascade mechanisms, and flow dynamics in laminar and turbulent regimes. In the laminar regime, viscous forces dominate, and the flow maintains its smooth structured nature with the decay of kinetic energy primarily driven by viscous dissipation. As the Reynolds number increases, the flow becomes unstable and transitions to turbulence, marked by vortex stretching and reconnection. The instabilities lead to the breakdown of large-scale vortices and the formation of small-scale flow structures, with energy dissipation dominated by these small-scale motions.

The Taylor--Green vortex is primarily studied in the incompressible or low Mach limit, where compressibility effects are minimal, such that the primary focus is on the interaction between viscous and inertial forces. However, with increasing Mach number, compressibility effects become significant, leading to additional phenomena like shocklets and strong variations in the thermodynamic fields. Energy cascade processes are influenced by these compressibility effects, and the dissipation of energy is affected by shocks and strong compressibility which induce nonlinear interactions between acoustic, vortical, and entropic modes~\citep{Unnikrishnan2019}. As such, the compressible Taylor--Green vortex serves as a more complex test case for studying high-speed flow phenomena.

In this work, we consider Reynolds numbers of 400 and 1600 and Mach numbers of 0.5 to 1.25, which span laminar to transitional/turbulent and subsonic to supersonic regimes. At the upper-end ($Re = 1600$, $M=1.25$), this flow configuration corresponds to a benchmark problem introduced by \citet{Lusher2021}, which has been used to study supersonic turbulent flow behavior and evaluate the accuracy and shock-resolving capabilities of high-fidelity numerical schemes (see, for example, \citet{Chapelier2024}, \citet{Witherden2024}, \citet{Wilde2021}). For this problem, the tri-periodic domain is set as $-\pi < x,y,z \leq \pi$, and the initial isothermal flow field is set as 
\begin{subequations}
    \begin{align}
        U_0(\mathbf{x}) &= U_\tref\sin(x)\cos(y)\cos(z), \\
        V_0(\mathbf{x}) &= -U_\tref\cos(x)\sin(y)\cos(z), \\
        W_0(\mathbf{x}) &= 0, \\
        P_0(\mathbf{x}) &= P_\tref + \frac{1}{16}\left[\cos(2x) + \cos(2y)\right] \left[2 +\cos(2z)\right], \\
        \rho_0(\mathbf{x}) &= \rho_\tref P_0(\mathbf{x})/P_\tref,
    \end{align}
\end{subequations}
where $\rho$ is the density, $\mathbf{U} = [U,V,W]^T$ are the macroscopic velocities, and $P$ is the pressure. The reference pressure is set as $P_\tref = 1/(\gamma M^2)$ to achieve the desired Mach number based on a unit reference velocity $U_\tref$ and density $\rho_\tref$ and a specific heat ratio of $\gamma = 1.4$. Furthermore, we define a scaled temperature as $\theta = P/\rho$, with its reference value taken as $\theta = P_\tref/\rho_\tref = P_\tref$. We additionally define the Reynolds number $Re = \rho_\tref U_\tref L_\tref/\mu_\tref$ based on a unit reference length $L_\tref$ and a reference dynamic viscosity $\mu_\tref$ (i.e., $Re = 1/\mu_\tref$). 

\section{Governing equations}\label{sec:equations}
The Boltzmann equation, which gives a deterministic description of molecular gas dynamics, can be given as
\begin{equation}\label{eq:boltzmann}
    \partial_t f (\mathbf{x}, \mathbf{u}, \zeta, t) + \mathbf{u} \cdot \nabla f = \mathcal C(f, f'),
\end{equation}
where $\mathbf{x} \in \mathbb R^3$ represents the physical space, $\mathbf{u} \in \mathbb R^3$ represents the molecular velocity space, $\zeta \in \mathbb R^+$ represents a scalar internal energy, $f (\mathbf{x}, \mathbf{u}, \zeta, t) \in \mathbb R$ is a scalar particle distribution function, and $\mathcal C(f, f')$ is some collision operator that accounts for intermolecular interactions~\citep{Cercignani1988}. The distribution function represents a probability measure for a particle existing at a given location $\mathbf{x}$ traveling at a given velocity $\mathbf{u}$ with an internal energy $\zeta$. The moments of this distribution function recover the macroscopic state of the system, i.e., \begin{equation}\label{eq:moments}
    \mathbf{Q}(\mathbf{x}, t) = \left[\rho, \mathbf{m}, E \right]^T = 
    \int_{\mathbb R^3} \int_{0}^{\infty} f (\mathbf{x}, \mathbf{u}, \zeta, t)\ \boldsymbol{\psi} (\mathbf{u}, \zeta) \ \mathrm{d}\zeta \mathrm{d}\mathbf{u} 
\end{equation}
where $\mathbf{m}$ is the momentum vector, $E$ is the total energy, and $\boldsymbol{\psi} (\mathbf{u}, \zeta) \coloneqq [1, \mathbf{u}, (\mathbf{u}\cdot\mathbf{u})/2 + \zeta]^T$ is the vector of collision invariants. From this, the macroscopic velocity components and pressure can be computed as $\mathbf{U} =  \mathbf{m}/\rho$ and $P = (\gamma - 1)(E - \rho\mathbf{U}\cdot\mathbf{U}/2)$, respectively. We use the notation $\mathbf{u}$ to denote the microscopic velocities and $\mathbf{U}$ to denote the macroscopic velocities. Furthermore, we sometimes use the notation $\langle \cdot \rangle$ to denote integration across the velocity and internal energy domains, e.g., 
\begin{equation}
    \langle f (\mathbf{x}, \mathbf{u}, \zeta, t) \rangle = \int_{\mathbb R^3} \int_{0}^{\infty} f (\mathbf{x}, \mathbf{u}, \zeta, t)\ \mathrm{d}\zeta \mathrm{d}\mathbf{u}.
\end{equation}

The collision operator is approximated using the Bhatnagar--Gross--Krook (BGK) model~\citep{Bhatnagar1954}, given as 
\begin{equation}
    \mathcal C (f, f') \approx \frac{\mathcal M[f] - f(\mathbf{x}, \mathbf{u}, \zeta, t)}{\tau},
\end{equation}
where $\mathcal M[f]$ is an equilibrium distribution function and $\tau$ is some collision time scale. The standard BGK model utilizes a Maxwell--Boltzmann distribution to approximate the equilibrium state $\mathcal M[f]$. To extend the model to non-unit Prandtl numbers and account for the effects of internal degrees of freedom, the equilibrium distribution function is computed through the ellipsoidal BGK (ES-BGK) model~\citep{Holway1966} in conjunction with the internal energy model of \citet{Baranger2020} as 
\begin{equation}
    \mathcal M[f] = M_{\mathbf{u}} (\mathbf{u})\times M_{\zeta}(\zeta),
\end{equation}
where
\begin{subequations}
    \begin{align}
        \mathcal M_{\mathbf{u}} &= \frac{\rho }{\sqrt{\text{det}(2 \pi \mathcal{T})}}\exp \left [-\frac{1}{2}\left[\mathbf{u} - \mathbf{U} \right]^T \mathcal{T}^{-1}\left[\mathbf{u} - \mathbf{U}\right]\right], \\
        \mathcal M_{\zeta} &= \Lambda(\delta) \left (\frac{\zeta}{\theta} \right)^{\frac{\delta}{2} - 1} \frac{1}{\theta}\exp \left(-\frac{\zeta}{\theta} \right).
    \end{align}
\end{subequations}

Here, the temperature tensor $\mathcal T$ is a convex combination of the macroscopic temperature $\theta$ and the density-normalized stress tensor $\sigma$, i.e.,
\begin{equation}\label{eq:stress}
    \mathcal{T} = \frac{1}{Pr} \theta \mathbf{I} + \left(1 - \frac{1}{Pr}\right) \frac{\sigma}{\rho}, \quad \text{where} \quad 
    \sigma_{ij} = \langle f c_i c_j \rangle,
\end{equation}
$\mathbf{I}$ is the identity matrix in $\mathbb{R}^{3\times 3}$, $\mathbf{c} = \mathbf{u} - \mathbf{U}$ is the peculiar velocity, $\delta = 2$ is the number of internal degrees of freedom, and $\Lambda (\delta) = 1/\Gamma(\delta/2)$, where $\Gamma$ is the gamma function. The collision time scale $\tau$ can be related to the dynamic viscosity, pressure, and Prandtl number $Pr$ as
\begin{equation}
    \tau = \frac{1}{Pr}\frac{\mu}{P}.
\end{equation}
In this work, the Prandtl number is set as $Pr = 0.71$, and the dynamic viscosity is computed using the Sutherland viscosity model~\citep{Sutherland1893}, given by the relation
\begin{equation}
    \mu = \mu_{\text{ref}}\left(\frac{\theta}{\theta_{\text{ref}}} \right)^{\frac{3}{2}}\frac{\theta_{\text{ref}} + \theta_S}{\theta + \theta_S},
\end{equation}
where $\theta_S$ denotes the (scaled) Sutherland temperature. The reference and Sutherland temperatures are set corresponding to dimensional values of $273K$ and $111.4K$, respectively (i.e., $\theta_S = \frac{111.4}{273}\theta_\tref$)

The symmetry of the full collision operator endows the Boltzmann model with a functional $H(f)$, defined as,
\begin{equation}
    H(f) =  \langle  f \log (f) \rangle,
\end{equation}
for which the inequality
\begin{equation}
    \partial_t H(f) + \mathbf{u} \cdot \nabla H(f) = \langle C(f, f') \log (f) \rangle = D(f) \leq 0
\end{equation}
holds locally. This differential inequality, formally known as Boltzmann's H-theorem, demonstrates that the Boltzmann model has an inherent irreversibility in the form of a local entropy dissipation for the molecular entropy $H(f)$~\citep{SaintRaymond2009}. We note here that the internal energy domain can be neglected for brevity since in the diatomic case the internal energy domain-integrated entropy is equivalent to the monatomic entropy~\citep{Baranger2020}. The H-theorem formalizes the notion that the (negative) entropy of a closed system is non-decreasing in time, driving the system toward thermodynamic equilibrium. It can be seen that the (negative) entropy acts as a Lyapunov functional for the Boltzmann equation, with the dynamical system in equilibrium when $f = \mathcal M[f]$, i.e., when $D(f) = 0$. As such, it is plausible that the behavior of this functional encodes useful information regarding the underlying dynamics and behavior of turbulent fluid flows. In particular, the notion of a measure of non-equilibrium (i.e., the deviation of $f$ from $\mathcal M[f]$) is of interest as these non-equilibrium effects are what drive viscous interactions in the flow. One form of quantifying this non-equilibrium is encoded in the local entropy dissipation rate $D(f)$. 

While the standard Boltzmann entropy $H(f)$ functional is well-studied in the context of statistical mechanics, it is also possible to define two additional functionals: a (negative) cross-entropy, given as
\begin{equation}
    H(f\ |\ \mathcal M[f]) :=  \langle f \log \mathcal M[f] \rangle, 
\end{equation}
and a (negative) relative entropy, given in the form of a Kullback--Leibler divergence as 
\begin{equation}
H_{KL}(f\ |\ \mathcal M[f]) =  \langle f \log (f/\mathcal M[f]) \rangle =  H(f) - H( f\ |\ \mathcal M[f]) \ge 0. 
\end{equation}
The latter is convex in the space of the distribution functions, i.e., 
\begin{equation}
    H_{KL}(\lambda f_1 + (1 - \lambda) f_2\ |\ \lambda \mathcal M[f_1] + (1 -\lambda) \mathcal M[f_2] ) \le \lambda H_{KL} ( f_1\ |\ \mathcal M[f_1]) + ( 1- \lambda) H_{KL} (f_2\ |\ \mathcal M[f_2]),
\end{equation}
for any $0 \leq \lambda \leq 1$. Moreover, the following Gibbs' inequality holds,
\begin{equation}
    H(f \ | \ \mathcal M[f]) \leq  H(f).
\end{equation}
We can therefore define an associated relative entropy dissipation rate as
\begin{equation}\label{eq:reldis}
    D_{KL}(f\ |\  \mathcal M[f]) = \left \langle C(f,f') \log\ (f/ \mathcal M[f])\ \right\rangle,
\end{equation}
which for relaxation-type collision models reduces to 
\begin{equation}
    D_{KL}(f\ |\  \mathcal M[f]) \approx \left \langle \frac{\mathcal M[f] - f}{\tau} \log\ (f/ \mathcal M[f])\ \right\rangle.
\end{equation}
We note here that certain properties of the relative entropy shown above are known for Gaussian relaxation models (e.g., the monatomic BGK model), and whether this behavior extends to more complex relaxation models (e.g., polyatomic ellipsoidal models) is an open problem. The idea of analyzing the behavior of a relative entropy initially stems from the work of \citet{Yau1991} for Ginzburg--Landau models. Extensions to the Boltzmann equation were performed in the form of a normalization of the distribution function $f$ by some equilibrium state (see, for example, \citet{Lions2001} and \citet{SaintRaymond2009}), although not necessarily by the local equilibrium state $\mathcal M[f]$. To the authors' knowledge, the use of the local equilibrium state $\mathcal M[f]$ for normalization was first formally introduced in \citet{Golse2014}.

The cross-entropy functional $H(f\ |\ \mathcal M[f]) $ is of particular interest for relaxation-type collision approximations such as the BGK model as it  represents, in some sense, the ``distance'' of the distribution function from its equilibrium, which is precisely what a relaxation operator attempts to minimize. While also a Lyapunov functional for the underlying dynamical system, the relative entropy exhibits some additional properties which are not present in the standard Boltzmann entropy. First, the relative entropy functional is identically zero at equilibrium ($f = \mathcal M[f]$), such that it becomes a more appropriate measure of non-equilibrium as $H_{KL}(f\ |\ \mathcal M[f]) = 0$ if and only if $f$ is in equilibrium. This is not the case for the Boltzmann entropy itself, only the dissipation rate $D(f)$. Second, the local relative entropy dissipation rate $D_{KL}(f\ |\ \mathcal M[f])$ is symmetric with respect to $f$ and $\mathcal M[f]$ for a BGK-type model and strictly positive for all $\mathbf{u}$, which is also not the case for the local Boltzmann entropy production rate $D(f)$.  
For these reasons, we  study both of these entropy measures in the context of their relation to the dynamics of the macroscopic flow field. 

\section{Numerical approach}\label{sec:numerical}
The numerical approach in this work consists of a high-order unstructured flux reconstruction scheme~\citep{Huynh2007} for the spatial domain paired with a discretely conservative/entropy-satisfying discrete velocity model~\citep{Mieussens2000} for the molecular velocity domain. This approach broadly follows the scheme initially introduced in \citet{Dzanic2023} with the minor modifications used in \citet{Dzanic2024POF, Dzanic2024AIAA, Dzanic2024CUFS}. We present here a brief overview of the numerical approach, but for more in-depth details, we refer the reader to the referenced works. 

\subsection{Discretization}

The numerical discretization consists of three components for the internal energy, velocity, and spatial domains. Beginning with the internal energy domain, we use a reduced distribution approach (see \citet{Baranger2020}, Section 4.5), which allows for efficiently modeling the internal energy domain through a coupled set of scalar fields $[F(\mathbf{x}, \mathbf{u}, t), G(\mathbf{x}, \mathbf{u}, t)]^T$ as 
\begin{equation}
    \partial_t 
    \begin{bmatrix}
        F(\mathbf{x}, \mathbf{u}, t) \\
        G(\mathbf{x}, \mathbf{u}, t)
    \end{bmatrix} 
    + \mathbf{u}\cdot\nabla
    \begin{bmatrix}
        F(\mathbf{x}, \mathbf{u}, t) \\
        G(\mathbf{x}, \mathbf{u}, t)
    \end{bmatrix} 
    = \frac{1}{\tau}
    \begin{bmatrix}
        \hphantom{\frac{\delta \theta}{2}}\mathcal M_{\mathbf{u}}[F] - F(\mathbf{x}, \mathbf{u}, t)  \\
        \frac{\delta \theta}{2} M_{\mathbf{u}}[F] - G(\mathbf{x}, \mathbf{u}, t)
    \end{bmatrix},
\end{equation}
where
\begin{subequations}
    \begin{align}
        F(\mathbf{x}, \mathbf{u}, t) &=  \int_{0}^{\infty} f (\mathbf{x}, \mathbf{u}, \zeta, t) \ \mathrm{d}\zeta , \\
        G(\mathbf{x}, \mathbf{u}, t) &= \int_{0}^{\infty} \zeta f (\mathbf{x}, \mathbf{u}, \zeta, t) \ \mathrm{d}\zeta.
    \end{align}
\end{subequations}
The velocity domain $\Omega^{\mathbf{u}}$ is approximated via a discrete velocity model, consisting of $N_u$ fixed discrete velocity nodes $\mathbf{u}_i \in \Omega^{\mathbf{u}}$ for $i \in \lbrace 1, \ldots, N_u\rbrace$. This reduces the coupled six dimensional transport equation to a set of $N_u$ three dimensional transport equations for each scalar field, i.e.,
\begin{equation}\label{eq:transport}
    \partial_t 
    \begin{bmatrix}
        F_i(\mathbf{x}, t) \\
        G_i(\mathbf{x}, t)
    \end{bmatrix} 
    + \mathbf{u}_i\cdot\nabla
    \begin{bmatrix}
        F_i(\mathbf{x}, t) \\
        G_i(\mathbf{x}, t)
    \end{bmatrix} 
    = \frac{1}{\tau}
    \begin{bmatrix}
        \hphantom{\frac{\delta \theta}{2}}g^\mathbf{u}_i (\mathbf{x}, t) - F_i(\mathbf{x}, t)  \\
        \frac{\delta \theta}{2} g^\mathbf{u}_i (\mathbf{x}, t) - G_i(\mathbf{x}, t)
    \end{bmatrix} \quad \forall \ i \in \lbrace 1, \ldots, N_u\rbrace,
\end{equation}
These three-dimensional transport equations are solved using the flux reconstruction scheme. For this scheme, the spatial domain $\Omega^{\mathbf{x}}$ is partitioned into $N_e$ elements $\Omega^{\mathbf{x}}_k $ such that $\Omega^{\mathbf{x}} = \bigcup_{N_e}\Omega^{\mathbf{x}}_k$ and $\Omega^{\mathbf{x}}_i\cap\Omega^{\mathbf{x}}_j=\emptyset$ for $i\neq j$. Each distribution function component within each element is approximated by an order $\mathbb P_p$ polynomial as
\begin{equation}
    F_i (\mathbf{x}) = \sum_{j = 1}^{N_s} F_{i,j} {\phi}_j (\mathbf{x}), \quad G_i (\mathbf{x}) = \sum_{j = 1}^{N_s} G_{i,j} {\phi}_j (\mathbf{x}),
\end{equation}
where $F_{i,j}$ and $G_{i,j}$ are sets of $N_s$ basis coefficients for all $i \in \{1,..., N_u\}$ and ${\phi}_j (\mathbf{x})$ is the corresponding set of nodal polynomial basis functions of maximal order $p$. The spatial gradient operator in \cref{eq:transport} is approximated via the standard flux reconstruction methodology for linear transport equations with upwind-biasing at the interfaces. Furthermore, we enforce discrete positivity in the distribution function via the positivity-preserving limiter of \citet{Zhang2010}, but no explicit numerical shock capturing is applied.

To ensure discrete conservation of the macroscopic conserved variables, we utilize the discrete velocity model of \citet{Mieussens2000} for approximating the equilibrium distribution function in the collision operator. Via a local nonlinear optimization problem at each spatial node $j$, we solve for a modified equilibrium distribution function $\mathcal M_{*,i}$ which satisfies the discrete compatibility condition, i.e.,
\begin{equation}
    \sum_i^{N_u} m_i F_{i,j} \boldsymbol{\psi}(\mathbf{u}_i) = \sum_i^{N_u} m_i \mathcal M_{*,i} \boldsymbol{\psi}(\mathbf{u}_i),
\end{equation}
and minimizes the discrete entropy, i.e.,
\begin{equation}
    \sum_i^{N_u} m_i F_{i,j} \log \left (F_{i,j} \right),
\end{equation}
where $\mathbf{m}$ is a discrete integration operator such that 
\begin{equation}
    \sum_i^{N_u} m_i F_{i,j} \approx \int_{\mathbb R^3} F_j \ \mathrm{d}\mathbf{u}.
\end{equation}

\subsection{Computational setup}
The numerical experiments consisted of periodic cubic spatial domains which were discretized using uniform hexahedral meshes with $\mathbb P_3$ approximations, resulting in a nominally fourth-order accurate spatial scheme. For the $Re = 400$ cases, we use $N_e = 32^3$ elements, resulting in $256^3$ spatial degrees of freedom, whereas for the $Re = 1600$ cases, we use $N_e = 64^3$ elements, resulting in $512^3$ spatial degrees of freedom. These resolution levels were taken from relevant simulations in the literature which showcase sufficient convergence of the relevant quantities of interest in this work (e.g., ~\citep{Lusher2021,Chapelier2024,Peng2018}. For all cases, we use a uniform Cartesian velocity domain with $N_u = 8^3$ degrees of freedom and a trapezoidal quadrature rule. Overall, the simulations consisted of $17{\times}10^9$ total degrees of freedom for the $Re = 400$ cases and $137{\times}10^9$ total degrees of freedom for the $Re = 1600$ cases. Computations were performed on Frontier at the Oak Ridge Leadership Computing Facility using up to 256 AMD MI250X GPUs. 

At initialization, the solution was set to the equilibrium distribution function corresponding to the macroscopic flow initial conditions. Time-stepping was performed using an explicit four-stage, fourth-order Runge--Kutta scheme, with a maximum time step limited by the minimum of the advective time step constraint and the collision time. The advective time step constraint was computed with a CFL factor of approximately 0.5 based on the maximum discrete particle velocity. The velocity domain was set as $\Omega^{\mathbf{u}} = [-u_{\max}, u_{\max}]^3$, where the extent was computed as 
\begin{equation}
    u_{\max} = U_\tref + k \sqrt{\theta_\tref}.
\end{equation}
In this work, we use $k = 3$, for which the truncated velocity domain encompasses at least $99.7\%$ of the initial distribution function.  

\section{Results}\label{sec:results}
In this section, we first present a verification of the present model against reference Navier--Stokes results in the literature at $Re = 1600$ and $M=1.25$ to demonstrate consistency with established benchmarks. Next, we present the results of the numerical experiments, starting with an analysis of the macroscopic flow characteristics, followed by insights derived from the kinetic representation. Finally, we examine the behavior of the kinetic distribution function in the context of subgrid-scale modeling.

\subsection{Verification}
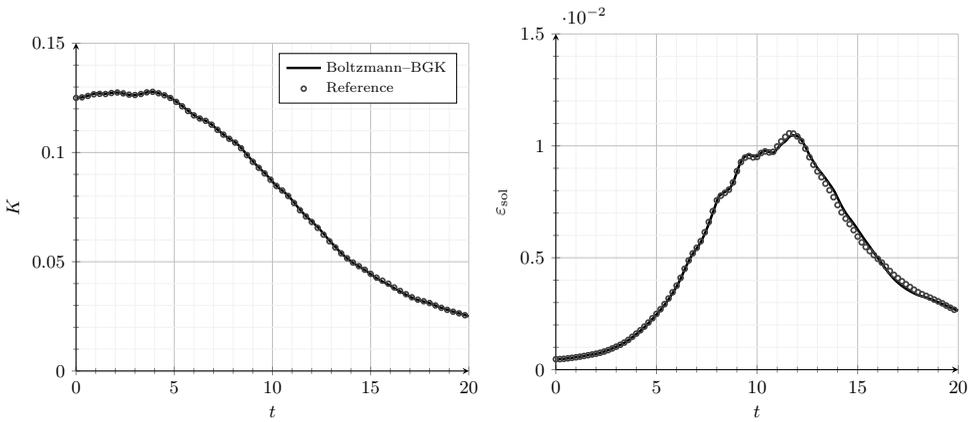
\begin{figure}
        \centering
        \subfloat[Kinetic energy]{\adjustbox{width=0.48\linewidth, valign=b}{\begin{tikzpicture}[spy using outlines={rectangle, height=3cm,width=2.5cm, magnification=3, connect spies}]
    \begin{axis}
    [
        axis line style={latex-latex},
        axis y line=left,
        axis x line=left,
        clip mode=individual,
        xmode=linear, 
        ymode=linear,
        xlabel = {$t$},
        ylabel = {$K$},
        xmin = 0, xmax = 20,
        ymin = 0.0, ymax = 0.15,
        legend cell align={left},
        legend style={font=\scriptsize, at={(0.97, 0.97)}, anchor=north east},
        x tick label style={/pgf/number format/.cd, fixed, fixed zerofill, precision=0, /tikz/.cd},
        y tick label style={/pgf/number format/.cd, fixed, precision=2, /tikz/.cd},	
        grid=both,
        grid style={line width=.1pt, draw=gray!10},
        major grid style={line width=.2pt,draw=gray!50},
        minor x tick num=4,
        minor y tick num=4,
    ]
        
        

        
        
        
        
        \addplot[color=black, style={very thick}] table[x expr=\thisrow{t}, y=ke, col sep=comma]{./figs/data/bgktgv-M1p25-Re1600-p3-n128-post.csv};
        \addlegendentry{Boltzmann--BGK};
        
        \addplot[color=black!75, style={thick}, only marks, mark=o, mark options={scale=0.6}, mark repeat = 6, mark phase = 0] 
        table[x expr=\thisrow{t}, y=ke, col sep=comma, mark=*]{./figs/data/lusher_m1p25.csv};
        \addlegendentry{Reference};

        
        
        

    \end{axis}
\end{tikzpicture}}}
        \subfloat[Solenoidal dissipation]{\adjustbox{width=0.48\linewidth, valign=b}{\begin{tikzpicture}[spy using outlines={rectangle, height=3cm,width=2.5cm, magnification=3, connect spies}]
    \begin{axis}
    [
        axis line style={latex-latex},
        axis y line=left,
        axis x line=left,
        clip mode=individual,
        xmode=linear, 
        ymode=linear,
        xlabel = {$t$},
        ylabel = {$\varepsilon_{\text{sol}}$},
        xmin = 0, xmax = 20,
        ymin = 0.0, ymax = 0.015,
        legend cell align={left},
        legend style={font=\scriptsize, at={(0.97, 0.97)}, anchor=north east},
        x tick label style={/pgf/number format/.cd, fixed, fixed zerofill, precision=0, /tikz/.cd},
        scale = 1,
        grid=both,
        grid style={line width=.1pt, draw=gray!10},
        major grid style={line width=.2pt,draw=gray!50},
        minor x tick num=4,
        minor y tick num=4,
    ]

        \addplot[color=black, style={very thick}] table[x=t, y=sol, col sep=comma]{./figs/data/bgktgv-M1p25-Re1600-p3-n128-post.csv};

        \addplot[color=black!75, style={thick}, only marks, mark=o, mark options={scale=0.6}, mark repeat = 4, mark phase = 0] 
        table[x=t, y=soldis, col sep=comma, mark=*]{./figs/data/lusher_m1p25.csv};

    \end{axis}
\end{tikzpicture}}}
        \caption{Temporal evolution of the kinetic energy (left) and solenoidal dissipation rate (right) for the $Re = 1600$, $M = 1.25$ case. }
        \label{fig:valid}  
    \end{figure}

To show that the kinetic approach predicts the correct hydrodynamic behavior with respect to the Navier--Stokes equations, we present a verification with respect to the DNS results of \citet{Lusher2021} at the highest Reynolds and Mach number ($Re = 1600$, $M=1.25$). The domain-averaged kinetic energy, computed as
\begin{equation}
    K = \frac{1}{\rho_\tref (2 \pi)^3 }\int_{\Omega^{\mathbf{x}}} \half \rho \mathbf{U} \cdot \mathbf{U} \ \mathrm{d}\mathbf{x},
\end{equation}
and the solenoidal component of the viscous dissipation rate, computed as
\begin{equation}
    \varepsilon_{\text{sol}} = \frac{1}{\rho_\tref (2 \pi)^3 }\int_{\Omega^{\mathbf{x}}} \mu \boldsymbol{\omega} \cdot \boldsymbol{\omega} \ \mathrm{d}\mathbf{x},
\end{equation}
where $\boldsymbol{\omega} = \boldsymbol{\nabla} \times \mathbf{U}$ is the vorticity, are shown in comparison to the reference results in \cref{fig:valid}. The predicted kinetic energy profiles were effectively identical, with no visually discernible deviations between the two. Furthermore, the predicted solenoidal dissipation profiles were also in excellent agreement, with both results showing nearly identical behavior prior to the dissipation peak ($t \approx 12$) and only minor deviations afterward.

   \begin{figure}
        \centering
        \subfloat[Mach number cross-section]{\adjustbox{width=0.48\linewidth, valign=b}{\begin{tikzpicture}[spy using outlines={rectangle, height=3cm,width=2.5cm, magnification=3, connect spies}]
    \begin{axis}
    [
        axis line style={latex-latex},
        axis y line=left,
        axis x line=left,
        clip mode=individual,
        xmode=linear, 
        ymode=linear,
        xlabel = {$y/L$},
        ylabel = {$M$},
        xmin = 0, xmax = 3.142,
        ymin = 0.0, ymax = 1.6,
        legend cell align={left},
        legend style={font=\scriptsize, at={(0.97, 0.97)}, anchor=north east},
        ytick = {0,0.4, 0.8, 1.2, 1.6},
        xtick = {0, 0.78539, 1.57079, 2.35619, 3.14159},
        xticklabels={0, $\sfrac{\pi}{4}$, $\sfrac{\pi}{2}$, $\sfrac{3\pi}{4}$, $\pi$},
        grid=both,
        grid style={line width=.1pt, draw=gray!10},
        major grid style={line width=.2pt,draw=gray!50},
        minor x tick num=3,
        minor y tick num=3,
    ]

        \addplot[color=black, style={very thick}, each nth point=10, filter discard warning=false,] table[x=y, y=mach, col sep=comma]{./figs/data/machprofile.csv};
        \addlegendentry{Boltzmann--BGK};
        
        \addplot[color=black!75, style={thick}, only marks, mark=o, mark options={scale=0.6}, mark repeat = 3, mark phase = 0] 
        table[x=y, y=mach, col sep=comma, mark=*]{./figs/data/ref_machprofile.csv};
        \addlegendentry{Reference};

    \end{axis}
\end{tikzpicture}}}
        \subfloat[Dilatational dissipation]{\adjustbox{width=0.48\linewidth, valign=b}{\begin{tikzpicture}[spy using outlines={rectangle, height=3cm,width=2.5cm, magnification=3, connect spies}]
    \begin{axis}
    [
        axis line style={latex-latex},
        axis y line=left,
        axis x line=left,
        clip mode=individual,
        xmode=linear, 
        ymode=linear,
        xlabel = {$t$},
        ylabel = {$\varepsilon_{\text{dil}}$},
        xmin = 0, xmax = 20,
        ymin = 0.0, ymax = 0.0012,
        legend cell align={left},
        legend style={font=\scriptsize, at={(0.97, 0.97)}, anchor=north east},
        ytick = {0,0.0004,0.0008,0.0012},
        x tick label style={/pgf/number format/.cd, fixed, fixed zerofill, precision=0, /tikz/.cd},
        grid=both,
        grid style={line width=.1pt, draw=gray!10},
        major grid style={line width=.2pt,draw=gray!50},
        minor x tick num=4,
        minor y tick num=3,
        reverse legend,
    ]

        \addplot[color=PlotColor1!80, style={thick}, only marks, mark=o, mark options={scale=0.6}, mark repeat = 1, mark phase = 0] 
        table[x=t, y=dildis, col sep=comma]{./figs/data/OpenSBLI_TENO6_N2048_M125_Re1600.csv};
        \addlegendentry{Reference ($2048^3$)};    

        \addplot[color=PlotColor2!80, style={thick}, only marks, mark=o, mark options={scale=0.6}, mark repeat = 1, mark phase = 0] 
        table[x=t, y=dildis, col sep=comma, mark=*]{./figs/data/OpenSBLI_TENO6_N1024_M125_Re1600.csv};
        \addlegendentry{Reference ($1024^3$)};
        
        \addplot[color=PlotColor3!80, style={thick}, only marks, mark=o, mark options={scale=0.6}, mark repeat = 1, mark phase = 0] 
        table[x=t, y=dildis, col sep=comma, mark=*]{./figs/data/OpenSBLI_TENO6_N512_M125_Re1600.csv};
        \addlegendentry{Reference ($512^3$)};
        
        \addplot[color=black, style={very thick}] table[x=t, y=dil, col sep=comma]{./figs/data/bgktgv-M1p25-Re1600-p3-n128-post.csv};
        \addlegendentry{Boltzmann--BGK};
    \end{axis}
\end{tikzpicture}}}
        \caption{Mach number cross-section on the line $x,z = \pi$ at $t=2.5$ (left) and temporal evolution of the dilatational dissipation rate (right) for the $Re = 1600$, $M = 1.25$ case.}
        \label{fig:valid2}  
    \end{figure}
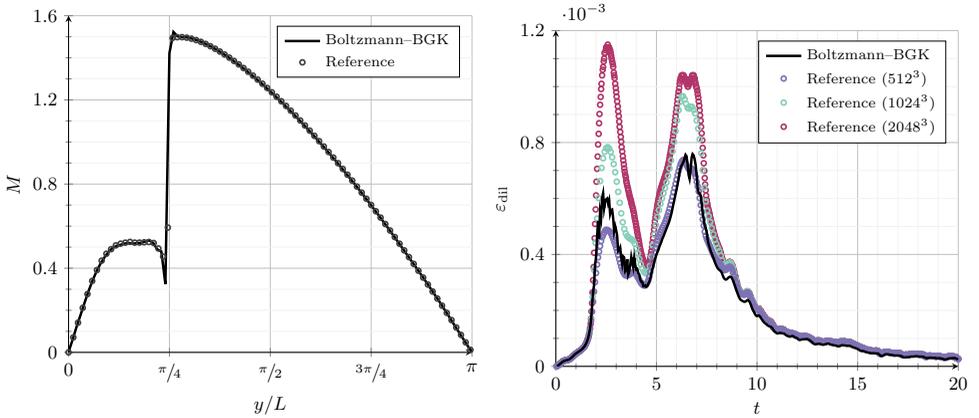
    
As a further verification of the shock-resolving capabilities of the high-order kinetic approach, we compare a cross-section of the local Mach number at $t=2.5$ on the line $x,z = \pi$ to the reference results in \cref{fig:valid2}. It can be seen that even without any explicit numerical shock capturing approach, the kinetic scheme shows very good resolution of the initial shock formation, with negligible overshoots/undershoots in the immediate vicinity of the shock. A more quantitative comparison is also shown in \cref{fig:valid2} in form of the dilatational dissipation rate, computed as
\begin{equation}
    \varepsilon_{\text{dil}} = \frac{4}{3\rho_\tref (2 \pi)^3 }\int_{\Omega^{\mathbf{x}}} \mu \left (\boldsymbol{\nabla} \cdot \mathbf{U} \right)^2 \ \mathrm{d}\mathbf{x},
\end{equation}
with respect to the convergence studies presented in \citet{Chapelier2024} using the identical numerical approach of \citet{Lusher2021}. We note here that the reference results on the $512^3$ grid are identical to the results in \citet{Lusher2021}. The present work shows dilatational dissipation results similar to the reference approach at the same resolution level, with marginally better prediction of the magnitude of the first dilatational dissipation peak, albeit with some minor oscillations. For this field, which is dependent on the divergence of velocity, mesh convergence is not achieved even at resolution levels of $2048^3$ as convergence is expected only when the shock thickness can be fully resolved by the numerical scheme. Nevertheless, the behavior is consistent with previous studies, demonstrating that the present model captures the general trends and physical characteristics of the dilatational dissipation accurately.

    \begin{figure}
        \centering
        \vspace{0.5em}
        \subfloat[$t = 5$]{
        \adjustbox{width=0.54\linewidth, valign=b}{\includegraphics[]{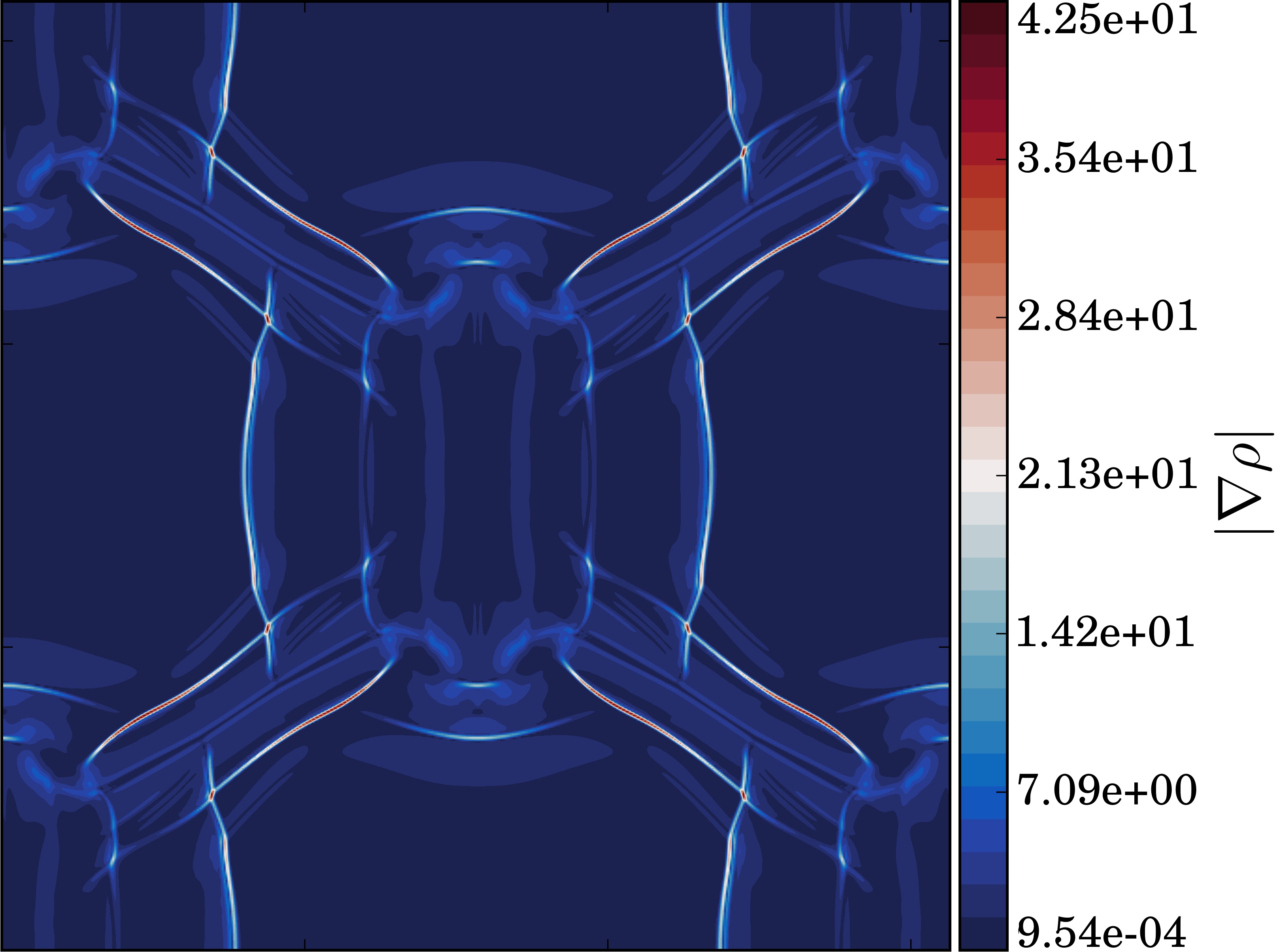}}
        \adjustbox{width=0.4\linewidth, valign=b}{\includegraphics[]{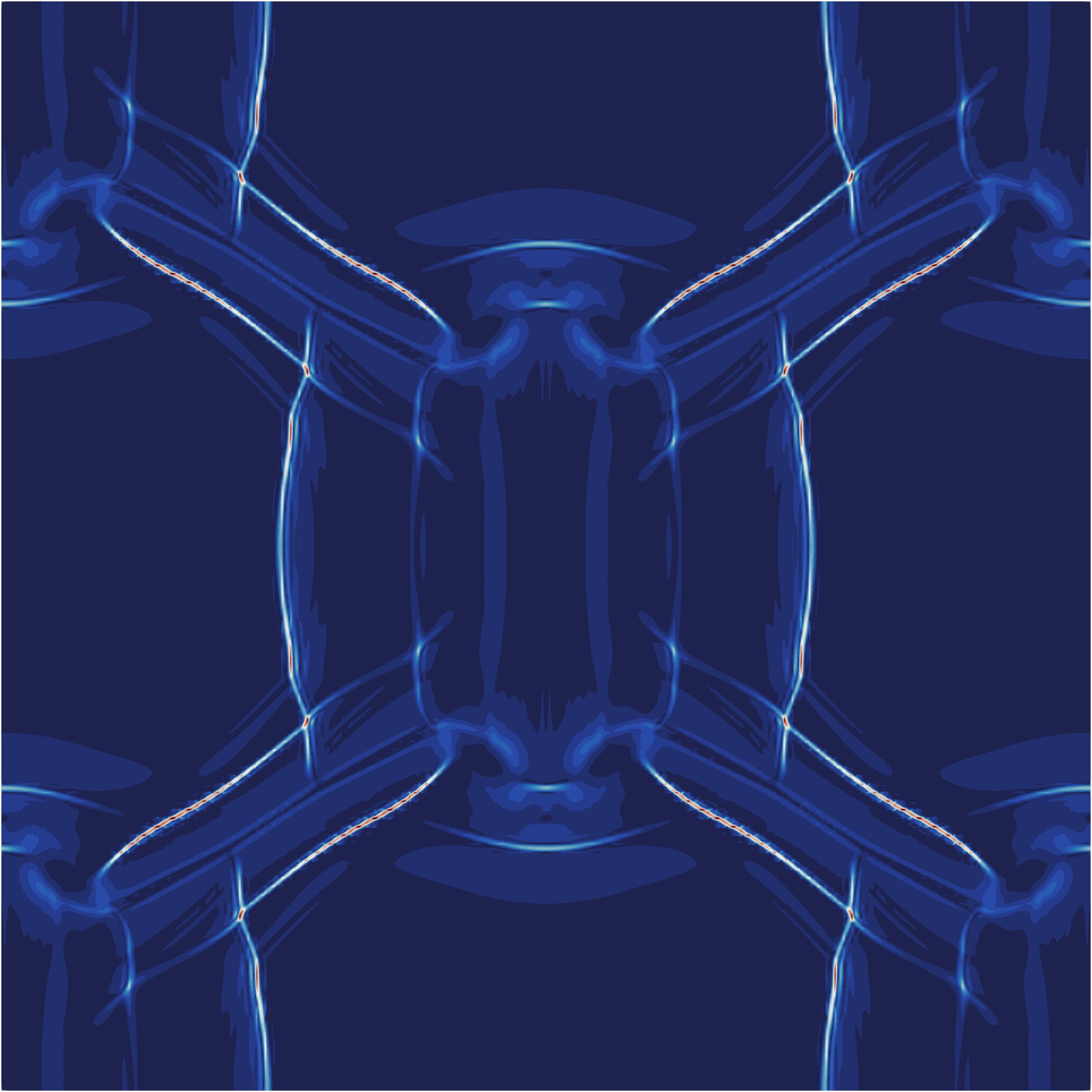}}}
        \vspace{0.5em}
        \newline
        \vspace{0.5em}
        \subfloat[$t = 7.5$]{
        \adjustbox{width=0.54\linewidth, valign=b}{\includegraphics[]{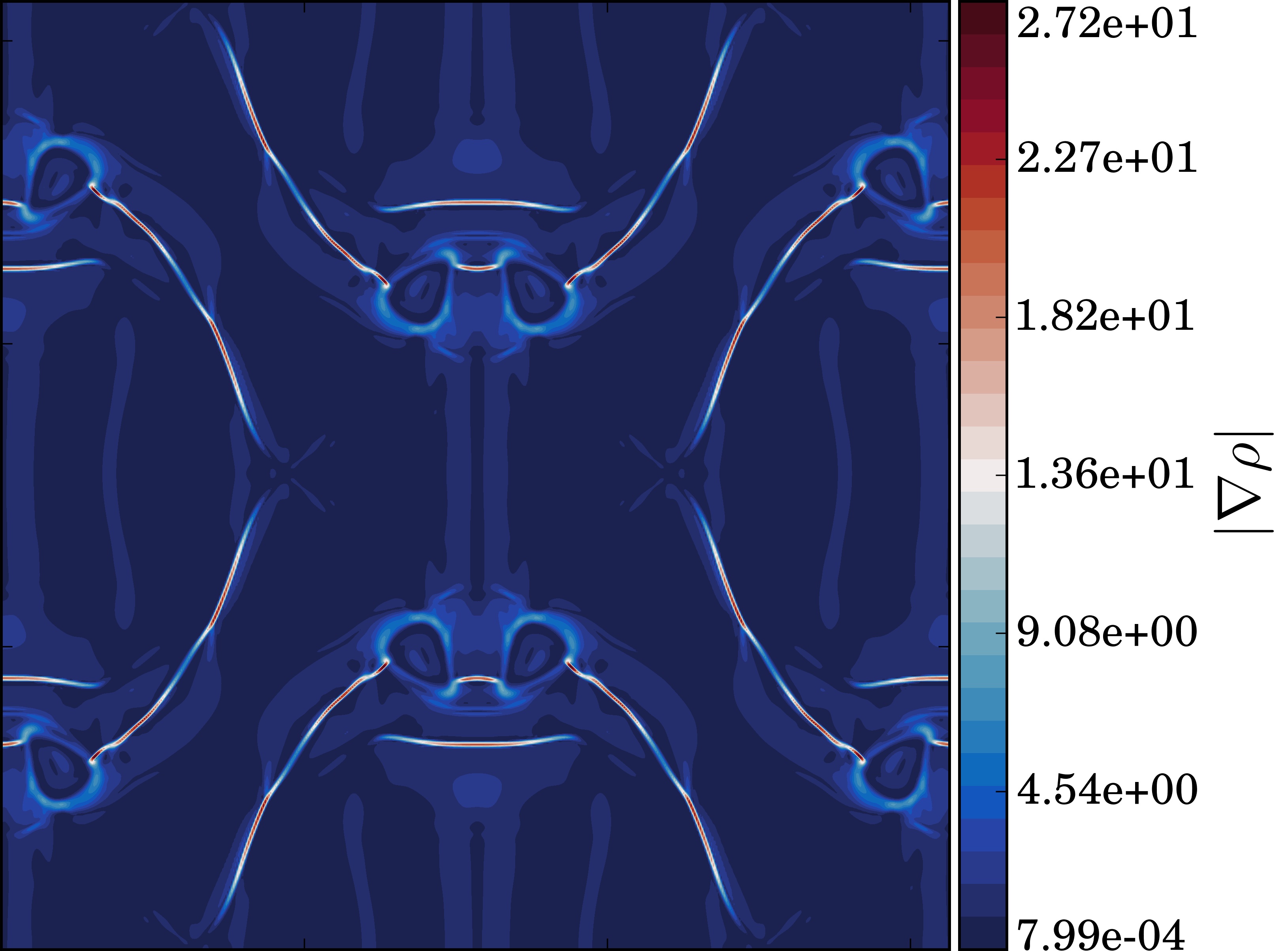}}
        \adjustbox{width=0.4\linewidth, valign=b}{\includegraphics[]{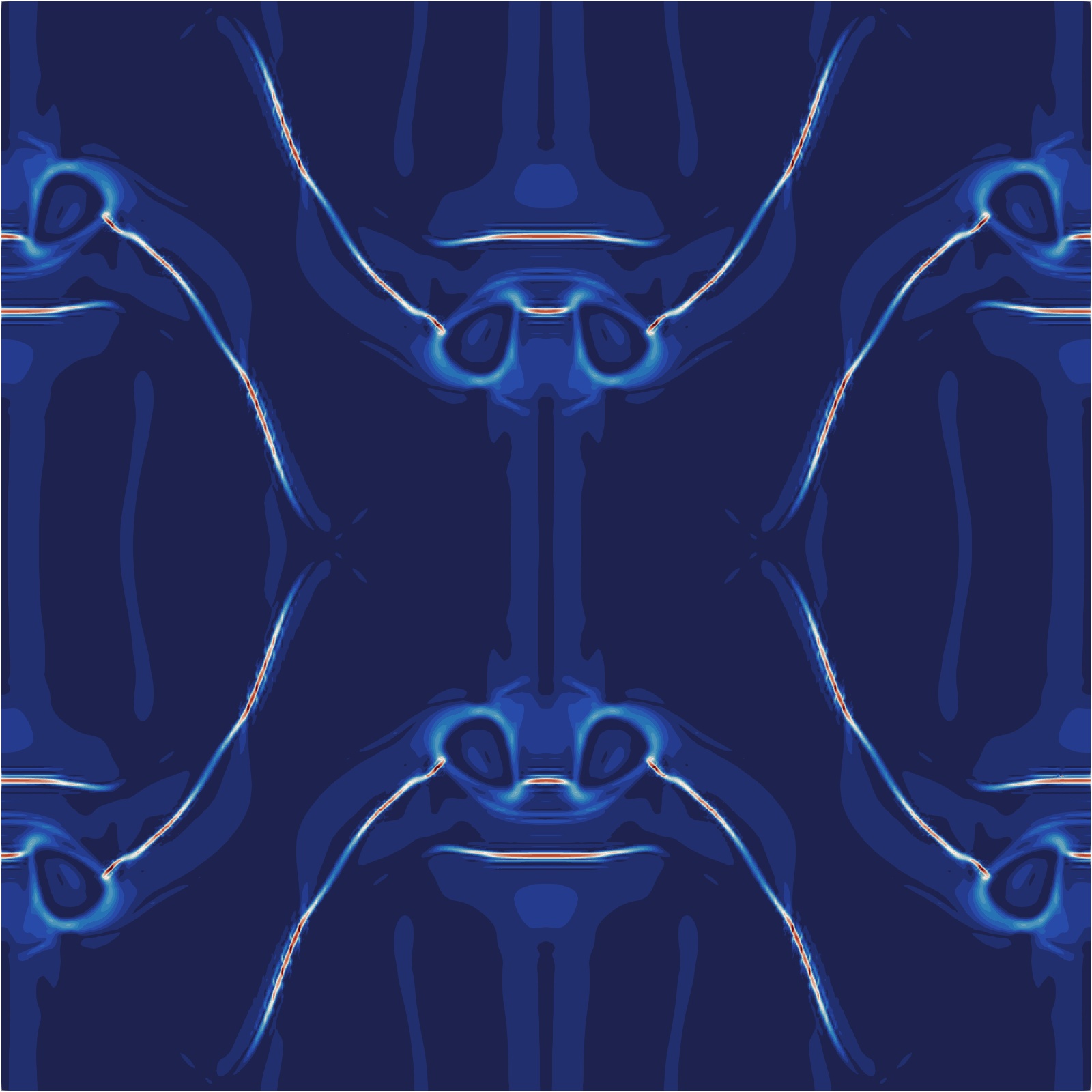}}}
        \newline
        \vspace{0.5em}
        \subfloat[$t = 10$]{
        \adjustbox{width=0.54\linewidth, valign=b}{\includegraphics[]{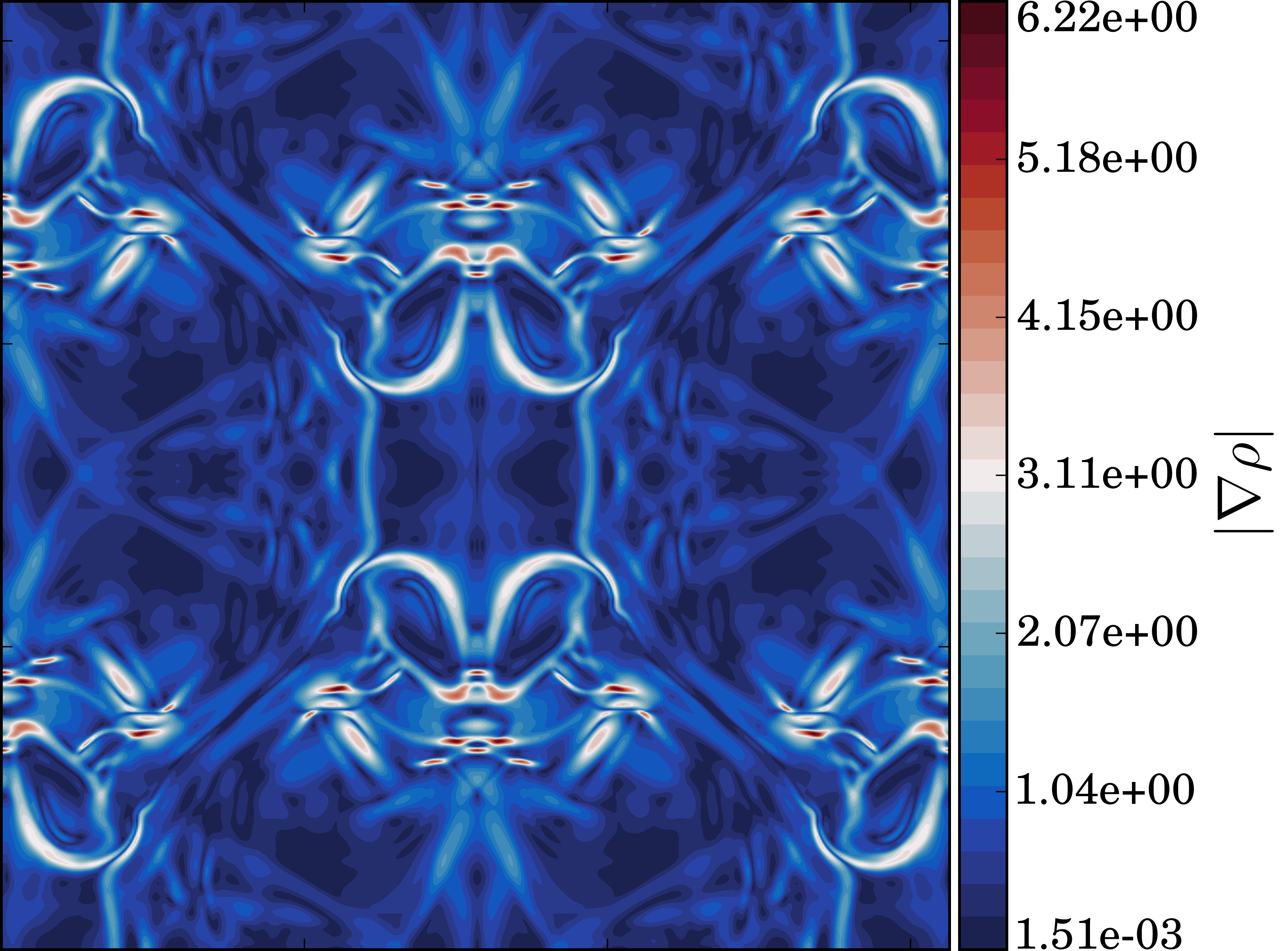}}
        \adjustbox{width=0.4\linewidth, valign=b}{\includegraphics[]{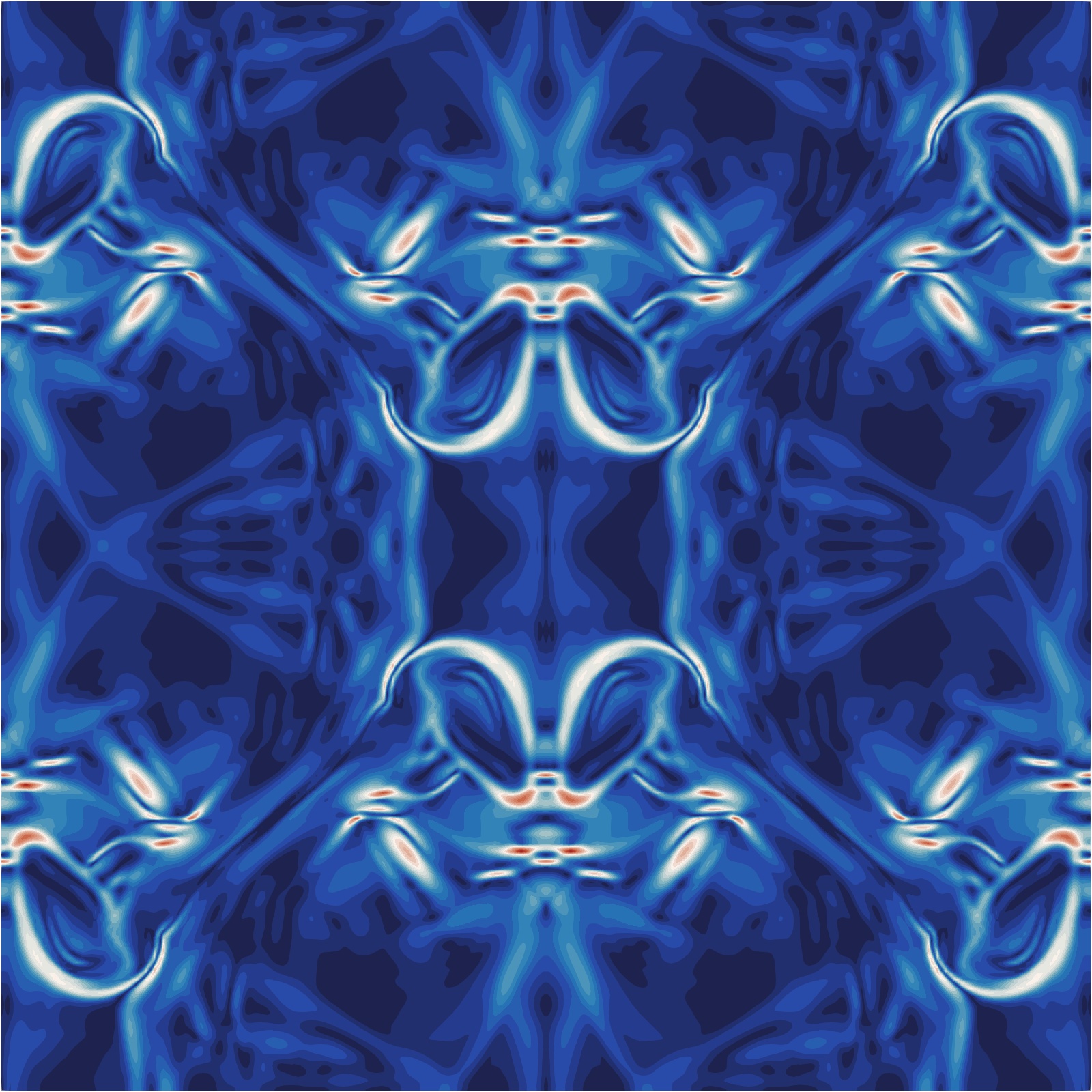}}}
        \newline
        \caption{Comparison of the density gradient norm on the plane $y = \pi$ at varying times between the results of \citet{Lusher2021} (left) and the present work (right) for the $Re = 1600$, $M = 1.25$ case. }\label{fig:validschlieren} 
    \end{figure}

A final comparison was performed with respect to the predicted flow structures at shock and vortex formation times. A Schlieren-type representation of the density gradient norm $\| \nabla \rho \|_2$ on the plane $y = \pi$ is shown at varying times in \cref{fig:validschlieren}. We note here that while effort was made to ensure that the predicted legends were identical between the reference results and the present work, minor discrepancies may be present. Despite the fact that the present work utilizes a different numerical to solve an entirely separate governing equation, the two sets of results were markedly similar, showing the initial formation of normal shocks and the rollup of distinct vortical structures interacting with these shocks. Even after the formation of shocks and vortices in the flow, the predicted flow fields were nearly identical, with only minor visible deviations at shock fronts and in the contour lines at later times. The results indicate that the kinetic scheme used in this work can appropriately model the underlying flow physics in the hydrodynamic limit.

\subsection{Macroscopic flow characteristics}
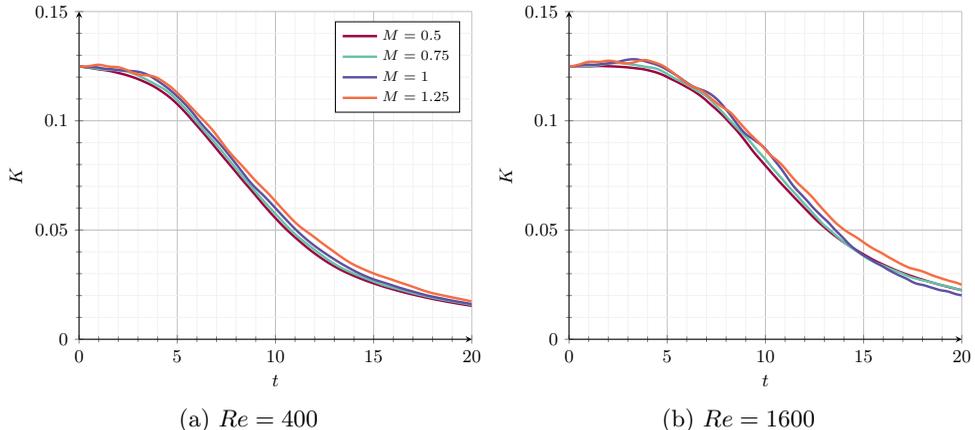
\begin{figure}
        \centering
        \subfloat[$Re = 400$]{\adjustbox{width=0.48\linewidth, valign=b}{\begin{tikzpicture}[spy using outlines={rectangle, height=3cm,width=2.5cm, magnification=3, connect spies}]
    \begin{axis}
    [
        axis line style={latex-latex},
        axis y line=left,
        axis x line=left,
        clip mode=individual,
        xmode=linear, 
        ymode=linear,
        xlabel = {$t$},
        ylabel = {$K$},
        xmin = 0, xmax = 20,
        ymin = 0.0, ymax = 0.15,
        legend cell align={left},
        legend style={font=\scriptsize, at={(0.97, 0.97)}, anchor=north east},
        x tick label style={/pgf/number format/.cd, fixed, fixed zerofill, precision=0, /tikz/.cd},
        y tick label style={/pgf/number format/.cd, fixed, precision=2, /tikz/.cd},	
        grid=both,
        grid style={line width=.1pt, draw=gray!10},
        major grid style={line width=.2pt,draw=gray!50},
        minor x tick num=4,
        minor y tick num=4,
    ]
        
        \addplot[color=PlotColor1, style={very thick}] table[x=t, y=ke, col sep=comma]{./figs/data/bgktgv-M0p5-Re400-p3-n64-post.csv};
        \addlegendentry{$M = 0.5$};
        
        \addplot[color=PlotColor2, style={very thick}] table[x=t, y=ke, col sep=comma]{./figs/data/bgktgv-M0p75-Re400-p3-n64-post.csv};
        \addlegendentry{$M = 0.75$};
        
        \addplot[color=PlotColor3, style={very thick}] table[x=t, y=ke, col sep=comma]{./figs/data/bgktgv-M1p0-Re400-p3-n64-post.csv};
        \addlegendentry{$M = 1$};
        
        \addplot[color=PlotColor4, style={very thick}] table[x=t, y=ke, col sep=comma]{./figs/data/bgktgv-M1p25-Re400-p3-n64-post.csv};
        \addlegendentry{$M = 1.25$};

    \end{axis}
\end{tikzpicture}}}
        \subfloat[$Re = 1600$]{\adjustbox{width=0.48\linewidth, valign=b}{\begin{tikzpicture}[spy using outlines={rectangle, height=3cm,width=2.5cm, magnification=3, connect spies}]
    \begin{axis}
    [
        axis line style={latex-latex},
        axis y line=left,
        axis x line=left,
        clip mode=individual,
        xmode=linear, 
        ymode=linear,
        xlabel = {$t$},
        ylabel = {$K$},
        xmin = 0, xmax = 20,
        ymin = 0.0, ymax = 0.15,
        legend cell align={left},
        legend style={font=\scriptsize, at={(0.97, 0.97)}, anchor=north east},
        x tick label style={/pgf/number format/.cd, fixed, fixed zerofill, precision=0, /tikz/.cd},
        y tick label style={/pgf/number format/.cd, fixed, precision=2, /tikz/.cd},		
        grid=both,
        grid style={line width=.1pt, draw=gray!10},
        major grid style={line width=.2pt,draw=gray!50},
        minor x tick num=4,
        minor y tick num=4,
    ]

        \addplot[color=PlotColor1, style={very thick}] table[x=t, y=ke, col sep=comma]{./figs/data/bgktgv-M0p5-Re1600-p3-n128-post.csv};
        
        \addplot[color=PlotColor2, style={very thick}] table[x=t, y=ke, col sep=comma]{./figs/data/bgktgv-M0p75-Re1600-p3-n128-post.csv};
        
        \addplot[color=PlotColor3, style={very thick}] table[x=t, y=ke, col sep=comma]{./figs/data/bgktgv-M1p0-Re1600-p3-n128-post.csv};
        
        \addplot[color=PlotColor4, style={very thick}] table[x=t, y=ke, col sep=comma]{./figs/data/bgktgv-M1p25-Re1600-p3-n128-post.csv};

    \end{axis}
\end{tikzpicture}}}
        \caption{Temporal evolution of the kinetic energy at varying Mach numbers for the $Re = 400$ (left) and $Re = 1600$ (right) cases. }
        \label{fig:ke} 
    \end{figure}

With the initial verification of the numerical approach performed, we now present the results of the numerical experiments, first in terms of the macroscopic flow features. For the volume-integrated quantities, the effects of increasing Reynolds and Mach numbers on the kinetic energy are shown in \cref{fig:ke}. At lower Mach numbers, the effects of increasing Reynolds number on the kinetic energy profiles were not nearly as pronounced as with higher Mach numbers. Both the $Re = 400$ and $Re = 1600$ cases show a constant decline in the kinetic energy at $M = 0.5$, with the viscous dissipation dominating the kinetic energy behavior. However, with increasing Mach number, the effects of pressure-work start to become non-negligible in comparison to the viscous dissipation, such that fluctuations and local increases in the kinetic energy could be observed. This behavior was most evident in the $M = 1.25$, $Re = 1600$ case (i.e., the case with the strongest compressibility and lowest viscous dissipation effects), where the conversion of internal energy to kinetic energy via the pressure-work term causes the kinetic energy to exceed its initial value over the interval $0 \leq t \leq 5$. The effect of increasing Mach number is also seen in the total kinetic energy dissipation, with higher Mach numbers resulting in less overall dissipation at both Reynolds numbers. 

A more in-depth view of the effects of Reynolds and Mach numbers on the vortical behavior of the flow can be seen in the solenoidal dissipation profiles in \cref{fig:soldis}. At $Re = 400$, where the flow behavior is predominantly laminar, the effects of increasing Mach number were primarily observed as a lag in the primary dissipation peak and a minor flattening of the primary and secondary dissipation peaks. At $Re = 1600$, the chaotic turbulent nature of the flow makes the effects of increasing Mach number much more drastic. In contrast to the lower Reynolds number case, the $M = 0.5$ results show a distinct primary dissipation peak at $t \approx 9$ without a distinct secondary peak. With increasing Mach number, this primary dissipation peak is significantly reduced, such that the $M=1.25$ results do not show a visible primary peak. However, the secondary dissipation peak at $t \approx 11.5$ is evident at higher Mach numbers. Similarly to the lower Reynolds number case, the results with increasing Mach number show as a lag and flattening in the peak dissipation rates.

   \begin{figure}
        \centering
        \subfloat[$Re = 400$]{\adjustbox{width=0.48\linewidth, valign=b}{\begin{tikzpicture}[spy using outlines={rectangle, height=3cm,width=2.5cm, magnification=3, connect spies}]
    \begin{axis}
    [
        axis line style={latex-latex},
        axis y line=left,
        axis x line=left,
        clip mode=individual,
        xmode=linear, 
        ymode=linear,
        xlabel = {$t$},
        ylabel = {$\varepsilon_{\text{sol}}$},
        xmin = 0, xmax = 20,
        ymin = 0.0, ymax = 0.012,
        legend cell align={left},
        legend style={font=\scriptsize, at={(0.97, 0.97)}, anchor=north east},
        ytick = {0,0.004,0.008,0.012},
        x tick label style={/pgf/number format/.cd, fixed, fixed zerofill, precision=0, /tikz/.cd},
        grid=both,
        grid style={line width=.1pt, draw=gray!10},
        major grid style={line width=.2pt,draw=gray!50},
        minor x tick num=4,
        minor y tick num=3,
    ]

        \addplot[color=PlotColor1, style={very thick}] table[x=t, y=sol, col sep=comma]{./figs/data/bgktgv-M0p5-Re400-p3-n64-post.csv};
        \addlegendentry{$M = 0.5$};
        
        \addplot[color=PlotColor2, style={very thick}] table[x=t, y=sol, col sep=comma]{./figs/data/bgktgv-M0p75-Re400-p3-n64-post.csv};
        \addlegendentry{$M = 0.75$};
        
        \addplot[color=PlotColor3, style={very thick}] table[x=t, y=sol, col sep=comma]{./figs/data/bgktgv-M1p0-Re400-p3-n64-post.csv};
        \addlegendentry{$M = 1$};
        
        \addplot[color=PlotColor4, style={very thick}] table[x=t, y=sol, col sep=comma]{./figs/data/bgktgv-M1p25-Re400-p3-n64-post.csv};
        \addlegendentry{$M = 1.25$};

    \end{axis}
\end{tikzpicture}}}
        \subfloat[$Re = 1600$]{\adjustbox{width=0.48\linewidth, valign=b}{\begin{tikzpicture}[spy using outlines={rectangle, height=3cm,width=2.5cm, magnification=3, connect spies}]
    \begin{axis}
    [
        axis line style={latex-latex},
        axis y line=left,
        axis x line=left,
        clip mode=individual,
        xmode=linear, 
        ymode=linear,
        xlabel = {$t$},
        ylabel = {$\varepsilon_{\text{sol}}$},
        xmin = 0, xmax = 20,
        ymin = 0.0, ymax = 0.012,
        legend cell align={left},
        legend style={font=\scriptsize, at={(0.97, 0.97)}, anchor=north east},
        ytick = {0,0.004,0.008,0.012},
        x tick label style={/pgf/number format/.cd, fixed, fixed zerofill, precision=0, /tikz/.cd},
        grid=both,
        grid style={line width=.1pt, draw=gray!10},
        major grid style={line width=.2pt,draw=gray!50},
        minor x tick num=4,
        minor y tick num=4,
    ]

        \addplot[color=PlotColor1, style={very thick}] table[x=t, y=sol, col sep=comma]{./figs/data/bgktgv-M0p5-Re1600-p3-n128-post.csv};
        
        \addplot[color=PlotColor2, style={very thick}] table[x=t, y=sol, col sep=comma]{./figs/data/bgktgv-M0p75-Re1600-p3-n128-post.csv};
        
        \addplot[color=PlotColor3, style={very thick}] table[x=t, y=sol, col sep=comma]{./figs/data/bgktgv-M1p0-Re1600-p3-n128-post.csv};
        
        \addplot[color=PlotColor4, style={very thick}] table[x=t, y=sol, col sep=comma]{./figs/data/bgktgv-M1p25-Re1600-p3-n128-post.csv};

    \end{axis}
\end{tikzpicture}}}
        \caption{Temporal evolution of the solenoidal dissipation rate at varying Mach numbers for the $Re = 400$ (left) and $Re = 1600$ (right) cases. }
        \label{fig:soldis} 
    \end{figure}
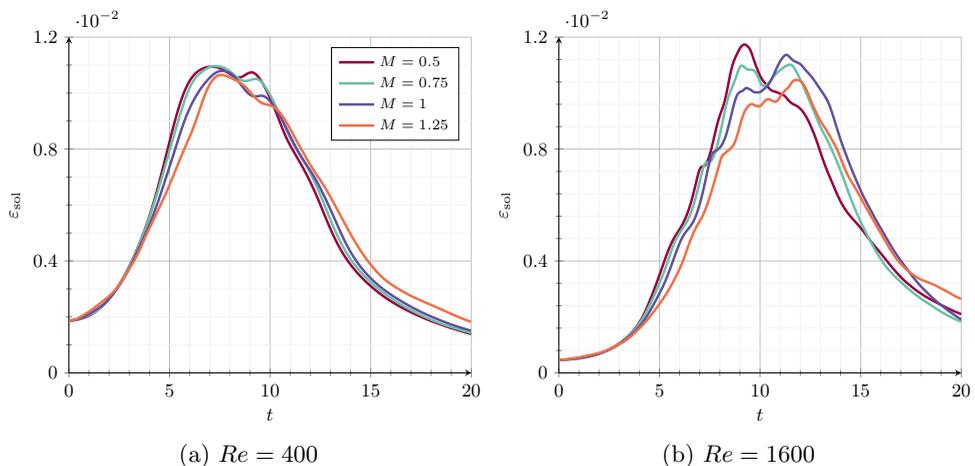

   \begin{figure}
        \centering
        \subfloat[$Re = 400$]{\adjustbox{width=0.48\linewidth, valign=b}{\begin{tikzpicture}[spy using outlines={rectangle, height=3cm,width=2.5cm, magnification=3, connect spies}]
    \begin{axis}
    [
        axis line style={latex-latex},
        axis y line=left,
        axis x line=left,
        clip mode=individual,
        xmode=linear, 
        ymode=linear,
        xlabel = {$t$},
        ylabel = {$\varepsilon_{\text{dil}}$},
        xmin = 0, xmax = 20,
        ymin = 0.0, ymax = 0.001,
        legend cell align={left},
        legend style={font=\scriptsize, at={(0.97, 0.97)}, anchor=north east},
        x tick label style={/pgf/number format/.cd, fixed, fixed zerofill, precision=0, /tikz/.cd},
        grid=both,
        grid style={line width=.1pt, draw=gray!10},
        major grid style={line width=.2pt,draw=gray!50},
        minor x tick num=4,
        minor y tick num=3,
    ]

        \addplot[color=PlotColor1, style={very thick}] table[x=t, y=dil, col sep=comma]{./figs/data/bgktgv-M0p5-Re400-p3-n64-post.csv};
        \addlegendentry{$M = 0.5$};
        
        \addplot[color=PlotColor2, style={very thick}] table[x=t, y=dil, col sep=comma]{./figs/data/bgktgv-M0p75-Re400-p3-n64-post.csv};
        \addlegendentry{$M = 0.75$};
        
        \addplot[color=PlotColor3, style={very thick}] table[x=t, y=dil, col sep=comma]{./figs/data/bgktgv-M1p0-Re400-p3-n64-post.csv};
        \addlegendentry{$M = 1$};
        
        \addplot[color=PlotColor4, style={very thick}] table[x=t, y=dil, col sep=comma]{./figs/data/bgktgv-M1p25-Re400-p3-n64-post.csv};
        \addlegendentry{$M = 1.25$};

    \end{axis}
\end{tikzpicture}}}
        \subfloat[$Re = 1600$]{\adjustbox{width=0.48\linewidth, valign=b}{\begin{tikzpicture}[spy using outlines={rectangle, height=3cm,width=2.5cm, magnification=3, connect spies}]
    \begin{axis}
    [
        axis line style={latex-latex},
        axis y line=left,
        axis x line=left,
        clip mode=individual,
        xmode=linear, 
        ymode=linear,
        xlabel = {$t$},
        ylabel = {$\varepsilon_{\text{dil}}$},
        xmin = 0, xmax = 20,
        ymin = 0.0, ymax = 0.001,
        legend cell align={left},
        legend style={font=\scriptsize, at={(0.97, 0.97)}, anchor=north east},
        x tick label style={/pgf/number format/.cd, fixed, fixed zerofill, precision=0, /tikz/.cd},
        grid=both,
        grid style={line width=.1pt, draw=gray!10},
        major grid style={line width=.2pt,draw=gray!50},
        minor x tick num=4,
        minor y tick num=3,
    ]

        \addplot[color=PlotColor1, style={very thick}] table[x=t, y=dil, col sep=comma]{./figs/data/bgktgv-M0p5-Re1600-p3-n128-post.csv};
        
        \addplot[color=PlotColor2, style={very thick}] table[x=t, y=dil, col sep=comma]{./figs/data/bgktgv-M0p75-Re1600-p3-n128-post.csv};
        
        \addplot[color=PlotColor3, style={very thick}] table[x=t, y=dil, col sep=comma]{./figs/data/bgktgv-M1p0-Re1600-p3-n128-post.csv};
        
        \addplot[color=PlotColor4, style={very thick}] table[x=t, y=dil, col sep=comma]{./figs/data/bgktgv-M1p25-Re1600-p3-n128-post.csv};

    \end{axis}
\end{tikzpicture}}}
        \caption{Temporal evolution of the dilatational dissipation rate at varying Mach numbers for the $Re = 400$ (left) and $Re = 1600$ (right) cases. }
        \label{fig:dildis} 
    \end{figure}
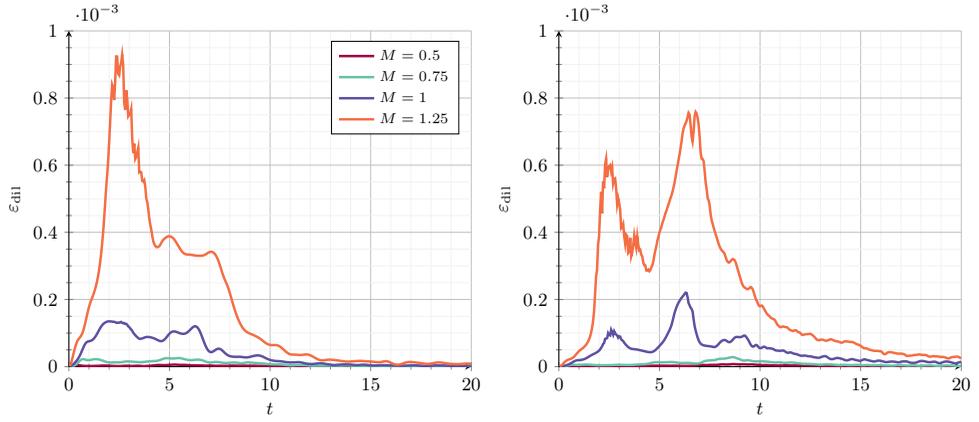

    \begin{figure}
        \centering
        \subfloat[$t = 4$]{\adjustbox{width=0.4\linewidth, valign=b}{\includegraphics[]{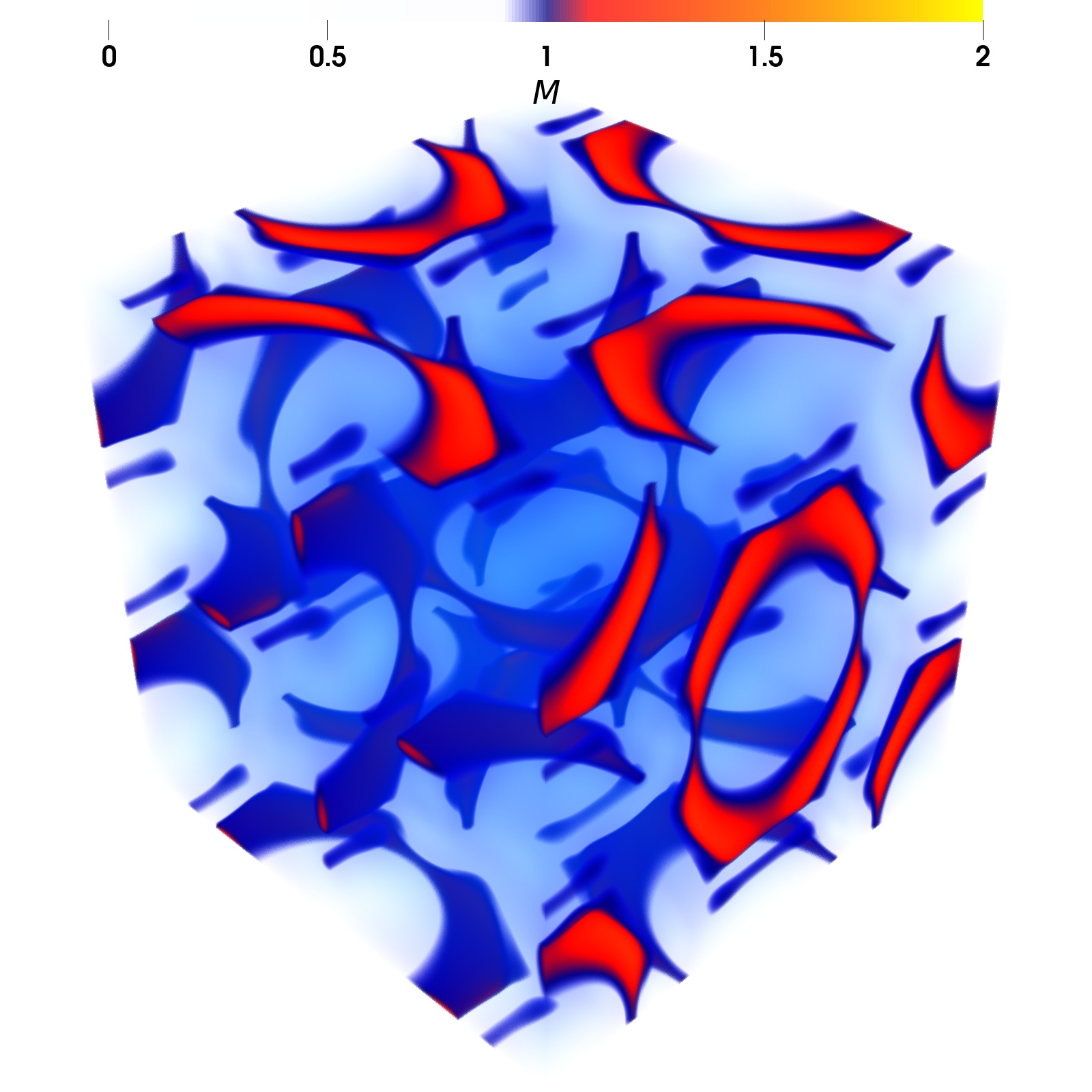}}}
        \subfloat[$t = 6$]{\adjustbox{width=0.4\linewidth, valign=b}{\includegraphics[]{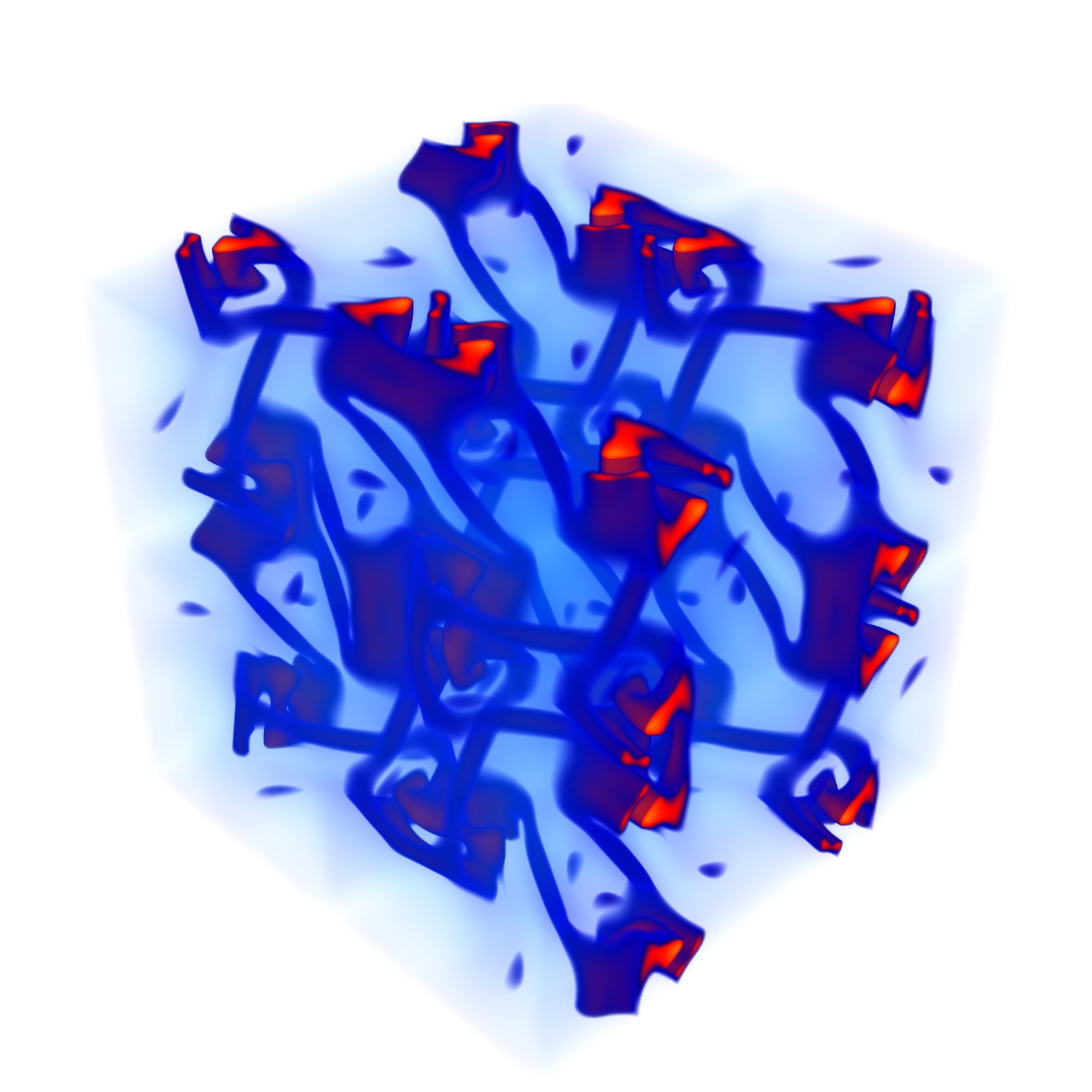}}}
        \newline
        \subfloat[$t = 8$]{\adjustbox{width=0.4\linewidth, valign=b}{\includegraphics[]{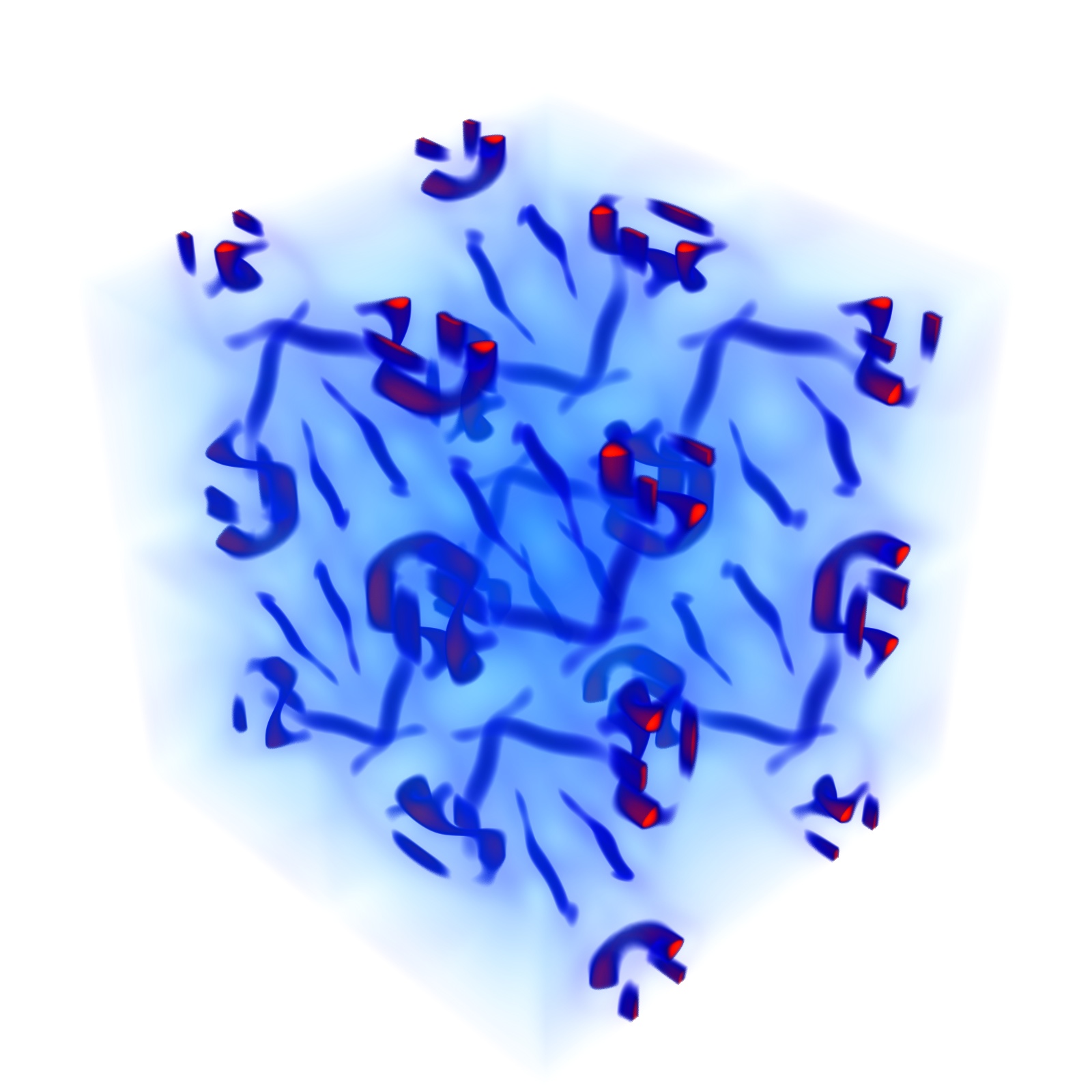}}}
        \subfloat[$t = 10$]{\adjustbox{width=0.4\linewidth, valign=b}{\includegraphics[]{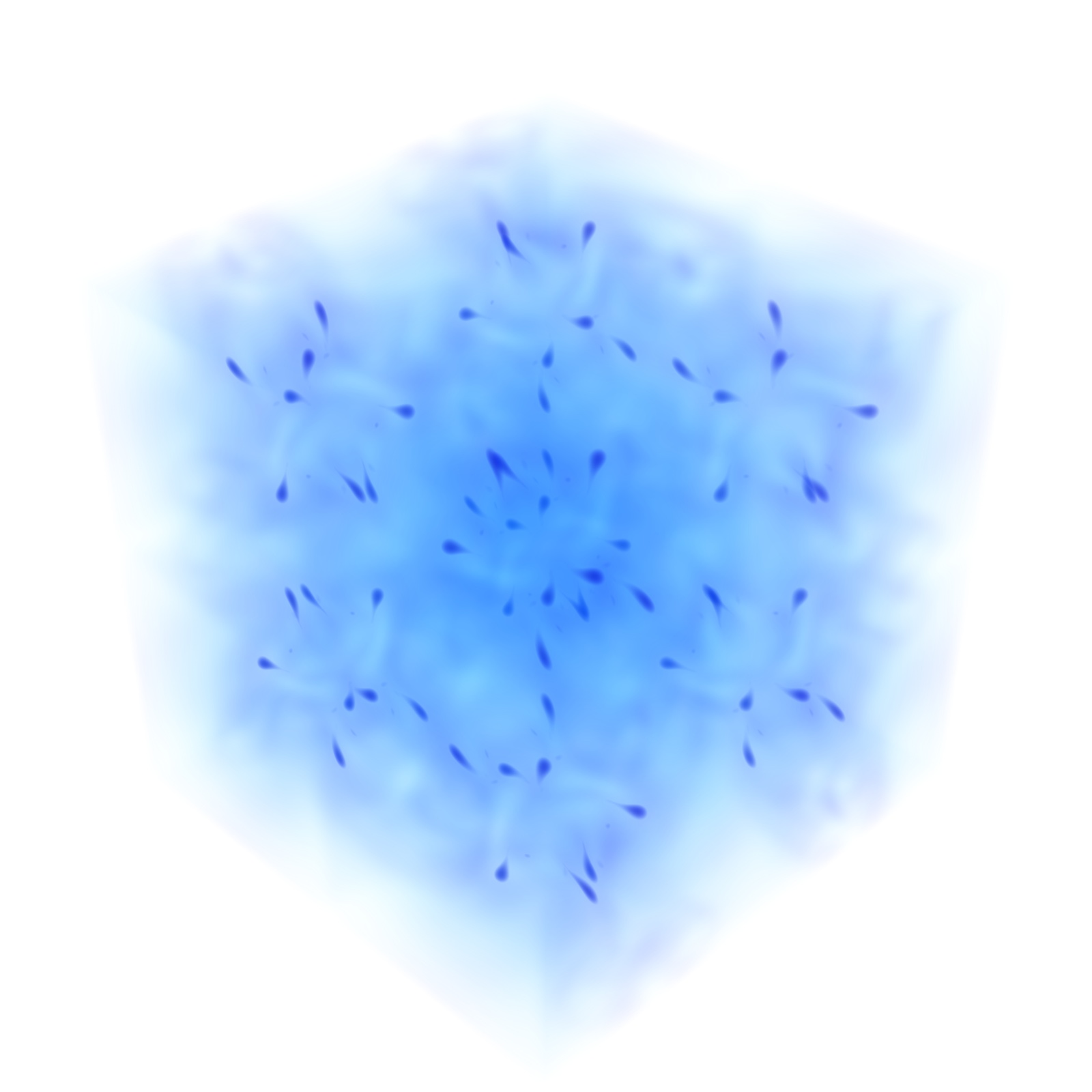}}}
        \newline
        \caption{Volume rendering of the Mach number for the $Re = 400$, $M = 1.25$ case at varying time intervals.}
        \label{fig:machRe400} 
    \end{figure}
    
We further explore the compressibility effects in the flow through the dilatational dissipation profiles in \cref{fig:dildis}. At all of the considered Mach and Reynolds numbers, the dilatational dissipation rates remained relatively small compared to solenoidal dissipation. As expected, the overall magnitude of the dilatational dissipation was strongly dependent on the Mach number, with the maximum dilatational dissipation increasing rapidly with increasing Mach number. At $M=0.5$ and $M=0.75$, the dilatational dissipation at both Reynolds numbers was relatively negligible compared to the higher Mach number cases, largely due to the predominantly subsonic nature of the flow. However, at $M=1$ and $M=1.25$, the dilatational dissipation profiles began to show notable differences due to the increased presence of shocklets in the flow. At $Re = 400$, the dilatational dissipation profile had a single distinct peak at $t \approx 2.5$, indicative of the initial shock formation, followed by a plateau and eventual decrease. This peak was more pronounced at $M = 1.25$ due to the increased strength of these shocks. At $Re = 1600$, the dilatational dissipation profile showed a similar peak at $t \approx 2.5$, although with a lower magnitude compared to the lower Reynolds number case as the effects of lower viscosity overshadowed the effects of the decreased shock thickness. However, the higher Reynolds number case showed a distinct second peak at $t \approx 6$ as shock-turbulence interactions in the flow started to dominate the dilatational dissipation rate. This second peak, which did not appear in the lower Reynolds number case, was of larger magnitude than the first peak, although the numerical experiments in \citet{Chapelier2024} indicate that if one were to fully resolve the flow down to the mean free path (i.e., fully resolve the shock structures in the flow), the first peak would be of similar magnitude or larger than the second peak. We note here that minor oscillations in the dilatational dissipation rate were observed for the highest Mach number cases as the squared gradient terms in the dilatational dissipation are highly numerically sensitive and the numerical approach does not apply any explicit shock capturing to dampen any minor oscillations. However, these do not appreciably affect the overall behavior of the dilatational dissipation profiles.

    \begin{figure}
        \centering
        \subfloat[$t = 4$]{\adjustbox{width=0.4\linewidth, valign=b}{\includegraphics[]{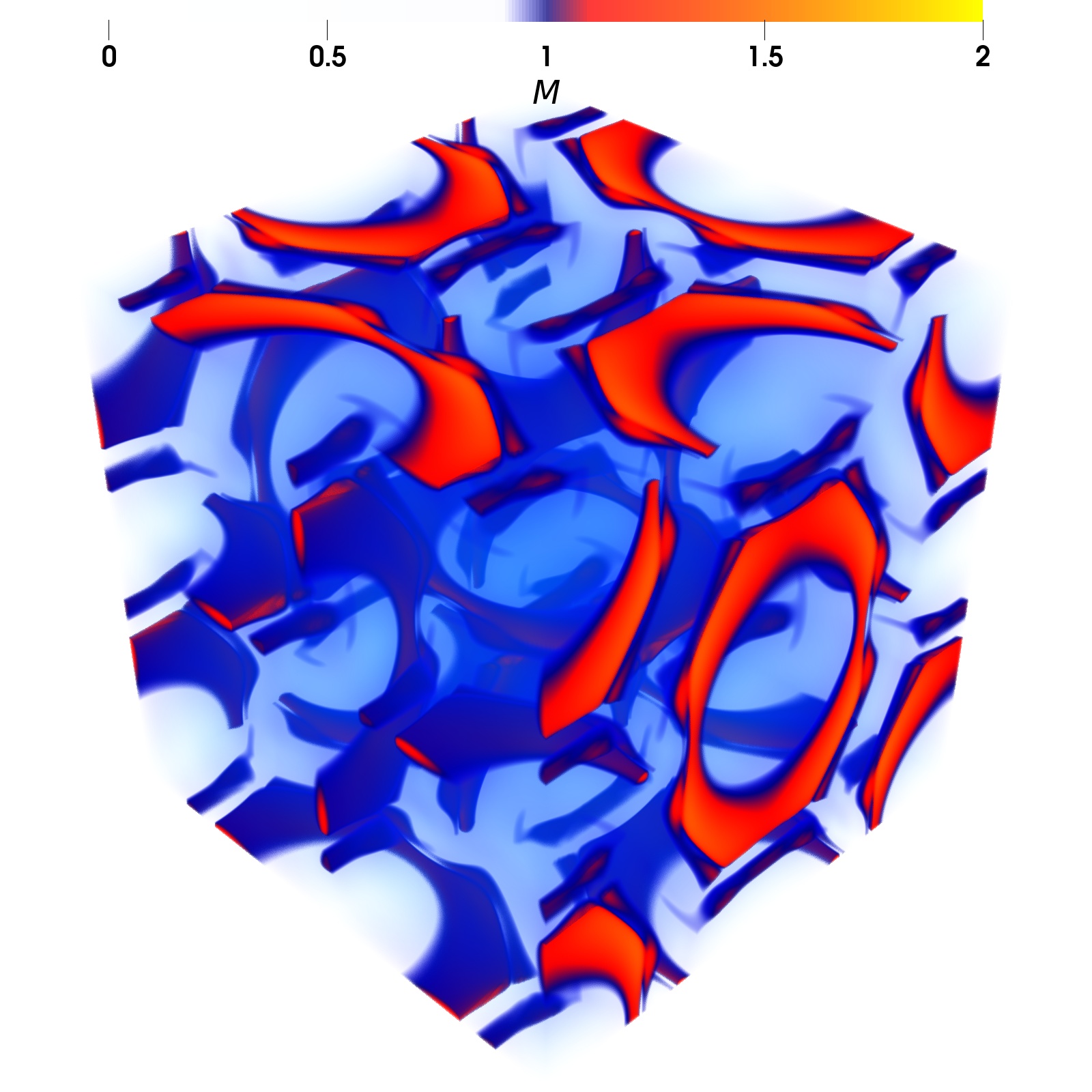}}}
        \subfloat[$t = 6$]{\adjustbox{width=0.4\linewidth, valign=b}{\includegraphics[]{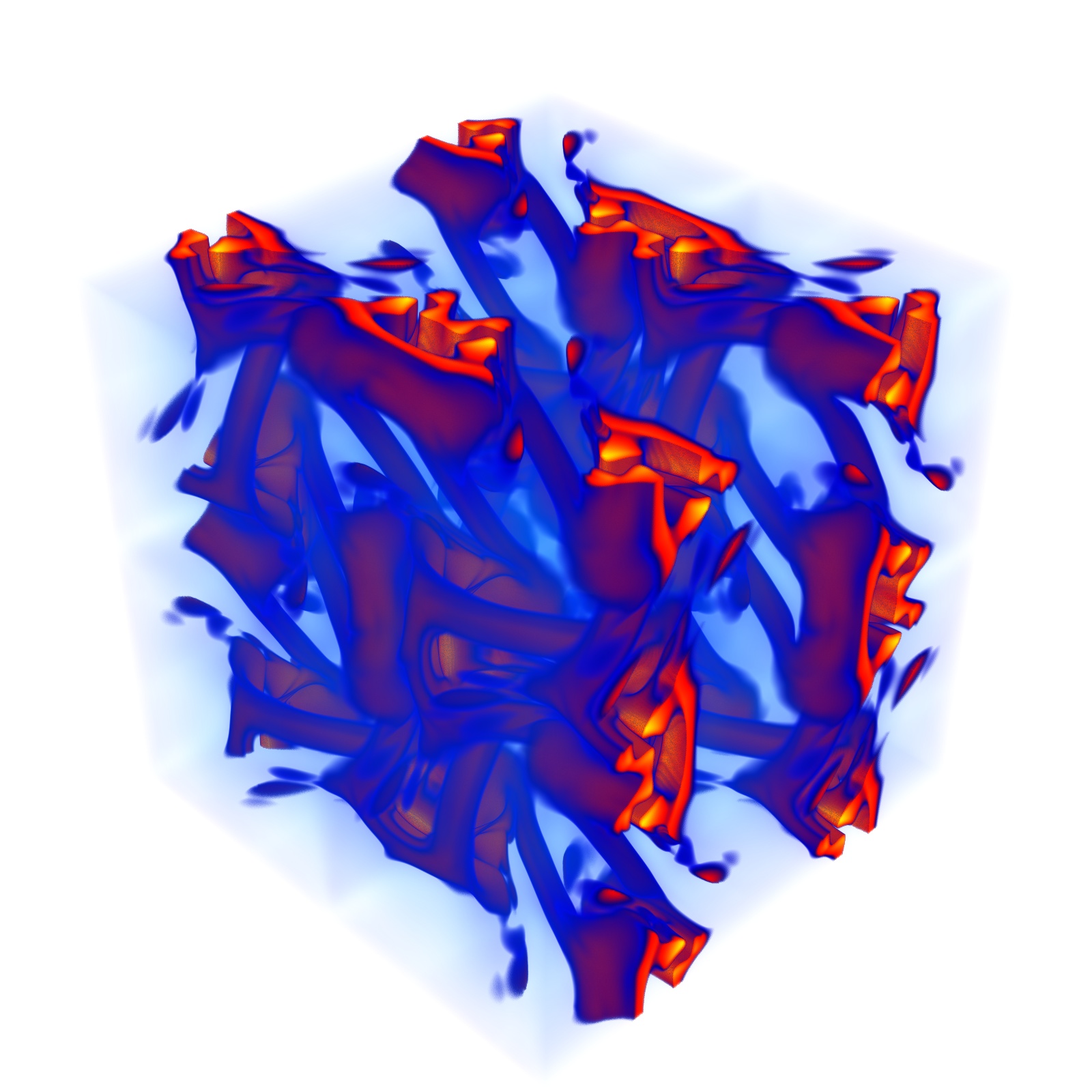}}}
        \newline
        \subfloat[$t = 8$]{\adjustbox{width=0.4\linewidth, valign=b}{\includegraphics[]{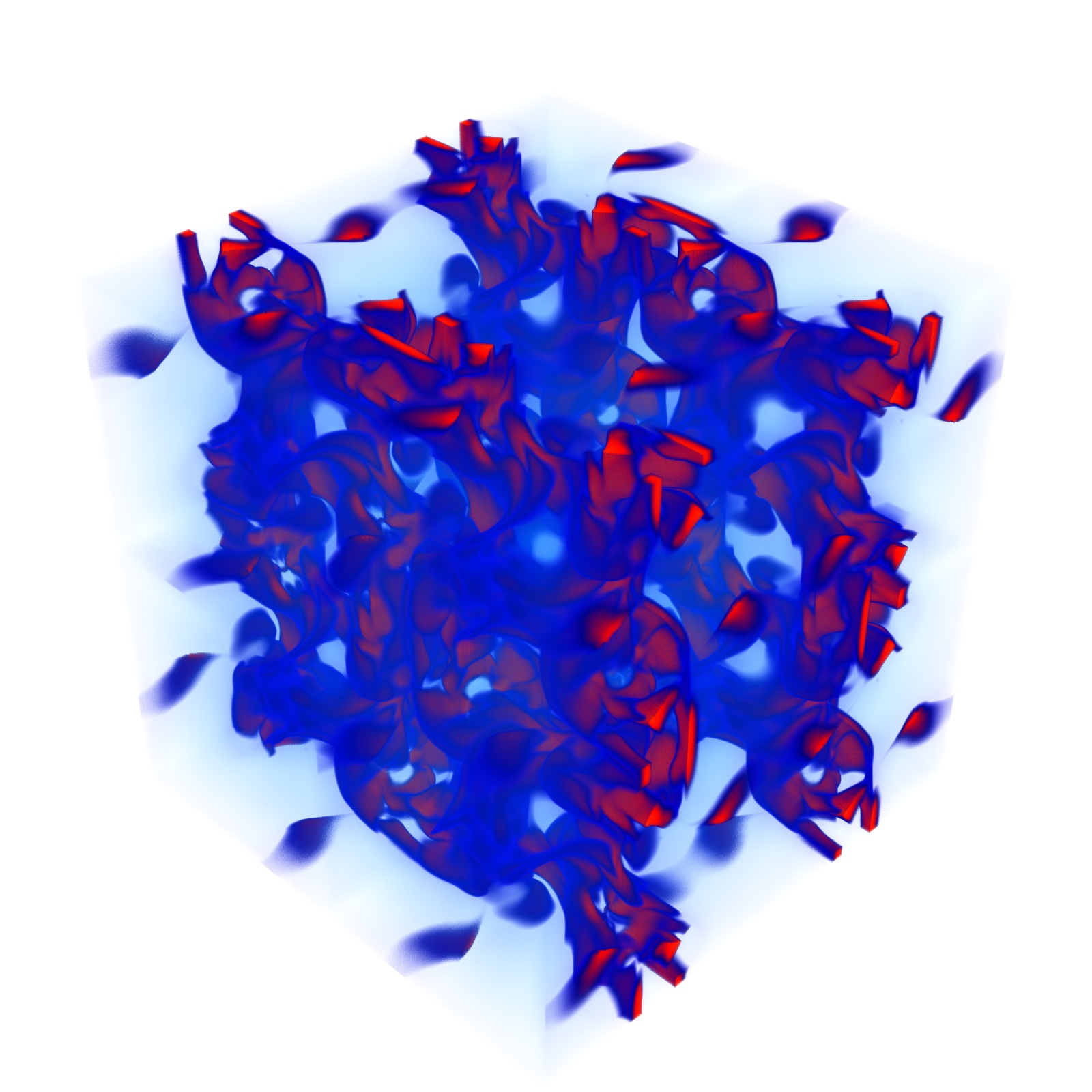}}}
        \subfloat[$t = 10$]{\adjustbox{width=0.4\linewidth, valign=b}{\includegraphics[]{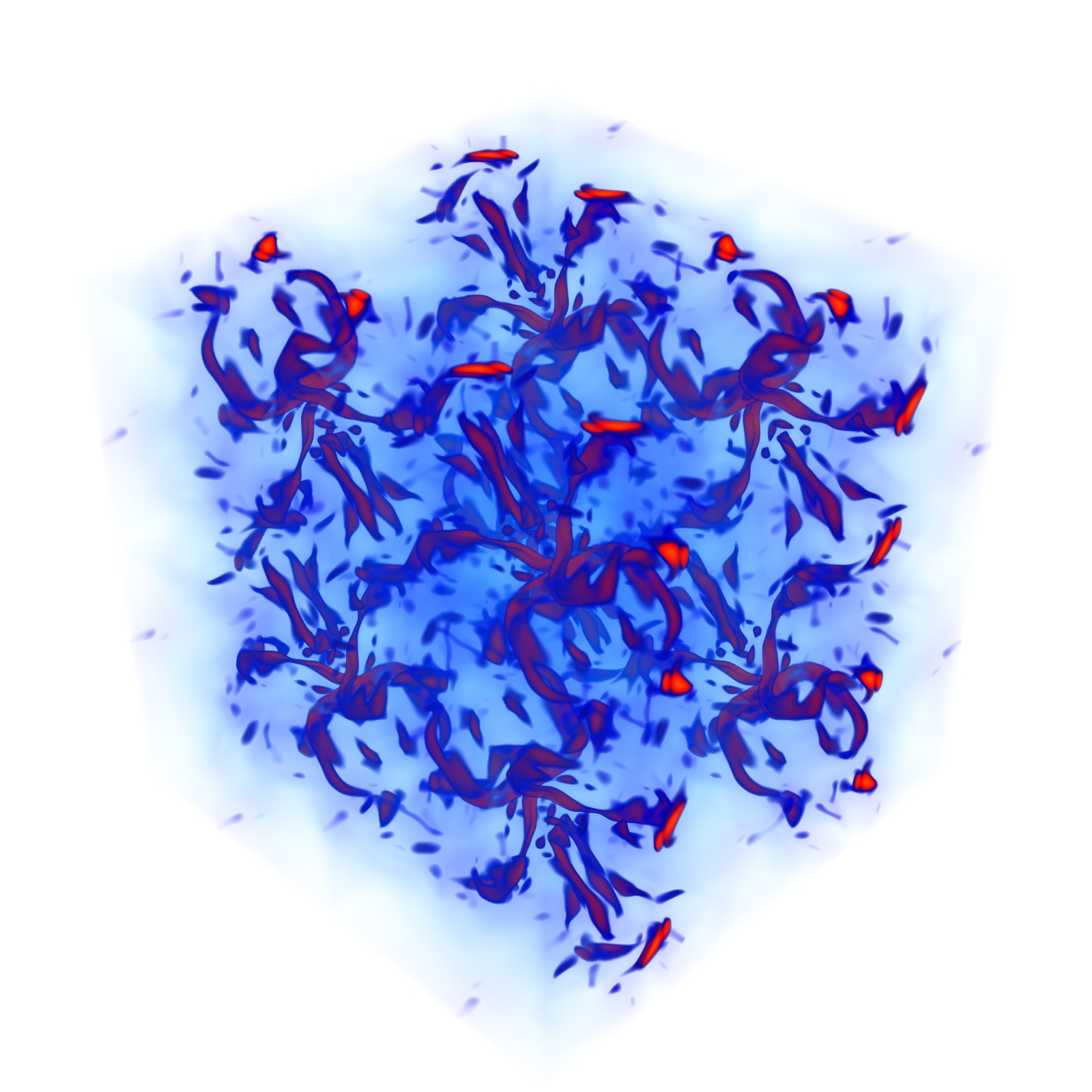}}}
        \newline
        \caption{Volume rendering of the Mach number for the $Re = 1600$, $M = 1.25$ case at varying time intervals.}
        \label{fig:machRe1600} 
    \end{figure}
    
A more detailed view of shock-vortex interactions in the flow can be seen through volume renderings of the local Mach number field. \cref{fig:machRe400} and \cref{fig:machRe1600} show these renderings at $M=1.25$ for $Re = 400$ and $Re=1600$, respectively. For the $Re=400$ case, the flow field showed the formation of well-defined shock structures at $t=4$. These shocks start to be substantially affected by the vortical nature of the flow, such that by $t=6$, the localized deformation and rollup of the shock surfaces was evident. At $t=8$, the shocks have gradually weakened, and the flow became increasingly dominated by viscous dissipation rather than compressibility effects, with the majority of shocks in the flow field dissipated away by $t=10$.

Similar behavior in the flow field was observed with the $Re=1600$ case at $t=4$, although with noticeably sharper shock fronts due to the reduced viscosity. By $t=6$, the interactions between shocks and vortices were much more pronounced than with $Re = 400$, with regions of higher local Mach number and more complex shock surfaces. At $t=8$, where for $Re = 400$ the shocks in the flow have largely dissipated away, the $Re=1600$ results still showed persistent shock surfaces which have been heavily deformed by vortex interactions. By $t=10$, the flow was dominated by a highly chaotic and turbulent field, with shocklets and small-scale vortices distributed throughout.

\subsection{Kinetic behavior}\label{ssec:kinetic}

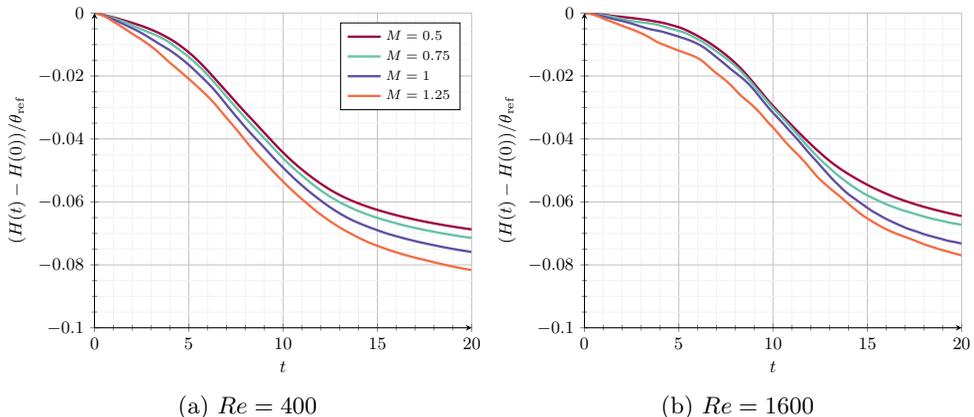
\begin{figure}
        \centering
        \subfloat[$Re = 400$]{\adjustbox{width=0.48\linewidth, valign=b}{\begin{tikzpicture}[spy using outlines={rectangle, height=3cm,width=2.5cm, magnification=3, connect spies}]
    \begin{axis}
    [
        axis line style={latex-latex},
        axis y line=left,
        axis x line=left,
        clip mode=individual,
        xmode=linear, 
        ymode=linear,
        xlabel = {$t$},
        ylabel = {$(H(t) - H(0))/ \theta_{\text{ref}}$},
        xmin = 0, xmax = 20,
        ymin = -0.10, ymax = 0,
        legend cell align={left},
        legend style={font=\scriptsize, at={(0.97, 0.97)}, anchor=north east},
        ytick = {0,-0.02,-0.04, -0.06, -0.08,-0.1},
        x tick label style={/pgf/number format/.cd, fixed, fixed zerofill, precision=0, /tikz/.cd},
        y tick label style={/pgf/number format/.cd, fixed, precision=2, /tikz/.cd},
        grid=both,
        grid style={line width=.1pt, draw=gray!10},
        major grid style={line width=.2pt,draw=gray!50},
        minor x tick num=4,
        minor y tick num=3,
    ]

        \addplot[color=PlotColor1, style={very thick}] table[x=t, y=h, col sep=comma]{./figs/data/bgktgv-M0p5-Re400-p3-n64-post.csv};
        \addlegendentry{$M = 0.5$};
        
        \addplot[color=PlotColor2, style={very thick}] table[x=t, y=h, col sep=comma]{./figs/data/bgktgv-M0p75-Re400-p3-n64-post.csv};
        \addlegendentry{$M = 0.75$};
        
        \addplot[color=PlotColor3, style={very thick}] table[x=t, y=h, col sep=comma]{./figs/data/bgktgv-M1p0-Re400-p3-n64-post.csv};
        \addlegendentry{$M = 1$};
        
        \addplot[color=PlotColor4, style={very thick}] table[x=t, y=h, col sep=comma]{./figs/data/bgktgv-M1p25-Re400-p3-n64-post.csv};
        \addlegendentry{$M = 1.25$};

    \end{axis}
\end{tikzpicture}}}
        \subfloat[$Re = 1600$]{\adjustbox{width=0.48\linewidth, valign=b}{\begin{tikzpicture}[spy using outlines={rectangle, height=3cm,width=2.5cm, magnification=3, connect spies}]
    \begin{axis}
    [
        axis line style={latex-latex},
        axis y line=left,
        axis x line=left,
        clip mode=individual,
        xmode=linear, 
        ymode=linear,
        xlabel = {$t$},
        ylabel = {$(H(t) - H(0))/ \theta_{\text{ref}}$},
        xmin = 0, xmax = 20,
        ymin = -0.10, ymax = 0,
        legend cell align={left},
        legend style={font=\scriptsize, at={(0.97, 0.97)}, anchor=north east},
        ytick = {0,-0.02,-0.04, -0.06, -0.08,-0.1},
        x tick label style={/pgf/number format/.cd, fixed, fixed zerofill, precision=0, /tikz/.cd},
        y tick label style={/pgf/number format/.cd, fixed, precision=2, /tikz/.cd},
        grid=both,
        grid style={line width=.1pt, draw=gray!10},
        major grid style={line width=.2pt,draw=gray!50},
        minor x tick num=4,
        minor y tick num=3,
    ]

        \addplot[color=PlotColor1, style={very thick}] table[x=t, y=h, col sep=comma]{./figs/data/bgktgv-M0p5-Re1600-p3-n128-post.csv};
        
        \addplot[color=PlotColor2, style={very thick}] table[x=t, y=h, col sep=comma]{./figs/data/bgktgv-M0p75-Re1600-p3-n128-post.csv};
        
        \addplot[color=PlotColor3, style={very thick}] table[x=t, y=h, col sep=comma]{./figs/data/bgktgv-M1p0-Re1600-p3-n128-post.csv};
        
        \addplot[color=PlotColor4, style={very thick}] table[x=t, y=h, col sep=comma]{./figs/data/bgktgv-M1p25-Re1600-p3-n128-post.csv};

    \end{axis}
\end{tikzpicture}}}
        \caption{Temporal evolution of the Boltzmann entropy at varying Mach numbers for the $Re = 400$ (left) and $Re = 1600$ (right) cases.}
        \label{fig:entropy} 
    \end{figure}

As the solution of the Boltzmann equations encodes significantly more information than just the macroscopic flow state, we can leverage this to extract more detailed insights into the molecular-scale dynamics and their connection to the overall flow behavior. We first consider the evolution of the evolution of the Boltzmann entropy $H(f)$, shown in \cref{fig:entropy} normalized by the reference temperature. For all cases, the entropy showed a constant decline as expected by the H-theorem. At both Reynolds numbers, the initial rate of entropy change increased with increasing Mach number. However, for the $Re = 400$ cases, the entropy profiles then showed a near constant rate of change for approximately $6 \leq t \leq 10$, independently of the Mach number. This effect was somewhat observed at $Re = 1600$ at a later and shorter time interval (approximately $9 \leq t \leq 11$), although not as pronounced. At later times, the entropy profiles smoothly began to taper off as viscous dissipation effects became dominant. At both Reynolds numbers, the $M=1.25$ results showed a more distinct behavior relative to the lower Mach number cases, seen in the form of an initially large slope with a somewhat abrupt change in the slope around $t=6$, with this change much more evident at $Re=1600$. This slope change lines up with the secondary dilatational dissipation peak in the $Re=1600$ case, where shock-turbulence interactions began to dominate the flow. 

\begin{figure}
        \centering
        \subfloat[$Re = 400$]{\adjustbox{width=0.48\linewidth, valign=b}{\begin{tikzpicture}[spy using outlines={rectangle, height=3cm,width=2.5cm, magnification=3, connect spies}]
    \begin{axis}
    [
        axis line style={latex-latex},
        axis y line=left,
        axis x line=left,
        clip mode=individual,
        xmode=linear, 
        ymode=linear,
        xlabel = {$t$},
        ylabel = {$- D (f) / \theta_{\text{ref}}$},
        xmin = 0, xmax = 20,
        ymin = 0, ymax = 0.015,
        ytick = {0,0.005,0.010,0.015},
        legend cell align={left},
        legend style={font=\scriptsize, at={(0.97, 0.97)}, anchor=north east},
        x tick label style={/pgf/number format/.cd, fixed, fixed zerofill, precision=0, /tikz/.cd},
        y tick label style={/pgf/number format/.cd, fixed, precision=2, /tikz/.cd},
        grid=both,
        grid style={line width=.1pt, draw=gray!10},
        major grid style={line width=.2pt,draw=gray!50},
        minor x tick num=4,
        minor y tick num=4,
    ]

        \addplot[color=PlotColor1, style={very thick}] table[x=t, y=cfflogf, col sep=comma]{./figs/data/bgktgv-M0p5-Re400-p3-n64-post.csv};
        \addlegendentry{$M = 0.5$};
        
        \addplot[color=PlotColor2, style={very thick}] table[x=t, y=cfflogf, col sep=comma]{./figs/data/bgktgv-M0p75-Re400-p3-n64-post.csv};
        \addlegendentry{$M = 0.75$};
        
        \addplot[color=PlotColor3, style={very thick}] table[x=t, y=cfflogf, col sep=comma]{./figs/data/bgktgv-M1p0-Re400-p3-n64-post.csv};
        \addlegendentry{$M = 1$};
        
        \addplot[color=PlotColor4, style={very thick}] table[x=t, y=cfflogf, col sep=comma]{./figs/data/bgktgv-M1p25-Re400-p3-n64-post.csv};
        \addlegendentry{$M = 1.25$};

    \end{axis}
\end{tikzpicture}}}
        \subfloat[$Re = 1600$]{\adjustbox{width=0.48\linewidth, valign=b}{\begin{tikzpicture}[spy using outlines={rectangle, height=3cm,width=2.5cm, magnification=3, connect spies}]
    \begin{axis}
    [
        axis line style={latex-latex},
        axis y line=left,
        axis x line=left,
        clip mode=individual,
        xmode=linear, 
        ymode=linear,
        xlabel = {$t$},
        ylabel = {$- D (f) / \theta_{\text{ref}}$},
        xmin = 0, xmax = 20,
        ymin = 0, ymax = 0.015,
        ytick = {0,0.005,0.010,0.015},
        legend cell align={left},
        legend style={font=\scriptsize, at={(0.97, 0.97)}, anchor=south east},
        x tick label style={/pgf/number format/.cd, fixed, fixed zerofill, precision=0, /tikz/.cd},
        y tick label style={/pgf/number format/.cd, fixed, precision=2, /tikz/.cd},
        grid=both,
        grid style={line width=.1pt, draw=gray!10},
        major grid style={line width=.2pt,draw=gray!50},
        minor x tick num=4,
        minor y tick num=4,
    ]

        \addplot[color=PlotColor1, style={very thick}] table[x=t, y=cfflogf, col sep=comma]{./figs/data/bgktgv-M0p5-Re1600-p3-n128-post.csv};
        
        \addplot[color=PlotColor2, style={very thick}] table[x=t, y=cfflogf, col sep=comma]{./figs/data/bgktgv-M0p75-Re1600-p3-n128-post.csv};
        
        \addplot[color=PlotColor3, style={very thick}] table[x=t, y=cfflogf, col sep=comma]{./figs/data/bgktgv-M1p0-Re1600-p3-n128-post.csv};
        
        \addplot[color=PlotColor4, style={very thick}] table[x=t, y=cfflogf, col sep=comma]{./figs/data/bgktgv-M1p25-Re1600-p3-n128-post.csv};

    \end{axis}
\end{tikzpicture}}}
        \caption{Temporal evolution of the Boltzmann entropy production rate at varying Mach numbers for the $Re = 400$ (left) and $Re = 1600$ (right) cases.}
        \label{fig:cfflogf} 
    \end{figure}
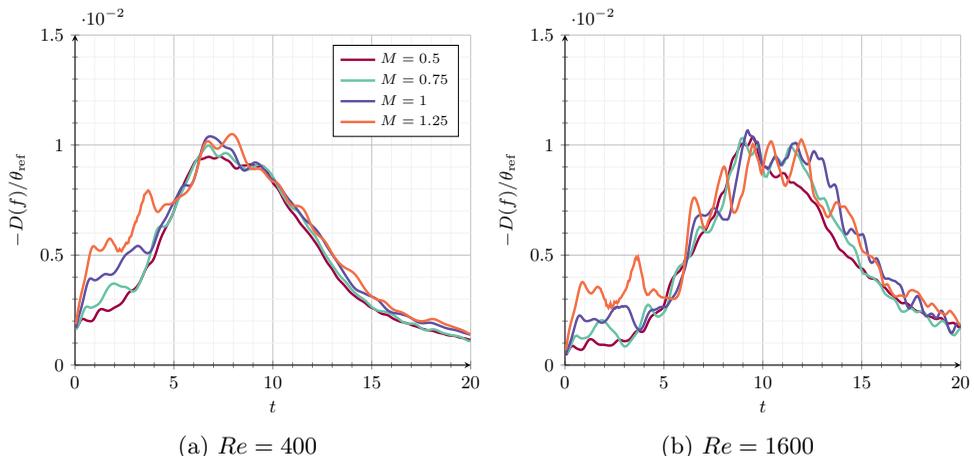

A secondary observation of the Boltzmann entropy profiles is that they share some resemblance to the kinetic energy profiles in \cref{fig:ke}. As such, it is of interest to compare the behavior of the entropy dissipation $D(f)$ (i.e., temporal rate of change of the entropy) to the viscous dissipation (solenoidal and dilatational) in the flow. The (negative) entropy dissipation profiles for the Boltzmann entropy are shown in \cref{fig:cfflogf}, normalized by the reference temperature. At lower Mach numbers, the Boltzmann entropy dissipation profiles exhibit similar behavior to the viscous dissipation profiles, taken as the sum of the profiles in \cref{fig:soldis} and \cref{fig:dildis}, although with a notable underprediction relative to the viscous dissipation peak. As the Mach number increased, the entropy dissipation profiles became more oscillatory and significantly deviated from the viscous dissipation profiles, particularly so at earlier times. These observations were consistent between both the $Re=400$ and $Re=1600$ cases.

   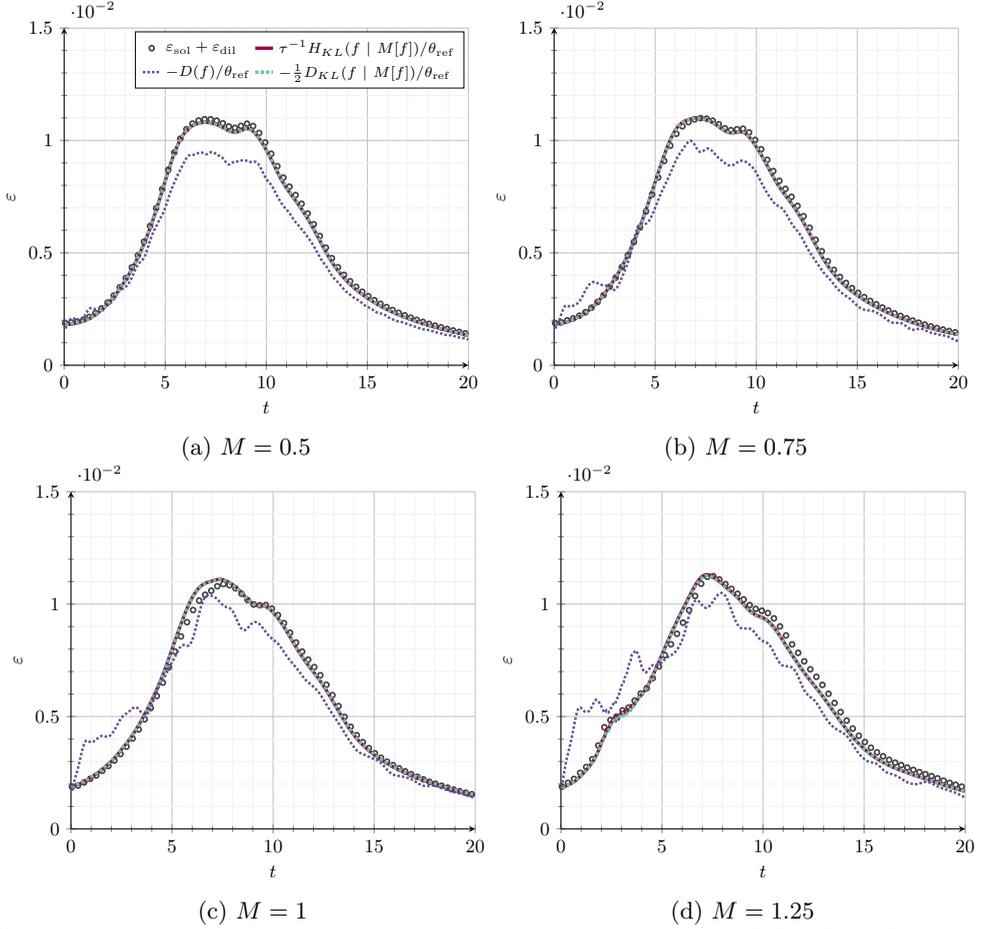
\begin{figure}
        \centering
        \subfloat[$M = 0.5$]{\adjustbox{width=0.48\linewidth, valign=b}{\begin{tikzpicture}[spy using outlines={rectangle, height=3cm,width=2.5cm, magnification=3, connect spies}]
    \begin{axis}
    [
        axis line style={latex-latex},
        axis y line=left,
        axis x line=left,
        clip mode=individual,
        xmode=linear, 
        ymode=linear,
        xlabel = {$t$},
        ylabel = {$\varepsilon$},
        xmin = 0, xmax = 20,
        ymin = 0, ymax = 0.015,
        legend cell align={left},
        legend columns=2, 
        transpose legend,
        legend style={font=\scriptsize, at={(0.99, 0.99)}, anchor=north east},
        x tick label style={/pgf/number format/.cd, fixed, fixed zerofill, precision=0, /tikz/.cd},
        scale = 1,
        grid=both,
        grid style={line width=.1pt, draw=gray!10},
        major grid style={line width=.2pt,draw=gray!50},
        minor x tick num=4,
        minor y tick num=4,
    ]
            \pgfplotsset{
        compat=1.11,
        legend image code/.code={
        \draw[mark repeat=2,mark phase=2]
        plot coordinates {
        (0cm,0cm)
        (0.15cm,0cm)        
        (0.3cm,0cm)         
        };%
        }
        }
        
        \addplot[color=black!75, style={thick}, only marks, mark=o, mark options={scale=0.6}, mark repeat = 6, mark phase = 0]  table[x=t, y=dis, col sep=comma]{./figs/data/bgktgv-M0p5-Re400-p3-n64-post.csv};
        \addlegendentry{$\varepsilon_{\text{sol}} + \varepsilon_{\text{dil}}$};
        
        \addplot[color=PlotColor3, style={very thick, densely dotted}] table[x=t, y=cfflogf, col sep=comma]{./figs/data/bgktgv-M0p5-Re400-p3-n64-post.csv};
        \addlegendentry{$-D(f)/\theta_\tref$};
        
        \addplot[color=PlotColor1, style={ultra thick}] table[x=t, y=flogfgrcptau, col sep=comma]{./figs/data/bgktgv-M0p5-Re400-p3-n64-post.csv};
        \addlegendentry{$\tau^{-1} H_{KL} (f\ | \ M[f])/\theta_\tref$};
        
        \addplot[color=PlotColor2, style={ultra thick, densely dotted}] table[x=t, y=cfflogfg, col sep=comma]{./figs/data/bgktgv-M0p5-Re400-p3-n64-post.csv};;
        \addlegendentry{$-\frac{1}{2}D_{KL} (f\ | \ M[f])/\theta_\tref$};

    \end{axis}
\end{tikzpicture}}}
        \subfloat[$M = 0.75$]{\adjustbox{width=0.48\linewidth, valign=b}{\begin{tikzpicture}[spy using outlines={rectangle, height=3cm,width=2.5cm, magnification=3, connect spies}]
    \begin{axis}
    [
        axis line style={latex-latex},
        axis y line=left,
        axis x line=left,
        clip mode=individual,
        xmode=linear, 
        ymode=linear,
        xlabel = {$t$},
        ylabel = {$\varepsilon$},
        xmin = 0, xmax = 20,
        ymin = 0, ymax = 0.015,
        legend cell align={left},
        legend style={font=\scriptsize, at={(0.97, 0.97)}, anchor=north east},
        x tick label style={/pgf/number format/.cd, fixed, fixed zerofill, precision=0, /tikz/.cd},
        scale = 1,
        grid=both,
        grid style={line width=.1pt, draw=gray!10},
        major grid style={line width=.2pt,draw=gray!50},
        minor x tick num=4,
        minor y tick num=4,
    ]
        
        \addplot[color=black!75, style={thick}, only marks, mark=o, mark options={scale=0.6}, mark repeat = 6, mark phase = 0]  table[x=t, y=dis, col sep=comma]{./figs/data/bgktgv-M0p75-Re400-p3-n64-post.csv};
        
        \addplot[color=PlotColor1, style={ultra thick}] table[x=t, y=flogfgrcptau, col sep=comma]{./figs/data/bgktgv-M0p75-Re400-p3-n64-post.csv};
        
        \addplot[color=PlotColor2, style={ultra thick, densely dotted}] table[x=t, y=cfflogfg, col sep=comma]{./figs/data/bgktgv-M0p75-Re400-p3-n64-post.csv};
        
        \addplot[color=PlotColor3, style={very thick, densely dotted}] table[x=t, y=cfflogf, col sep=comma]{./figs/data/bgktgv-M0p75-Re400-p3-n64-post.csv};
        
    \end{axis}
\end{tikzpicture}}}
        \newline
        \subfloat[$M = 1$]{\adjustbox{width=0.48\linewidth, valign=b}{\begin{tikzpicture}[spy using outlines={rectangle, height=3cm,width=2.5cm, magnification=3, connect spies}]
    \begin{axis}
    [
        axis line style={latex-latex},
        axis y line=left,
        axis x line=left,
        clip mode=individual,
        xmode=linear, 
        ymode=linear,
        xlabel = {$t$},
        ylabel = {$\varepsilon$},
        xmin = 0, xmax = 20,
        ymin = 0, ymax = 0.015,
        legend cell align={left},
        legend style={font=\scriptsize, at={(0.97, 0.97)}, anchor=north east},
        x tick label style={/pgf/number format/.cd, fixed, fixed zerofill, precision=0, /tikz/.cd},
        scale = 1,
        grid=both,
        grid style={line width=.1pt, draw=gray!10},
        major grid style={line width=.2pt,draw=gray!50},
        minor x tick num=4,
        minor y tick num=4,
    ]
        
        \addplot[color=black!75, style={thick}, only marks, mark=o, mark options={scale=0.6}, mark repeat = 6, mark phase = 0]  table[x=t, y=dis, col sep=comma]{./figs/data/bgktgv-M1p0-Re400-p3-n64-post.csv};
        
        \addplot[color=PlotColor1, style={ultra thick}] table[x=t, y=flogfgrcptau, col sep=comma]{./figs/data/bgktgv-M1p0-Re400-p3-n64-post.csv};
        
        \addplot[color=PlotColor2, style={ultra thick, densely dotted}] table[x=t, y=cfflogfg, col sep=comma]{./figs/data/bgktgv-M1p0-Re400-p3-n64-post.csv};
        
        \addplot[color=PlotColor3, style={very thick, densely dotted}] table[x=t, y=cfflogf, col sep=comma]{./figs/data/bgktgv-M1p0-Re400-p3-n64-post.csv};
    \end{axis}
\end{tikzpicture}}}
        \subfloat[$M = 1.25$]{\adjustbox{width=0.48\linewidth, valign=b}{\begin{tikzpicture}[spy using outlines={rectangle, height=3cm,width=2.5cm, magnification=3, connect spies}]
    \begin{axis}
    [
        axis line style={latex-latex},
        axis y line=left,
        axis x line=left,
        clip mode=individual,
        xmode=linear, 
        ymode=linear,
        xlabel = {$t$},
        ylabel = {$\varepsilon$},
        xmin = 0, xmax = 20,
        ymin = 0, ymax = 0.015,
        legend cell align={left},
        legend style={font=\scriptsize, at={(0.97, 0.97)}, anchor=north east},
        x tick label style={/pgf/number format/.cd, fixed, fixed zerofill, precision=0, /tikz/.cd},
        scale = 1,
        grid=both,
        grid style={line width=.1pt, draw=gray!10},
        major grid style={line width=.2pt,draw=gray!50},
        minor x tick num=4,
        minor y tick num=4,
    ]
        \addplot[color=black!75, style={thick}, only marks, mark=o, mark options={scale=0.6}, mark repeat = 6, mark phase = 0]  table[x=t, y=dis, col sep=comma]{./figs/data/bgktgv-M1p25-Re400-p3-n64-post.csv};
        
        \addplot[color=PlotColor1, style={ultra thick}] table[x=t, y=flogfgrcptau, col sep=comma]{./figs/data/bgktgv-M1p25-Re400-p3-n64-post.csv};
        
        \addplot[color=PlotColor2, style={ultra thick, densely dotted}] table[x=t, y=cfflogfg, col sep=comma]{./figs/data/bgktgv-M1p25-Re400-p3-n64-post.csv};
        
        \addplot[color=PlotColor3, style={very thick, densely dotted}] table[x=t, y=cfflogf, col sep=comma]{./figs/data/bgktgv-M1p25-Re400-p3-n64-post.csv};
    \end{axis}
\end{tikzpicture}}}
        \caption{Comparison of the temporal evolution of the total dissipation, Boltzmann entropy production rate, relative entropy, and relative entropy production rate at varying Mach numbers for the $Re = 400$ case.}
        \label{fig:norment400}  
    \end{figure}

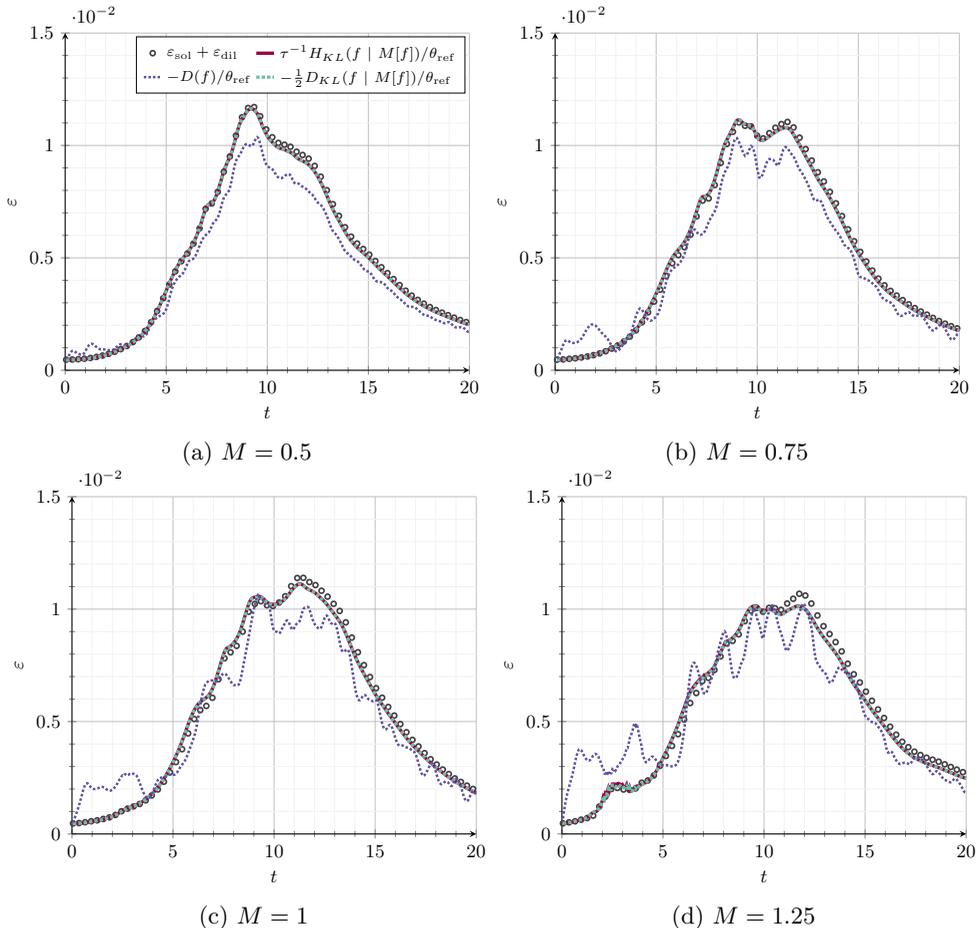
\begin{figure}
        \centering
        \subfloat[$M = 0.5$]{\adjustbox{width=0.48\linewidth, valign=b}{\begin{tikzpicture}[spy using outlines={rectangle, height=3cm,width=2.5cm, magnification=3, connect spies}]
    \begin{axis}
    [
        axis line style={latex-latex},
        axis y line=left,
        axis x line=left,
        clip mode=individual,
        xmode=linear, 
        ymode=linear,
        xlabel = {$t$},
        ylabel = {$\varepsilon$},
        xmin = 0, xmax = 20,
        ymin = 0, ymax = 0.015,
        legend cell align={left},
        legend columns=2, 
        transpose legend,
        legend style={font=\scriptsize, at={(0.99, 0.99)}, anchor=north east},
        x tick label style={/pgf/number format/.cd, fixed, fixed zerofill, precision=0, /tikz/.cd},
        scale = 1,
        grid=both,
        grid style={line width=.1pt, draw=gray!10},
        major grid style={line width=.2pt,draw=gray!50},
        minor x tick num=4,
        minor y tick num=4,
    ]
                \pgfplotsset{
        compat=1.11,
        legend image code/.code={
        \draw[mark repeat=2,mark phase=2]
        plot coordinates {
        (0cm,0cm)
        (0.15cm,0cm)        
        (0.3cm,0cm)         
        };%
        }
        }    
        \addplot[color=black!75, style={thick}, only marks, mark=o, mark options={scale=0.6}, mark repeat = 6, mark phase = 0]  table[x=t, y=dis, col sep=comma]{./figs/data/bgktgv-M0p5-Re1600-p3-n128-post.csv};
        \addlegendentry{$\varepsilon_{\text{sol}} + \varepsilon_{\text{dil}}$};
        
        \addplot[color=PlotColor3, style={very thick, densely dotted}] table[x=t, y=cfflogf, col sep=comma]{./figs/data/bgktgv-M0p5-Re1600-p3-n128-post.csv};
        \addlegendentry{$-D(f)/\theta_\tref$};
        
        \addplot[color=PlotColor1, style={ultra thick}] table[x=t, y=flogfgrcptau, col sep=comma]{./figs/data/bgktgv-M0p5-Re1600-p3-n128-post.csv};
        \addlegendentry{$\tau^{-1} H_{KL} (f\ | \ M[f])/\theta_\tref$};
        
        \addplot[color=PlotColor2, style={ultra thick, densely dotted}] table[x=t, y=cfflogfg, col sep=comma]{./figs/data/bgktgv-M0p5-Re1600-p3-n128-post.csv};
        \addlegendentry{$-\frac{1}{2}D_{KL} (f\ | \ M[f])/\theta_\tref$};

    \end{axis}
\end{tikzpicture}}}
        \subfloat[$M = 0.75$]{\adjustbox{width=0.48\linewidth, valign=b}{\begin{tikzpicture}[spy using outlines={rectangle, height=3cm,width=2.5cm, magnification=3, connect spies}]
    \begin{axis}
    [
        axis line style={latex-latex},
        axis y line=left,
        axis x line=left,
        clip mode=individual,
        xmode=linear, 
        ymode=linear,
        xlabel = {$t$},
        ylabel = {$\varepsilon$},
        xmin = 0, xmax = 20,
        ymin = 0, ymax = 0.015,
        legend cell align={left},
        legend style={font=\scriptsize, at={(0.97, 0.97)}, anchor=north east},
        x tick label style={/pgf/number format/.cd, fixed, fixed zerofill, precision=0, /tikz/.cd},
        scale = 1,
        grid=both,
        grid style={line width=.1pt, draw=gray!10},
        major grid style={line width=.2pt,draw=gray!50},
        minor x tick num=4,
        minor y tick num=4,
    ]
        
        \addplot[color=black!75, style={thick}, only marks, mark=o, mark options={scale=0.6}, mark repeat = 3, mark phase = 0]  table[x=t, y=dis, col sep=comma]{./figs/data/bgktgv-M0p75-Re1600-p3-n128-post.csv};
        
        \addplot[color=PlotColor1, style={ultra thick}] table[x=t, y=flogfgrcptau, col sep=comma]{./figs/data/bgktgv-M0p75-Re1600-p3-n128-post.csv};
        
        \addplot[color=PlotColor2, style={ultra thick, densely dotted}] table[x=t, y=cfflogfg, col sep=comma]{./figs/data/bgktgv-M0p75-Re1600-p3-n128-post.csv};
        
        \addplot[color=PlotColor3, style={very thick, densely dotted}] table[x=t, y=cfflogf, col sep=comma]{./figs/data/bgktgv-M0p75-Re1600-p3-n128-post.csv};
        
    \end{axis}
\end{tikzpicture}}}
        \newline
        \subfloat[$M = 1$]{\adjustbox{width=0.48\linewidth, valign=b}{\begin{tikzpicture}[spy using outlines={rectangle, height=3cm,width=2.5cm, magnification=3, connect spies}]
    \begin{axis}
    [
        axis line style={latex-latex},
        axis y line=left,
        axis x line=left,
        clip mode=individual,
        xmode=linear, 
        ymode=linear,
        xlabel = {$t$},
        ylabel = {$\varepsilon$},
        xmin = 0, xmax = 20,
        ymin = 0, ymax = 0.015,
        legend cell align={left},
        legend style={font=\scriptsize, at={(0.97, 0.97)}, anchor=north east},
        x tick label style={/pgf/number format/.cd, fixed, fixed zerofill, precision=0, /tikz/.cd},
        scale = 1,
        grid=both,
        grid style={line width=.1pt, draw=gray!10},
        major grid style={line width=.2pt,draw=gray!50},
        minor x tick num=4,
        minor y tick num=4,
    ]
        
        \addplot[color=black!75, style={thick}, only marks, mark=o, mark options={scale=0.6}, mark repeat = 6, mark phase = 0]  table[x=t, y=dis, col sep=comma]{./figs/data/bgktgv-M1p0-Re1600-p3-n128-post.csv};
        
        \addplot[color=PlotColor1, style={ultra thick}] table[x=t, y=flogfgrcptau, col sep=comma]{./figs/data/bgktgv-M1p0-Re1600-p3-n128-post.csv};
        
        \addplot[color=PlotColor2, style={ultra thick, densely dotted}] table[x=t, y=cfflogfg, col sep=comma]{./figs/data/bgktgv-M1p0-Re1600-p3-n128-post.csv};
        
        \addplot[color=PlotColor3, style={very thick, densely dotted}] table[x=t, y=cfflogf, col sep=comma]{./figs/data/bgktgv-M1p0-Re1600-p3-n128-post.csv};
    \end{axis}
\end{tikzpicture}}}
        \subfloat[$M = 1.25$]{\adjustbox{width=0.48\linewidth, valign=b}{\begin{tikzpicture}[spy using outlines={rectangle, height=3cm,width=2.5cm, magnification=3, connect spies}]
    \begin{axis}
    [
        axis line style={latex-latex},
        axis y line=left,
        axis x line=left,
        clip mode=individual,
        xmode=linear, 
        ymode=linear,
        xlabel = {$t$},
        ylabel = {$\varepsilon$},
        xmin = 0, xmax = 20,
        ymin = 0, ymax = 0.015,
        legend cell align={left},
        legend style={font=\scriptsize, at={(0.97, 0.97)}, anchor=north east},
        x tick label style={/pgf/number format/.cd, fixed, fixed zerofill, precision=0, /tikz/.cd},
        scale = 1,
        grid=both,
        grid style={line width=.1pt, draw=gray!10},
        major grid style={line width=.2pt,draw=gray!50},
        minor x tick num=4,
        minor y tick num=4,
    ]
        
        \addplot[color=black!75, style={thick}, only marks, mark=o, mark options={scale=0.6}, mark repeat = 6, mark phase = 0]  table[x=t, y=dis, col sep=comma]{./figs/data/bgktgv-M1p25-Re1600-p3-n128-post.csv};
        
        \addplot[color=PlotColor1, style={ultra thick}] table[x=t, y=flogfgrcptau, col sep=comma]{./figs/data/bgktgv-M1p25-Re1600-p3-n128-post.csv};
        
        \addplot[color=PlotColor2, style={ultra thick, densely dotted}] table[x=t, y=cfflogfg, col sep=comma]{./figs/data/bgktgv-M1p25-Re1600-p3-n128-post.csv};
        
        \addplot[color=PlotColor3, style={very thick, densely dotted}] table[x=t, y=cfflogf, col sep=comma]{./figs/data/bgktgv-M1p25-Re1600-p3-n128-post.csv};
    \end{axis}
\end{tikzpicture}}}
        \caption{Comparison of the temporal evolution of the total dissipation, Boltzmann entropy production rate, relative entropy, and relative entropy production rate at varying Mach numbers for the $Re = 1600$ case.}
        \label{fig:norment1600}  
    \end{figure}

The behavior of the relative entropy dissipation $D_{KL}(f\ | \ M[f])$ in \cref{eq:reldis} can also be compared to the viscous dissipation and Boltzmann entropy dissipation. The relative entropy dissipation profiles, normalized by the reference temperature, are shown for the $Re = 400$ and $Re = 1600$ cases in \cref{fig:norment400} and \cref{fig:norment1600}, respectively. \emph{The key observation is that the relative entropy dissipation profiles are markedly similar to the viscous dissipation profiles}, with excellent agreement between the two across all considered Mach and Reynolds numbers. At lower Mach number, better agreement was observed, although this effect was only marginal. This behavior is in contrast to the behavior of the Boltzmann entropy dissipation, which was significantly more oscillatory and deviated from the viscous dissipation profiles. 

We note here that a scaling factor of $\half$ was applied to the relative entropy $D_{KL}(f\ | \ M[f])$ which was not applied to the Boltzmann entropy dissipation $D(f)$. It appears that the relative entropy dissipation obeys a symmetry in the form of $D_{KL}(f\ |\ M[f]) \approx -D_{KL}(M[f]\ |\ f)$, which somewhat mimics the symmetry of the full Boltzmann collision operator. This symmetry leads to the approximation where the normalized relative entropy (normalized by the time scale $\tau$), $\tau^{-1}H_{KL}(f\ |\ M[f])$, is closely related to the relative entropy dissipation by the relation $\tau^{-1}H_{KL}(f\ |\ M[f]) \approx - \half D_{KL}(f\ |\ M[f])$, which can evidently be seen in \cref{fig:norment400} and \cref{fig:norment1600}. \emph{Therefore, this relation implies that the relative entropy itself serves as a proxy for the relative entropy dissipation rate which, by extension, serves as a proxy for the viscous dissipation rate.} This observation establishes a connection between different levels of statistical description of the system. Specifically, the relative entropy, being a zeroth-order quantity of the distribution function with respect to both space and time, is closely linked to relative entropy dissipation and viscous dissipation rates, both higher-order quantities with respect to space and/or time, such that even at the level of coarse-grained macroscopic quantities, there is a clear correspondence to the underlying statistical dynamics governed by the Boltzmann equation.

\subsection{Subgrid-scale features}
Following the analysis of the integrated macroscopic and kinetic quantities, we now consider the subgrid-scale behavior in the flow. In the realm of turbulence modeling, it is often desirable to only resolve scales larger than the minimum turbulent length scales and model the effects of the unresolved scales. As such, it is often the case that spatial averaging or filtering is employed to separate the resolved and unresolved scales of the flow. The cutoff scale corresponds to the grid resolution in a computational simulation, with smaller scales being classified as subgrid-scale features. The present goal is to analyze the effects of subgrid-scale features of interest to continuum models (e.g., subgrid-scale dissipation) \emph{arising from spatial filtering of the distribution function itself} which have a different mathematical structure in comparison to subgrid-scale features arising from spatial filtering of the Navier--Stokes equations. 

We consider a filtering operation of the form
\begin{equation}
    \overline{f}(\mathbf{x}, \mathbf{u}, t) = \frac{\int_{K(\mathbf{x})} f (\mathbf{x}, \mathbf{u}, t)\ \mathrm{d}\mathbf{x}}{\int_{K(\mathbf{x})}  \mathrm{d}\mathbf{x}},
\end{equation}
where $K(\mathbf{x})$ represents a local filter width. Furthermore, we consider a filtered form of an arbitrary quantity of interest $Q\left(f(\mathbf{x}, \mathbf{u}, t)\right)$ as
\begin{equation}
    \overline{Q\left(f(\mathbf{x}, \mathbf{u}, t)\right)} = \frac{\int_{K(\mathbf{x})} Q\left(f(\mathbf{x}, \mathbf{u}, t)\right)\ \mathrm{d}\mathbf{x}}{\int_{K(\mathbf{x})}  \mathrm{d}\mathbf{x}}. 
\end{equation}
As these quantities are typically nonlinear with respect to $f$, the filtering operation is noncommutative, such that there exists a residual (i.e., subgrid) term, defined as 
\begin{equation}\label{eq:residual}
    R_Q\left(f\right) = \overline{Q\left(f\right)} - Q\left(\overline{f}\right),
\end{equation}
representing the difference between the filtered quantity, which is not known \emph{a priori}, and the quantity as computed from the filtered distribution function, which can be readily calculated. Appropriate modeling of these residual subgrid terms is critical in underresolved simulations of turbulent flows to account for the effects of subgrid-scale turbulent flow dynamics. In particular, in the realm of large eddy simulation, the filter width is ostensibly chosen such as to encompass the energy-containing eddies of the flow while the effects of the smaller unresolved scales, which correspond to the scales at which dissipation effects are dominant, are typically modeled via additional numerical dissipation mechanisms in the governing equations. As such, it is of interest to see if the additional information encoded in the distribution function can help with modeling the unresolved (subgrid-scale) energy dissipation rates and shear stresses. 

Given that the task of modeling Reynolds stresses (i.e., subgrid momentum transport) with the Boltzmann equation is ostensibly as challenging as in the Navier--Stokes framework, the possibilities of alternate approaches to subgrid-scale turbulence modeling through the Boltzmann equation are not immediately evident. However, the observations in \cref{ssec:kinetic} indicate that the relative entropy functional is strongly linked to the macroscopic viscous dissipation, such that information in the distribution function linked to entropy measures may be useful for modeling subgrid-scale energy dissipation and shear stresses (e.g., one may attempt to model subgrid relative entropy as a surrogate for subgrid dissipation). We consider the behavior of the relative entropy functional in terms of filtering ($Q(f) = D_{KL}(f)$) to qualitatively study its behavior in comparison to the subgrid dissipation ($Q(f) = \varepsilon(f) = \varepsilon_s(f) + \varepsilon_d(f)$) to qualitatively identify any potential mechanisms for modeling subgrid dissipation. The filter width $K(\mathbf{x})$ here is taken as the element-width $\Delta x$, such that the filtered quantities were computed as an element-wise average of the high-order solution.


    \begin{figure}
        \centering
        \subfloat[Filtered relative entropy]{
        \adjustbox{width=0.35\linewidth, valign=b}{\includegraphics[]{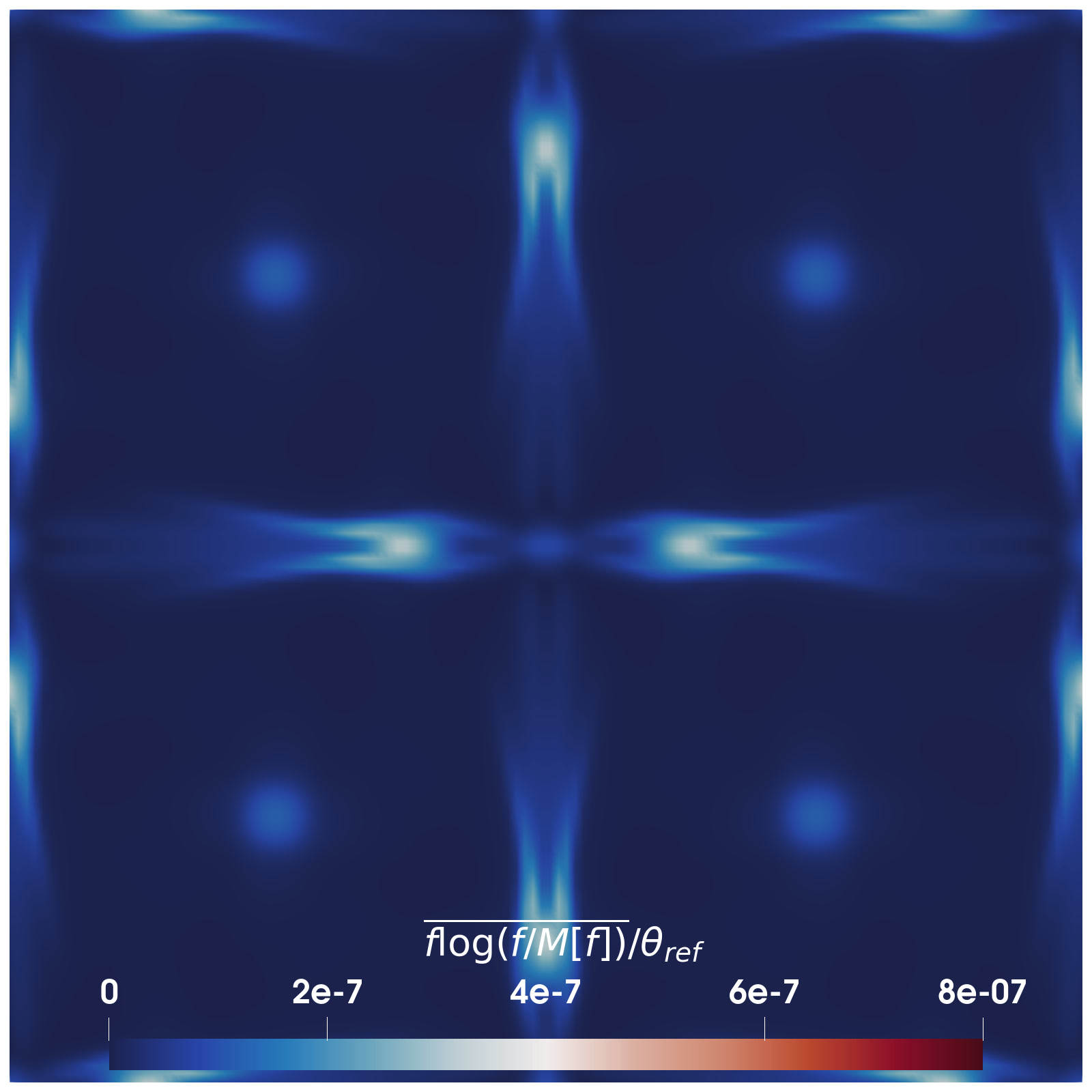}}}
        \hspace{2em}
        \subfloat[Filtered dissipation]{
        \adjustbox{width=0.35\linewidth, valign=b}{\includegraphics[]{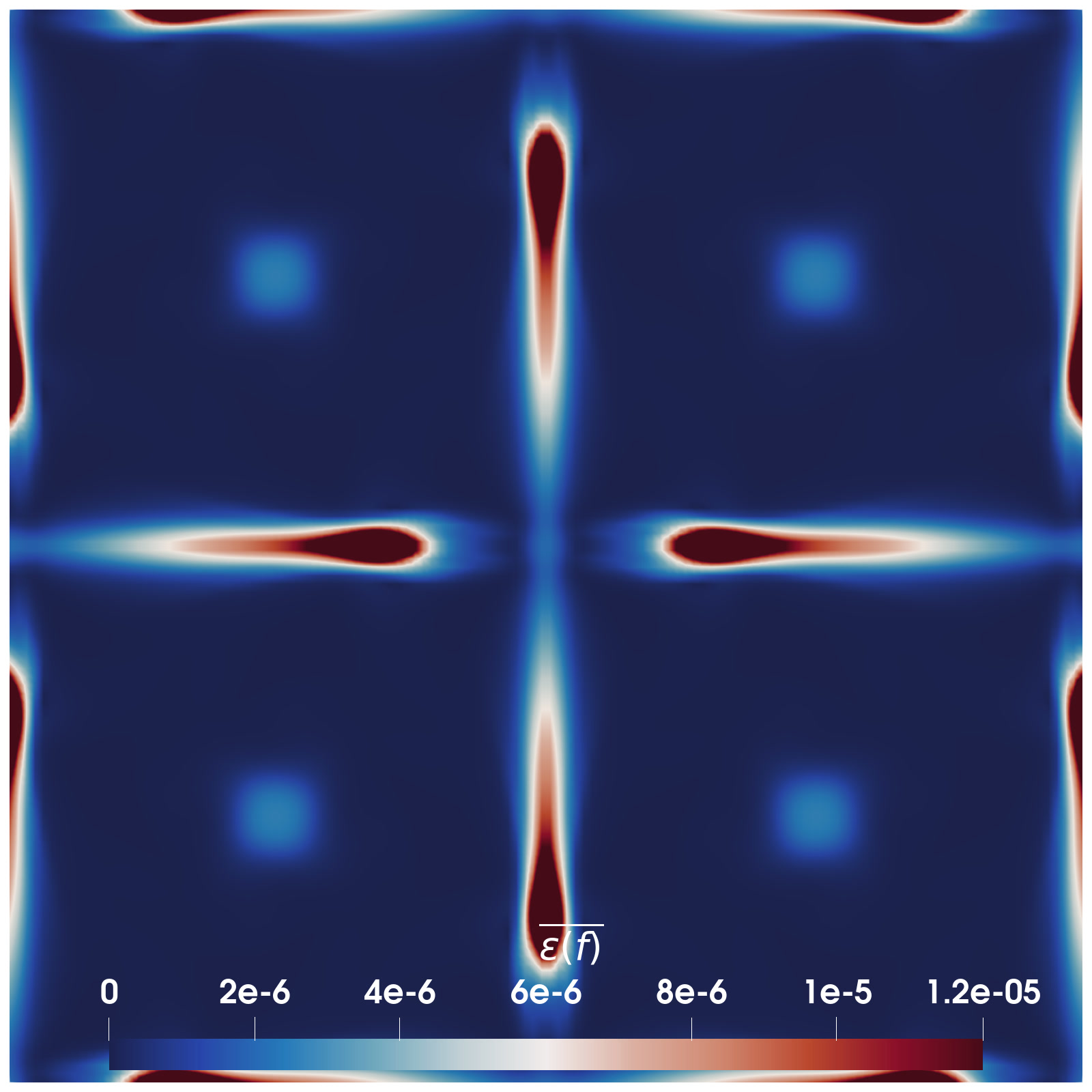}}}
        \newline
        \subfloat[Relative entropy of filtered solution]{
        \adjustbox{width=0.35\linewidth, valign=b}{\includegraphics[]{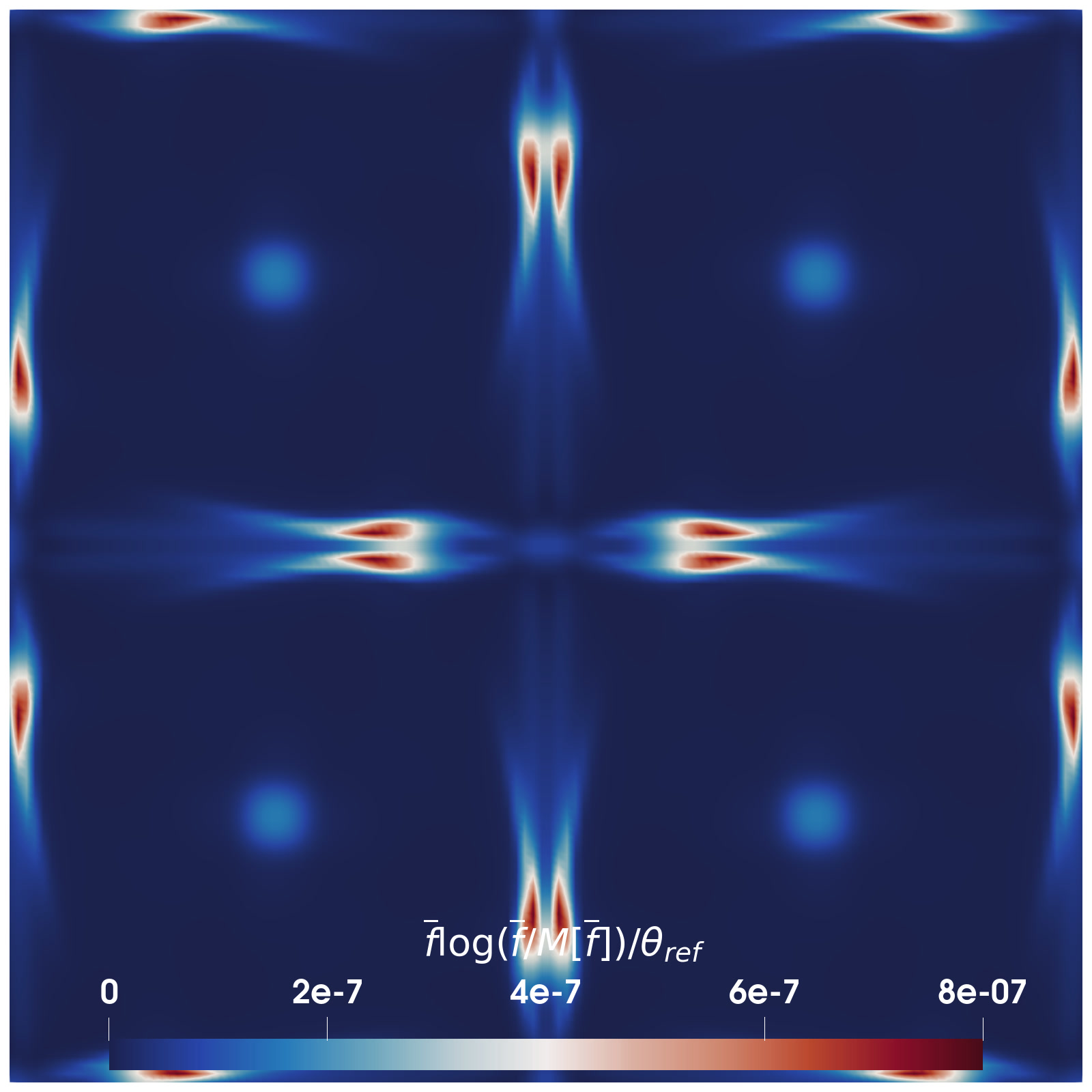}}}
        \hspace{2em}
        \subfloat[Dissipation of filtered solution]{
        \adjustbox{width=0.35\linewidth, valign=b}{\includegraphics[]{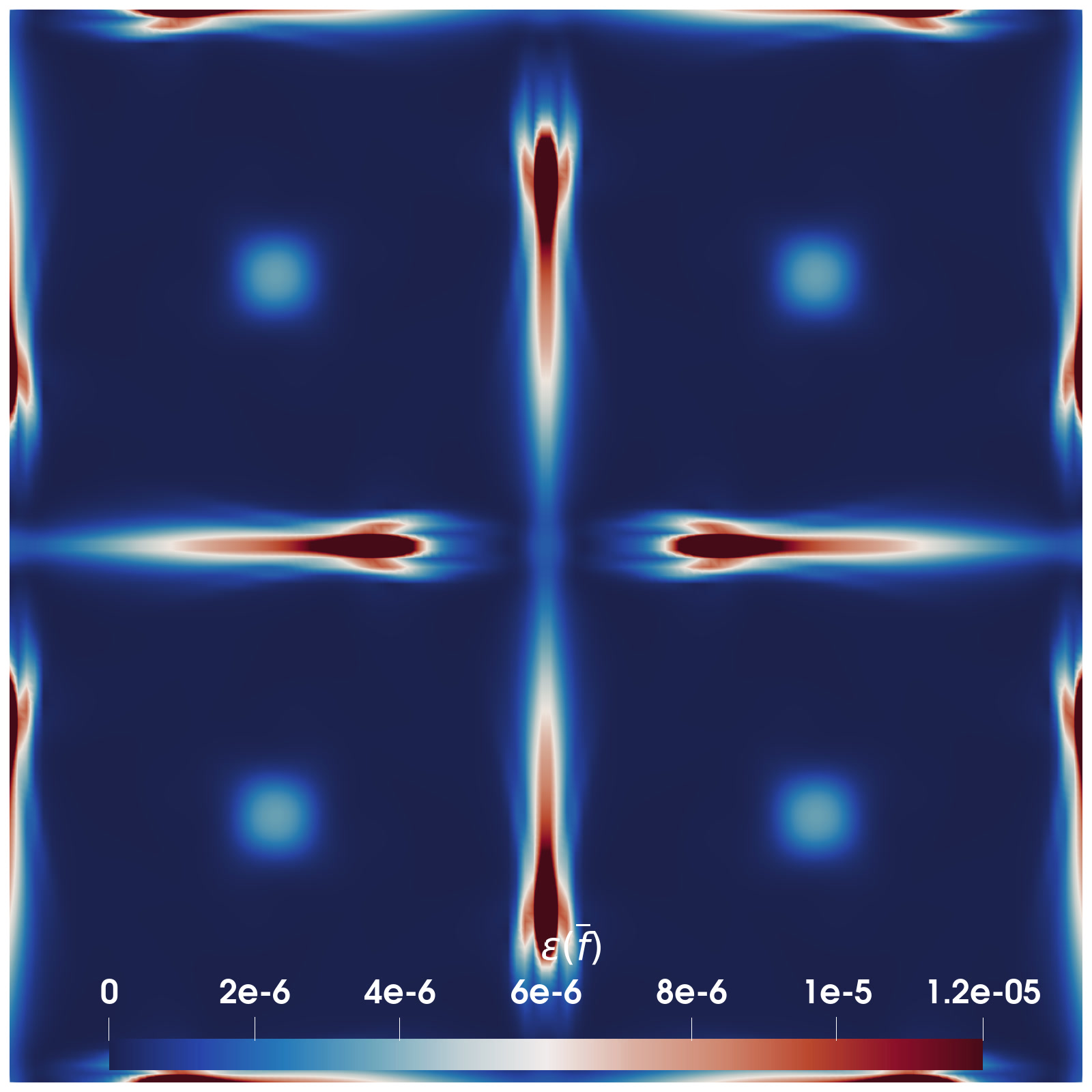}}}
        \newline
        \subfloat[Subgrid relative entropy]{
        \adjustbox{width=0.35\linewidth, valign=b}{\includegraphics[]{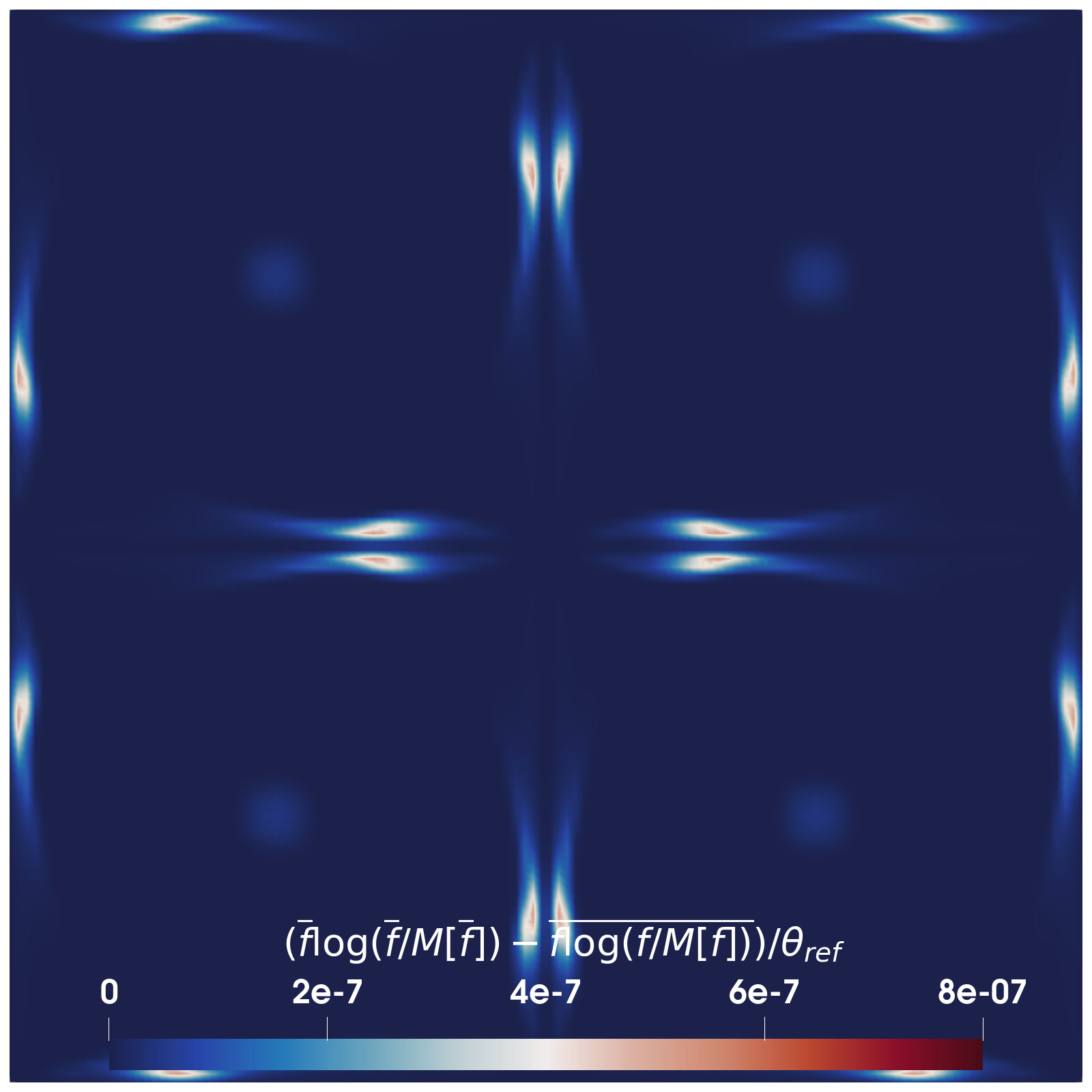}}}
        \hspace{2em}
        \subfloat[Subgrid dissipation]{
        \adjustbox{width=0.35\linewidth, valign=b}{\includegraphics[]{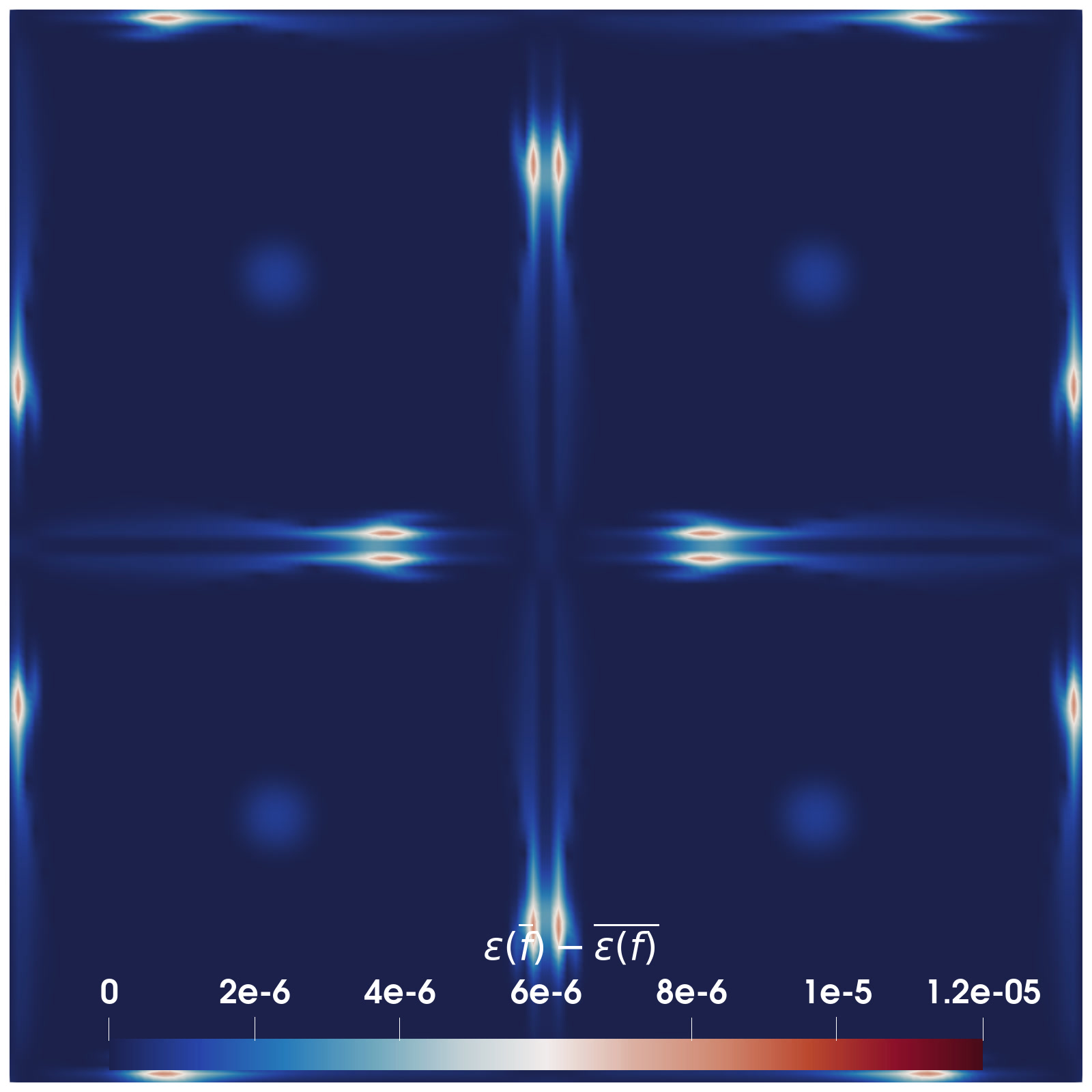}}}
        \newline
        \caption{
        Comparison of the filtered relative entropy/dissipation (top row), relative entropy/dissipation of the filtered solution (middle row), and relative entropy/dissipation residual (bottom row)  for the $M=0.5$, $Re = 1600$ case at $t=10$ on the plane $z = \pi$.}
        \label{fig:entropyfilter_M0p5Re1600z}
    \end{figure}

    \begin{figure}
        \centering
        \subfloat[Filtered relative entropy]{
        \adjustbox{width=0.35\linewidth, valign=b}{\includegraphics[]{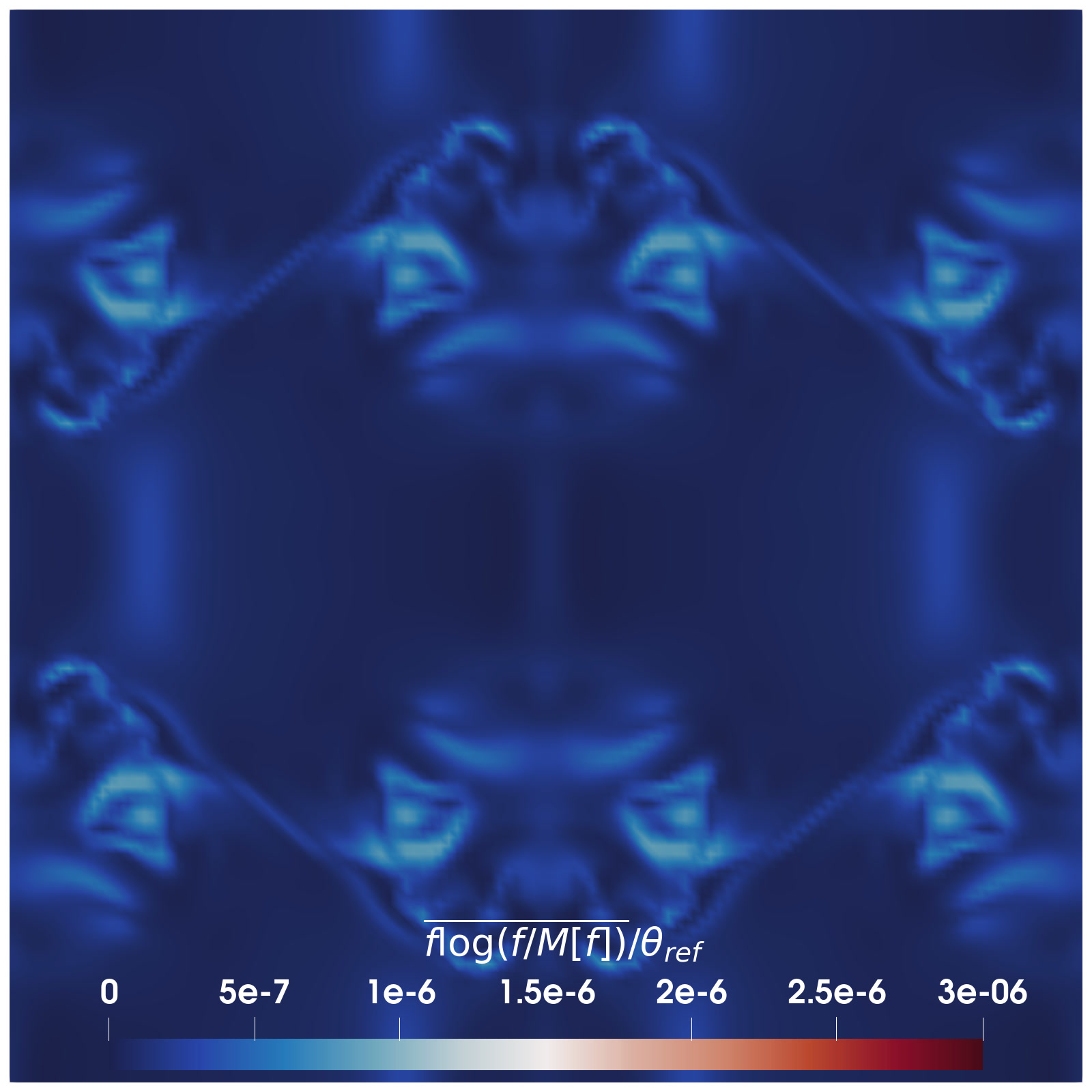}}}
        \hspace{2em}
        \subfloat[Filtered dissipation]{
        \adjustbox{width=0.35\linewidth, valign=b}{\includegraphics[]{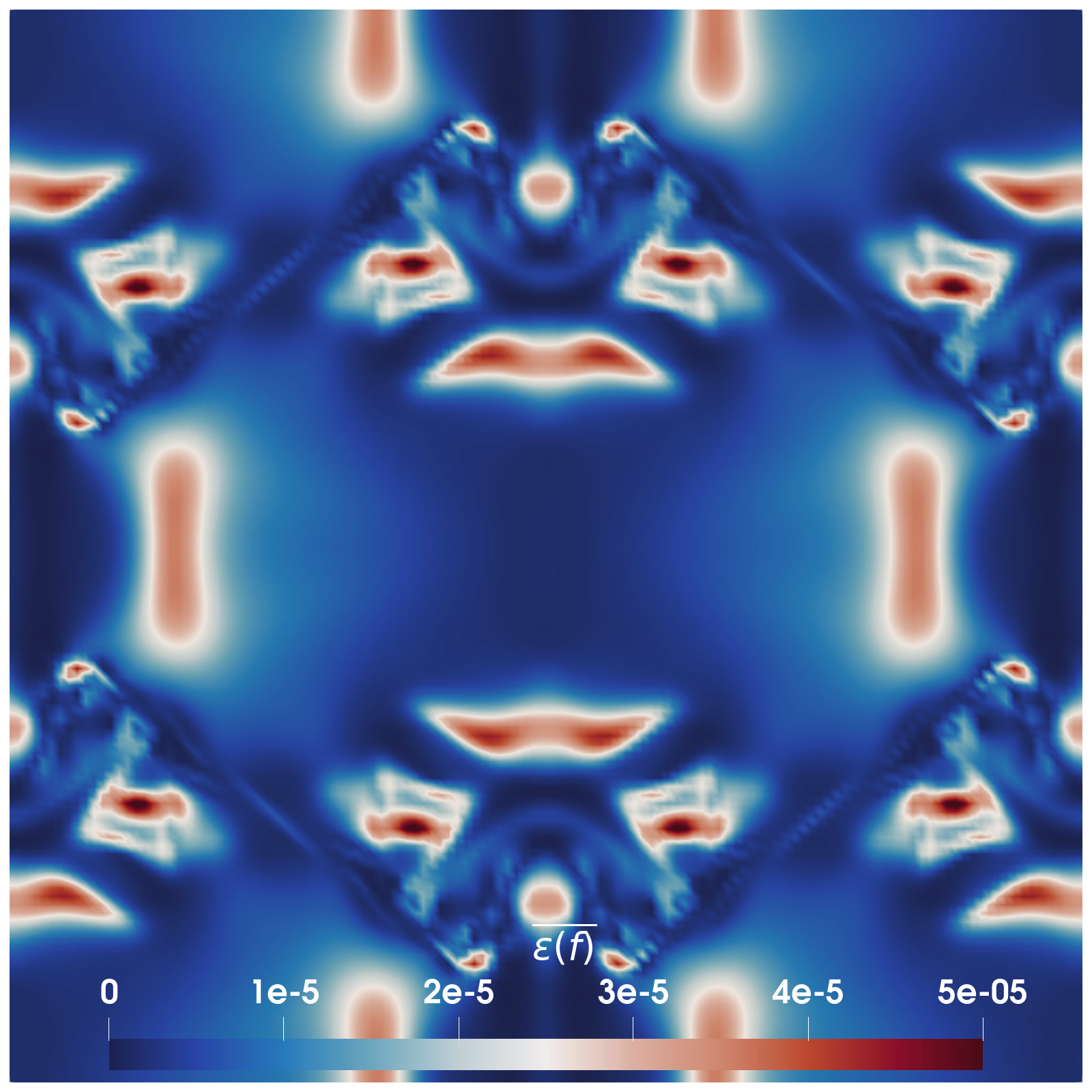}}}
        \newline
        \subfloat[Relative entropy of filtered solution]{
        \adjustbox{width=0.35\linewidth, valign=b}{\includegraphics[]{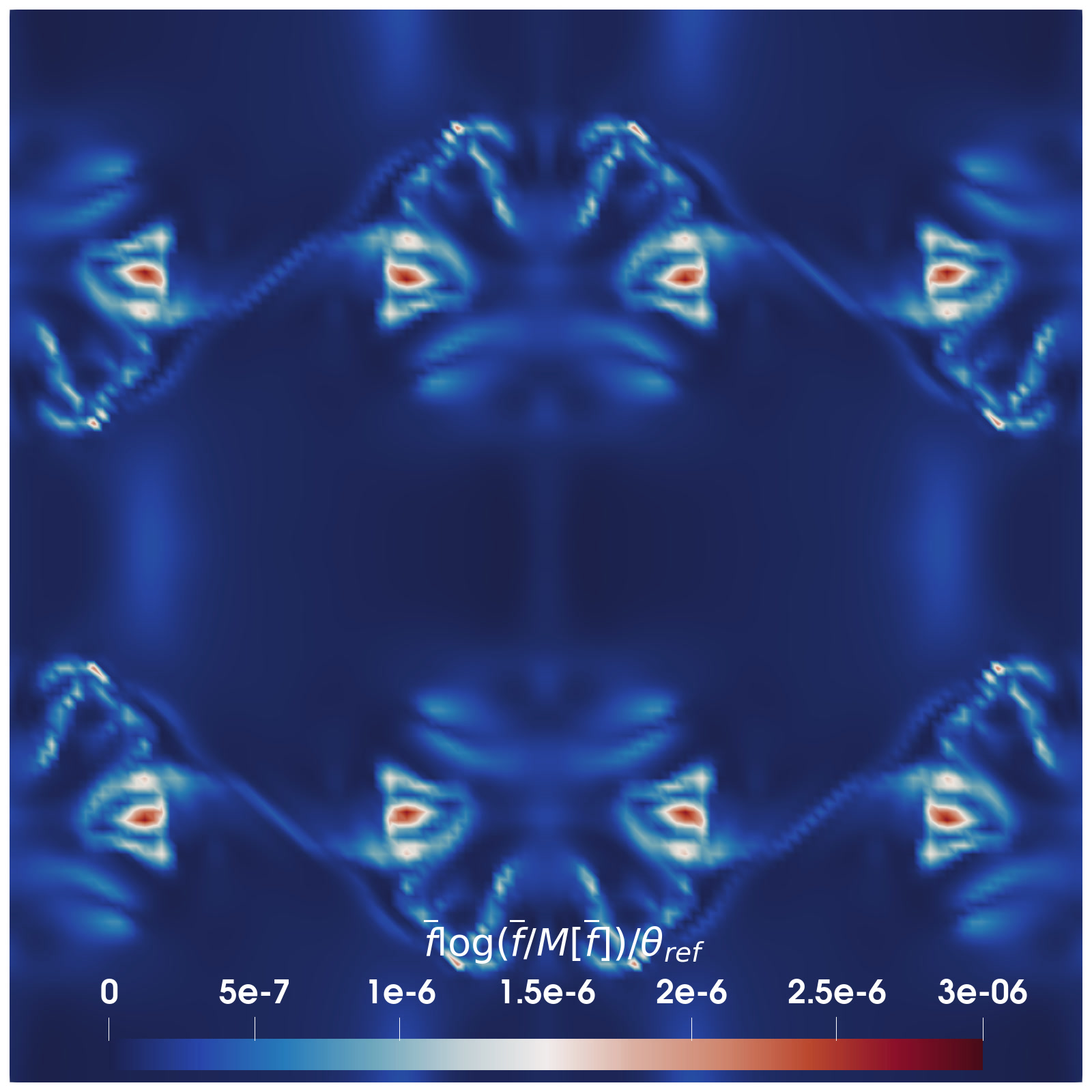}}}
        \hspace{2em}
        \subfloat[Dissipation of filtered solution]{
        \adjustbox{width=0.35\linewidth, valign=b}{\includegraphics[]{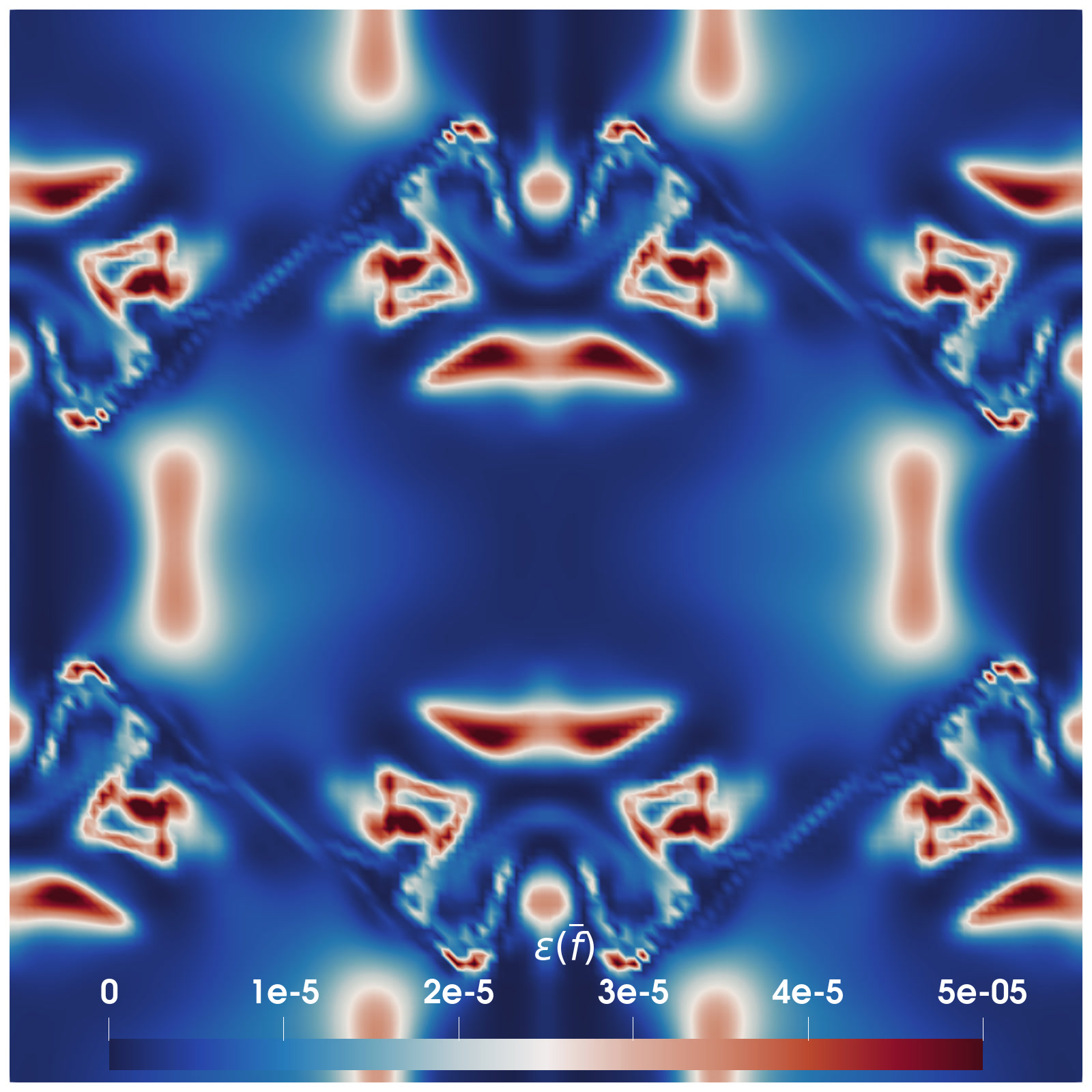}}}
        \newline
        \subfloat[Subgrid relative entropy]{
        \adjustbox{width=0.35\linewidth, valign=b}{\includegraphics[]{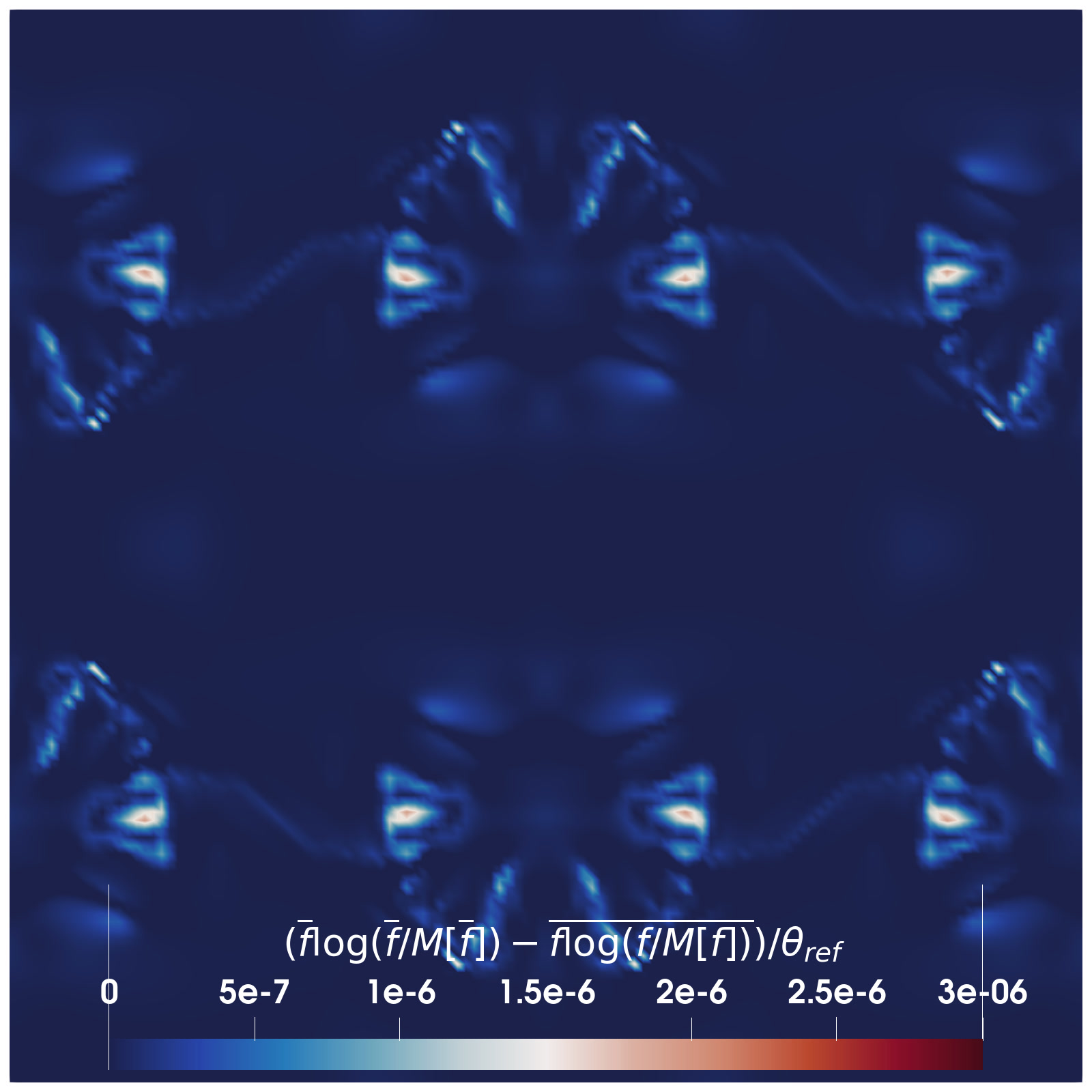}}}
        \hspace{2em}
        \subfloat[Subgrid dissipation]{
        \adjustbox{width=0.35\linewidth, valign=b}{\includegraphics[]{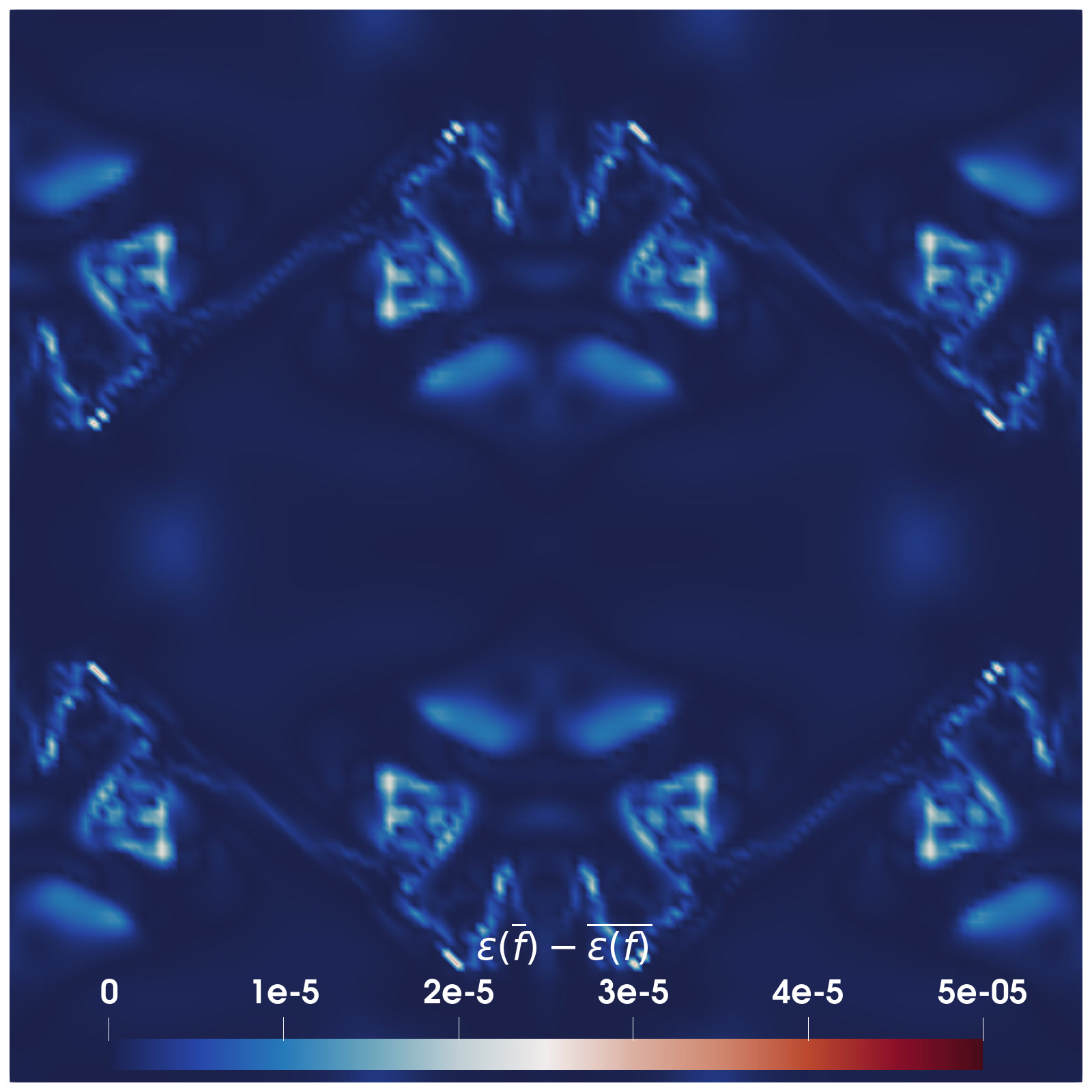}}}
        \newline
        \caption{
        Comparison of the filtered relative entropy/dissipation (top row), relative entropy/dissipation of the filtered solution (middle row), and relative entropy/dissipation residual (bottom row)  for the $M=0.5$, $Re = 1600$ case at $t=10$ on the plane $x = \pi$.}
        \label{fig:entropyfilter_M0p5Re1600x}
    \end{figure}

We first show the effects of filtering on the relative entropy and dissipation for the case of $Re = 1600$ and $M=0.5$. This comparison is shown in \cref{fig:entropyfilter_M0p5Re1600z} on the plane $z = \pi$ and \cref{fig:entropyfilter_M0p5Re1600x} on the plane $x = \pi$ at $t=10$ for the case of $M=0.5$. The first observation is that the relative entropy of the filtered distribution is generally of higher magnitude than the filtered relative entropy. The largest difference in magnitude appeared to be in regions characterized by high gradients in the velocity field (e.g., vortex cores). The second observation is that, at least qualitatively, the subgrid relative entropy appears to be reasonably well-correlated with the resolved relative entropy, such that one may attempt a Boussinesq-type approximation to model subgrid relative entropy as a function of the resolved relative entropy.

    \begin{figure}
        \centering
        \subfloat[Filtered relative entropy]{
        \adjustbox{width=0.35\linewidth, valign=b}{\includegraphics[]{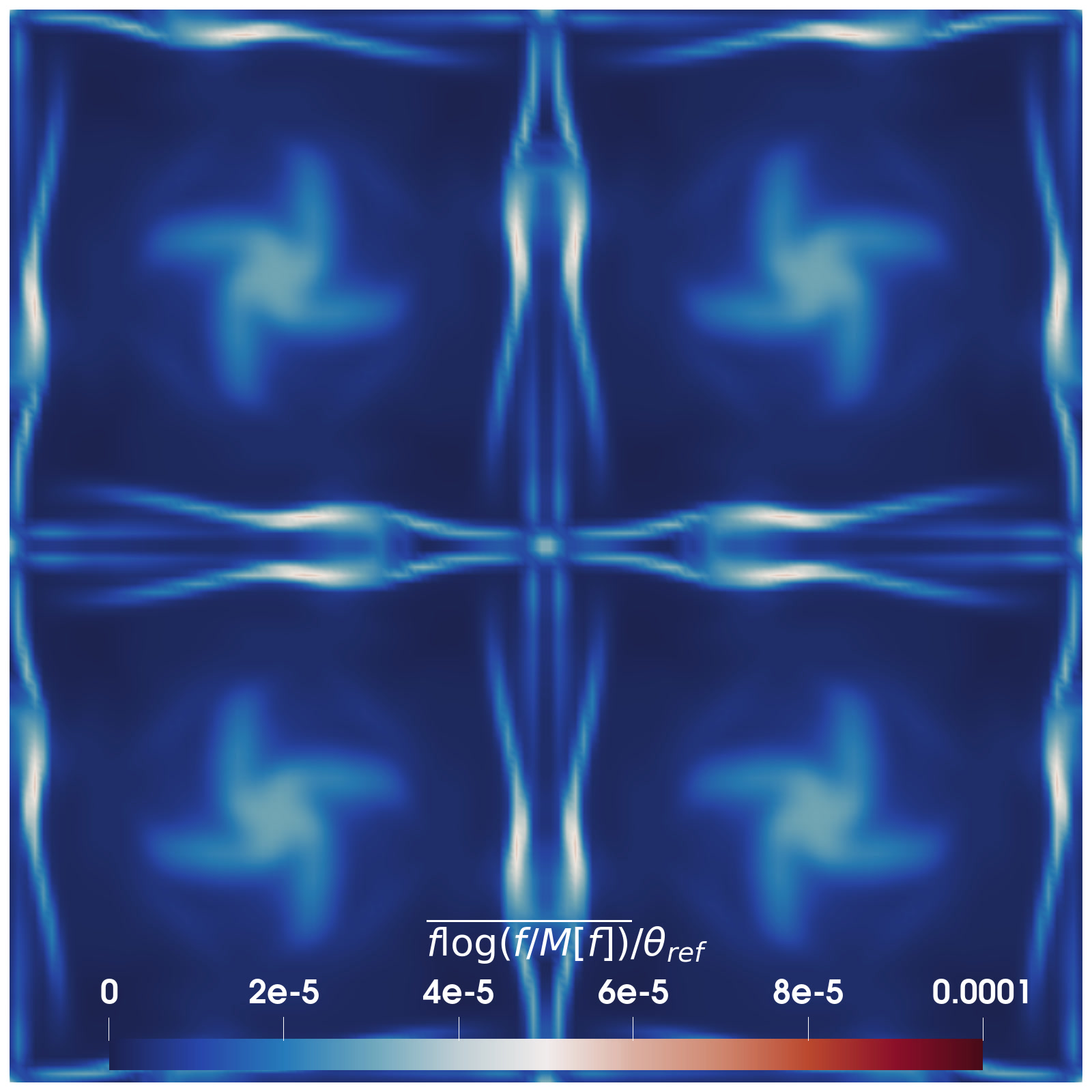}}}
        \hspace{2em}
        \subfloat[Filtered dissipation]{
        \adjustbox{width=0.35\linewidth, valign=b}{\includegraphics[]{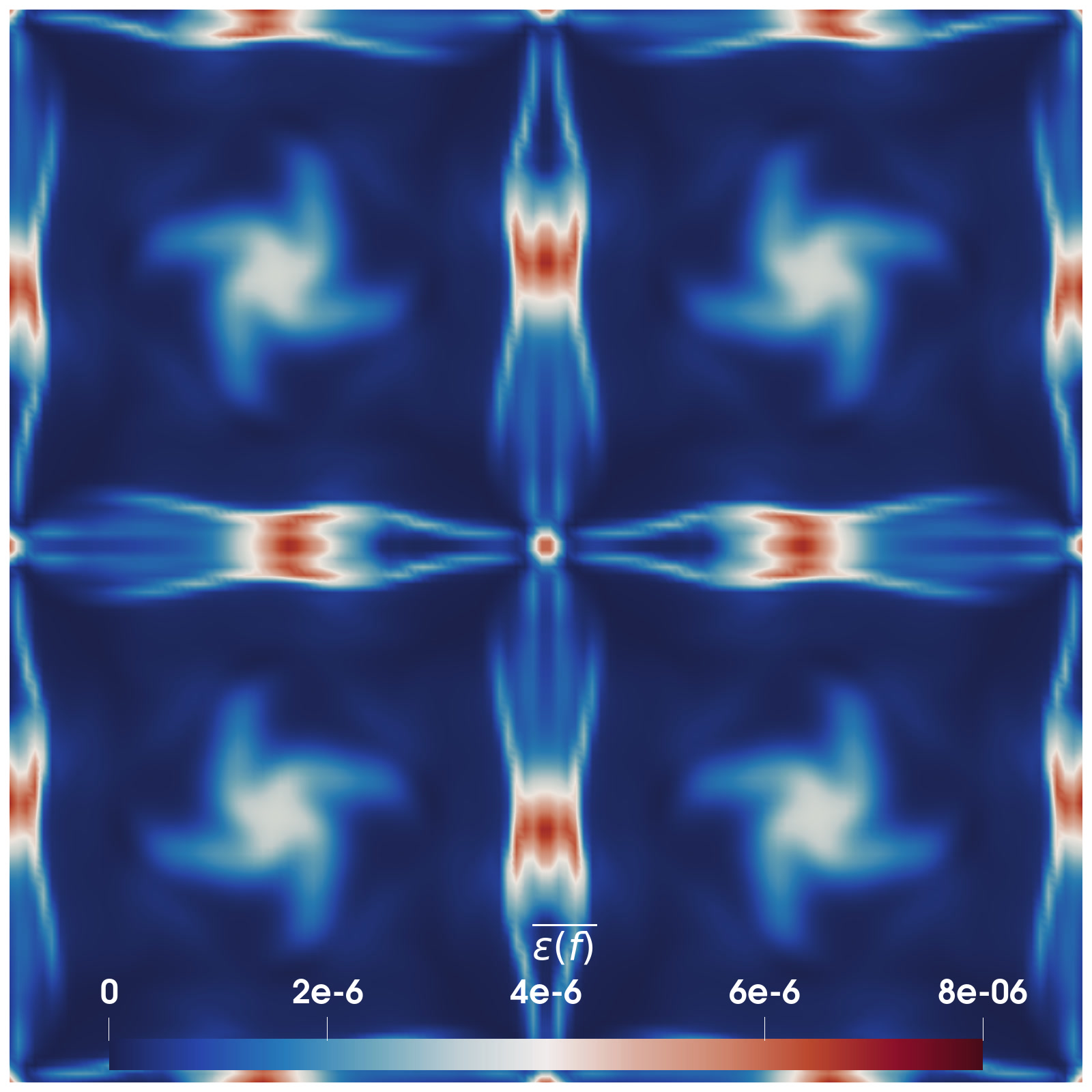}}}
        \newline
        \subfloat[Relative entropy of filtered solution]{
        \adjustbox{width=0.35\linewidth, valign=b}{\includegraphics[]{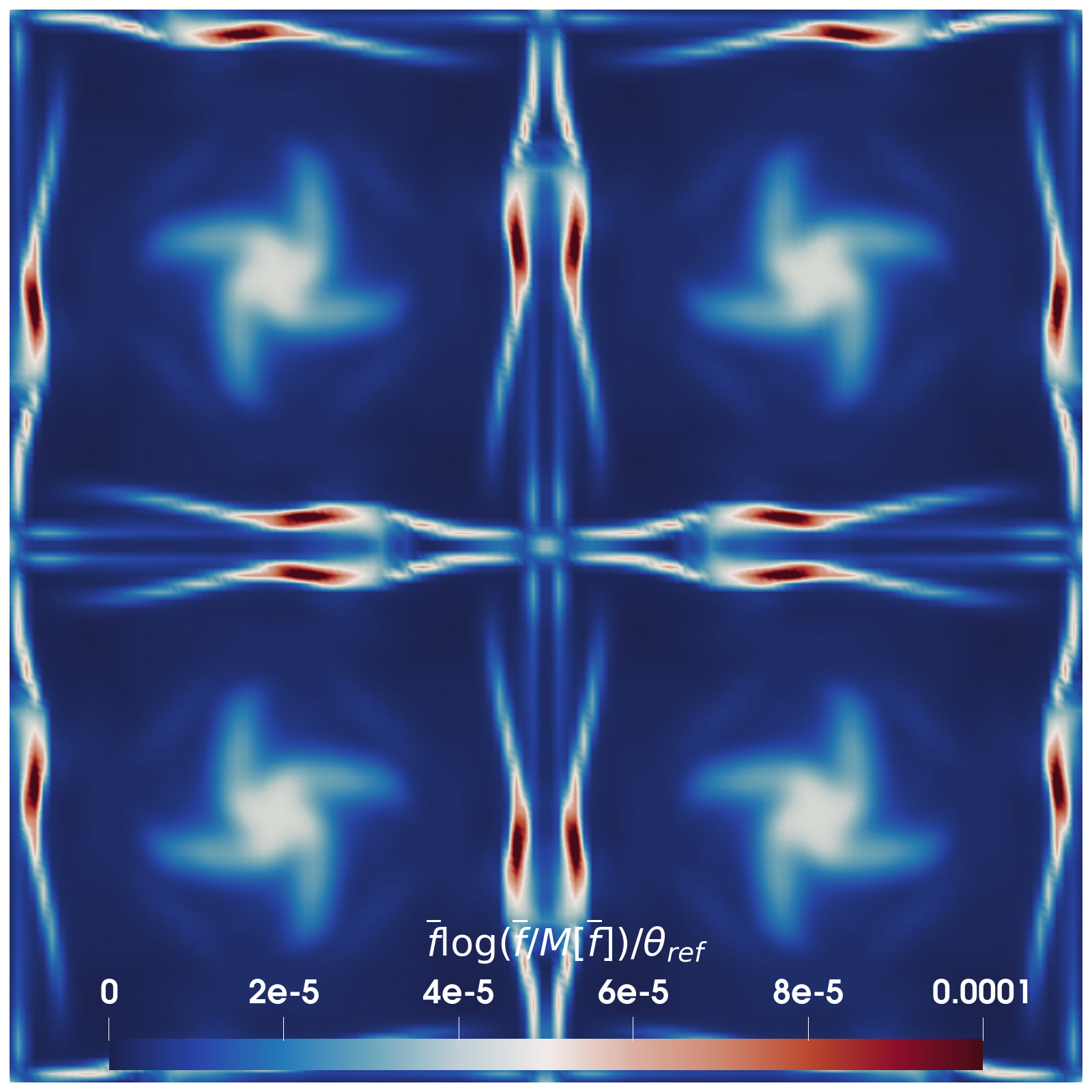}}}
        \hspace{2em}
        \subfloat[Dissipation of filtered solution]{
        \adjustbox{width=0.35\linewidth, valign=b}{\includegraphics[]{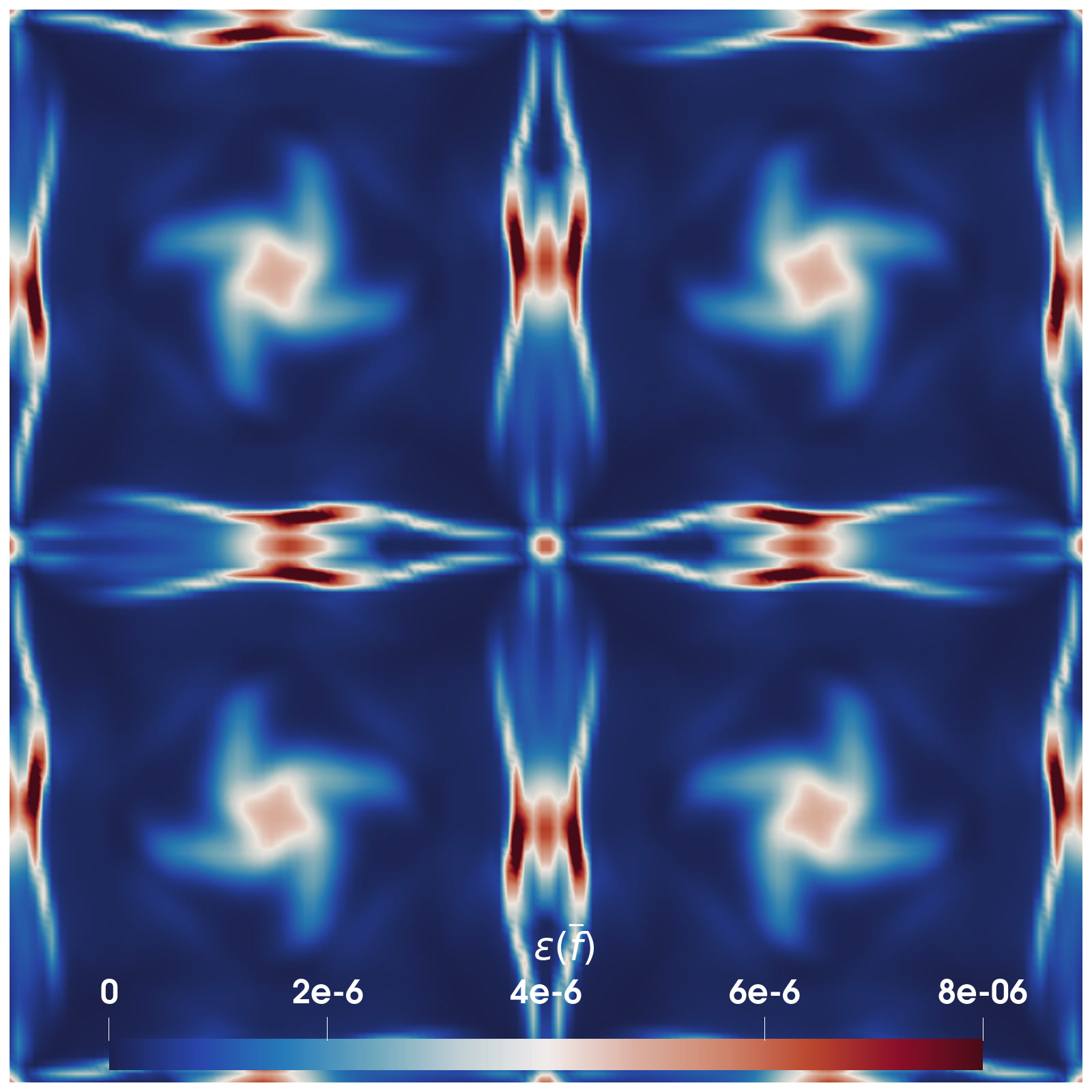}}}
        \newline
        \subfloat[Subgrid relative entropy]{
        \adjustbox{width=0.35\linewidth, valign=b}{\includegraphics[]{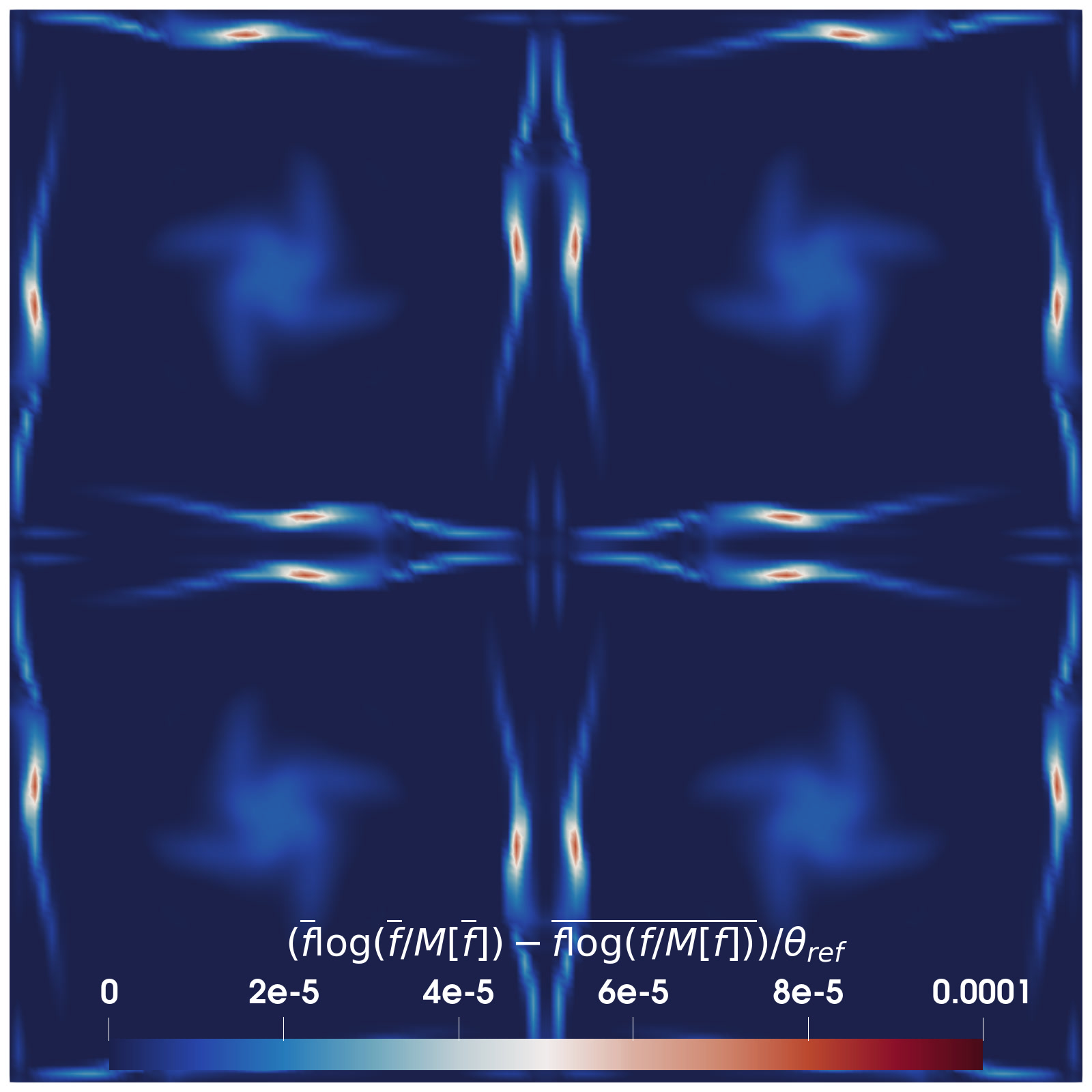}}}
        \hspace{2em}
        \subfloat[Subgrid dissipation]{
        \adjustbox{width=0.35\linewidth, valign=b}{\includegraphics[]{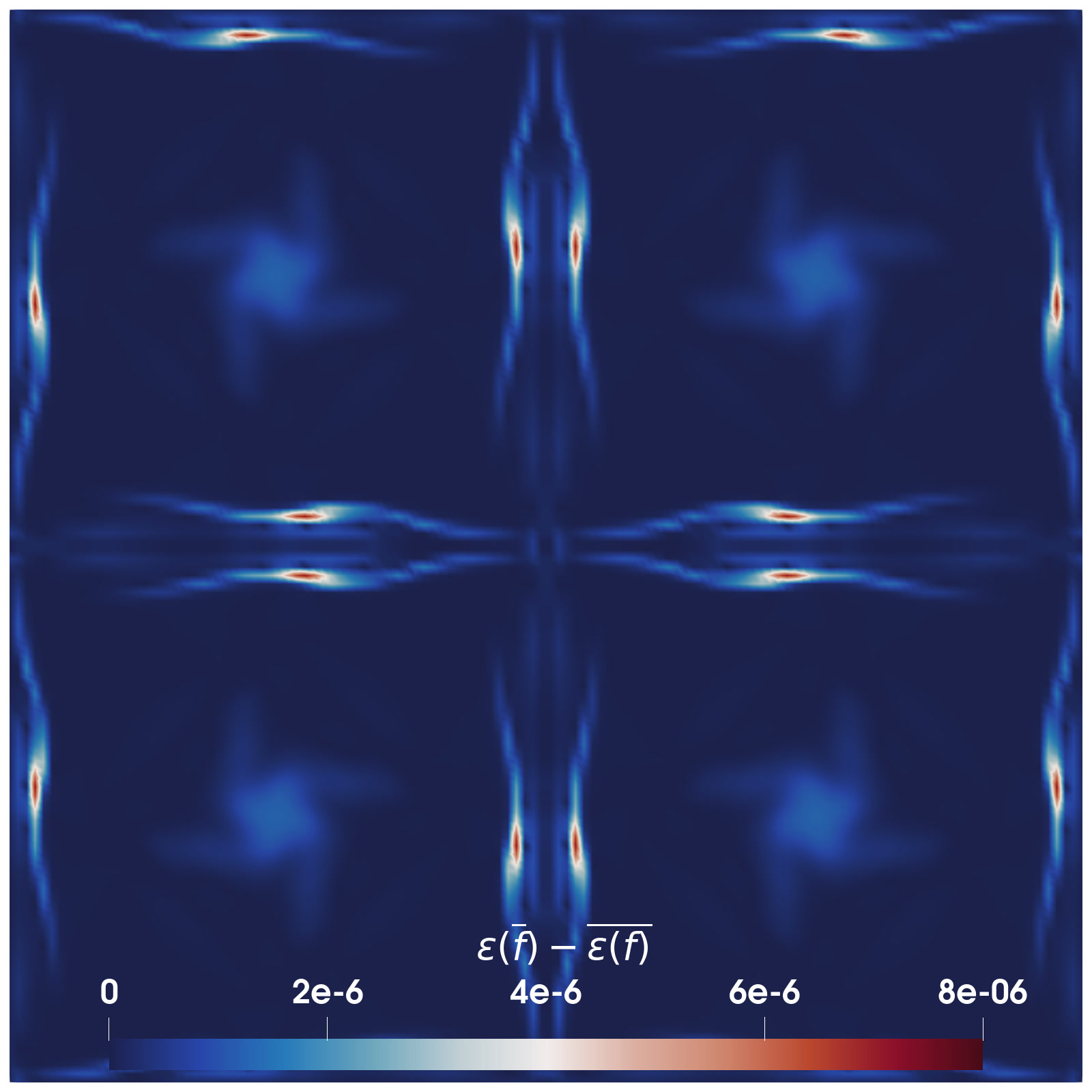}}}
        \newline
        \caption{
        Comparison of the filtered relative entropy/dissipation (top row), relative entropy/dissipation of the filtered solution (middle row), and relative entropy/dissipation residual (bottom row)  for the $M=0.5$, $Re = 1600$ case at $t=10$ on the plane $z = \pi$.}
        \label{fig:entropyfilter_M1p25Re1600z}
    \end{figure}

    \begin{figure}
        \centering
        \subfloat[Filtered relative entropy]{
        \adjustbox{width=0.35\linewidth, valign=b}{\includegraphics[]{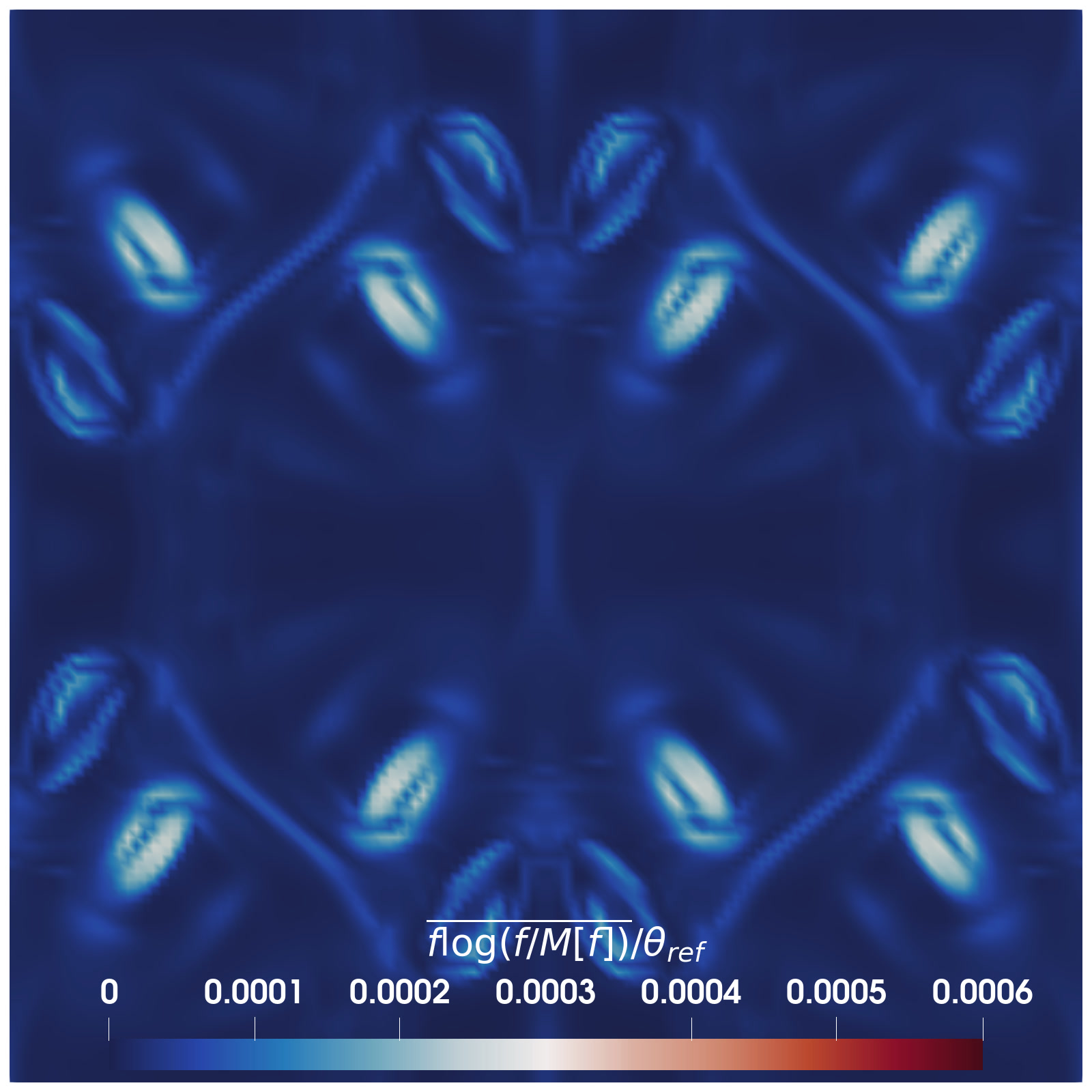}}}
        \hspace{2em}
        \subfloat[Filtered dissipation]{
        \adjustbox{width=0.35\linewidth, valign=b}{\includegraphics[]{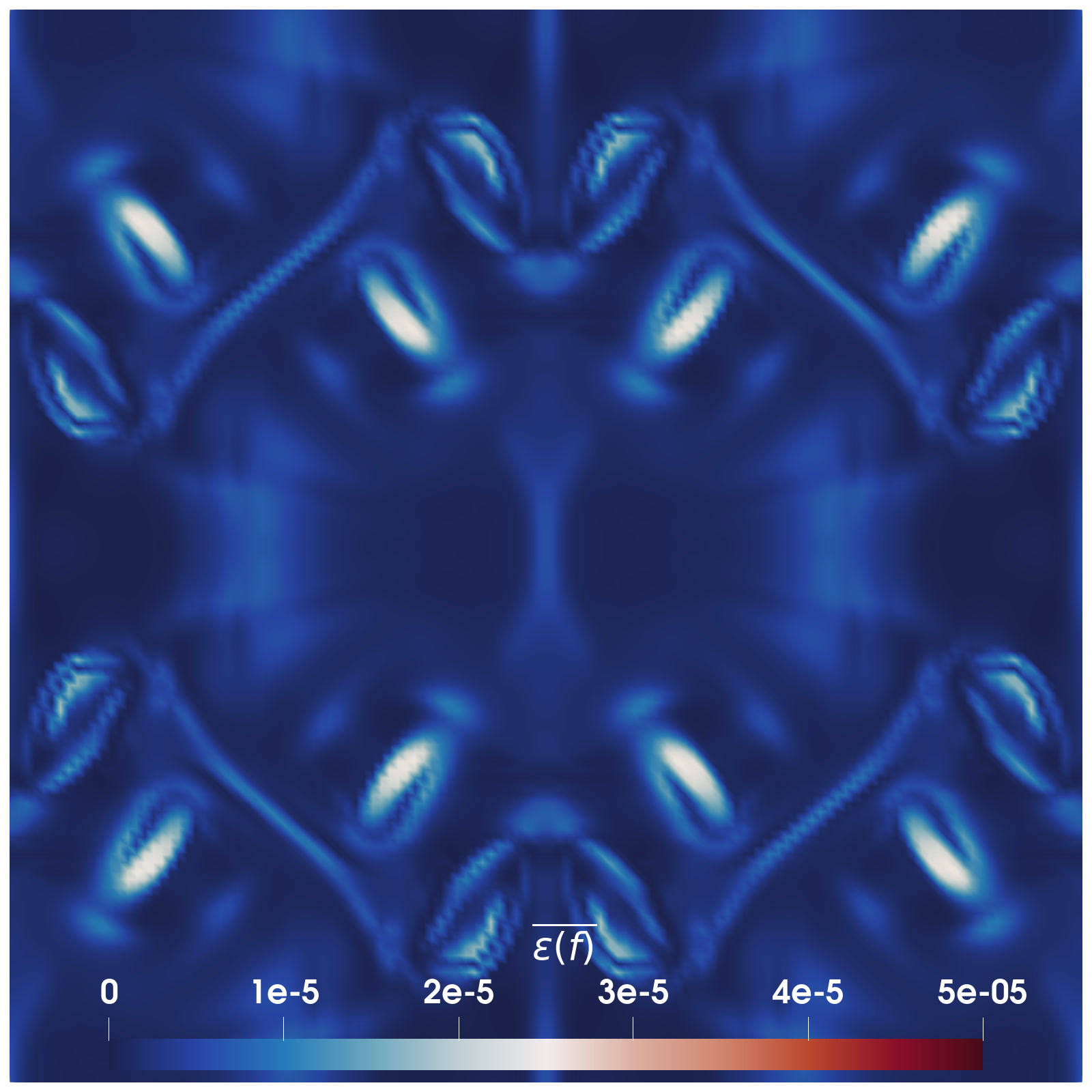}}}
        \newline
        \subfloat[Relative entropy of filtered solution]{
        \adjustbox{width=0.35\linewidth, valign=b}{\includegraphics[]{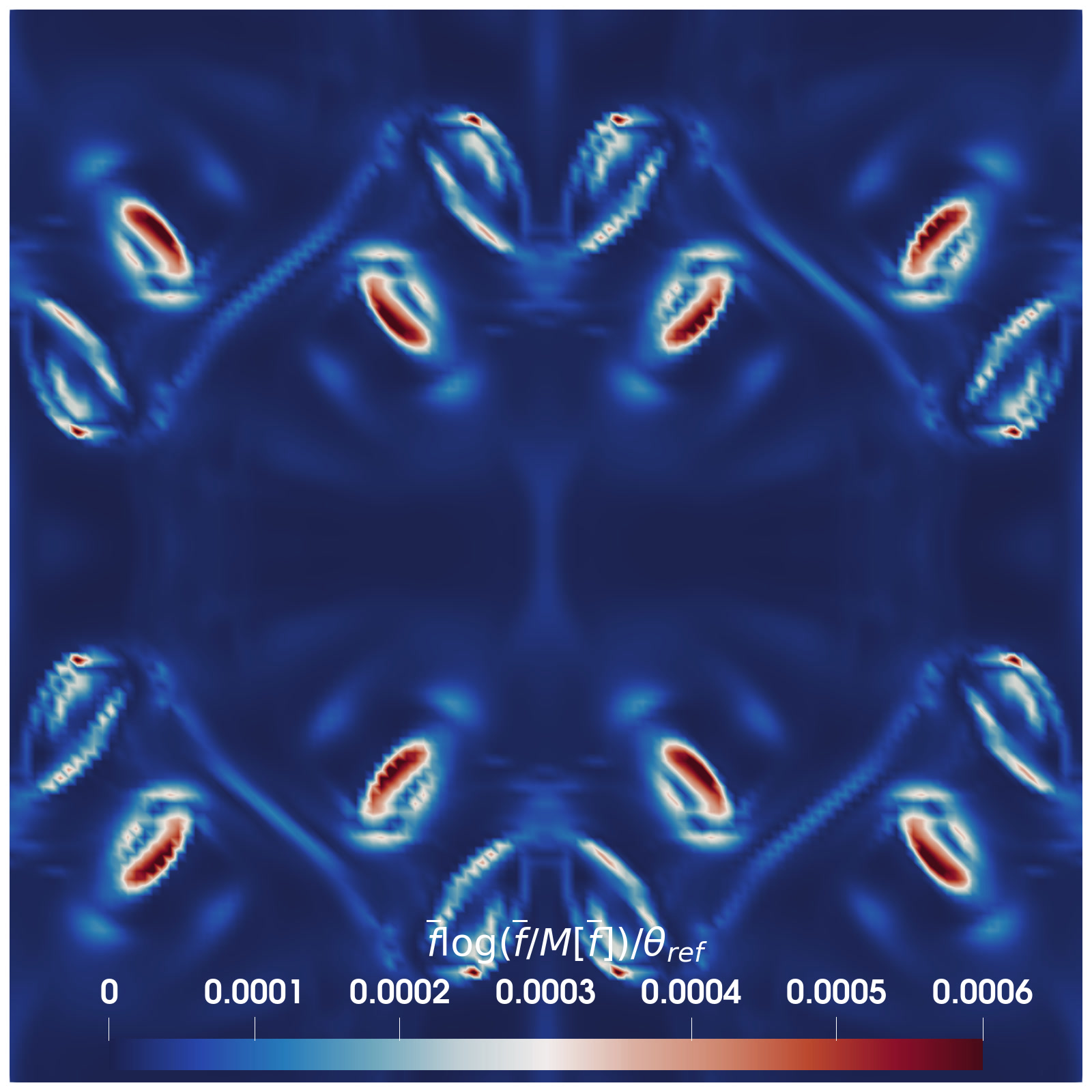}}}
        \hspace{2em}
        \subfloat[Dissipation of filtered solution]{
        \adjustbox{width=0.35\linewidth, valign=b}{\includegraphics[]{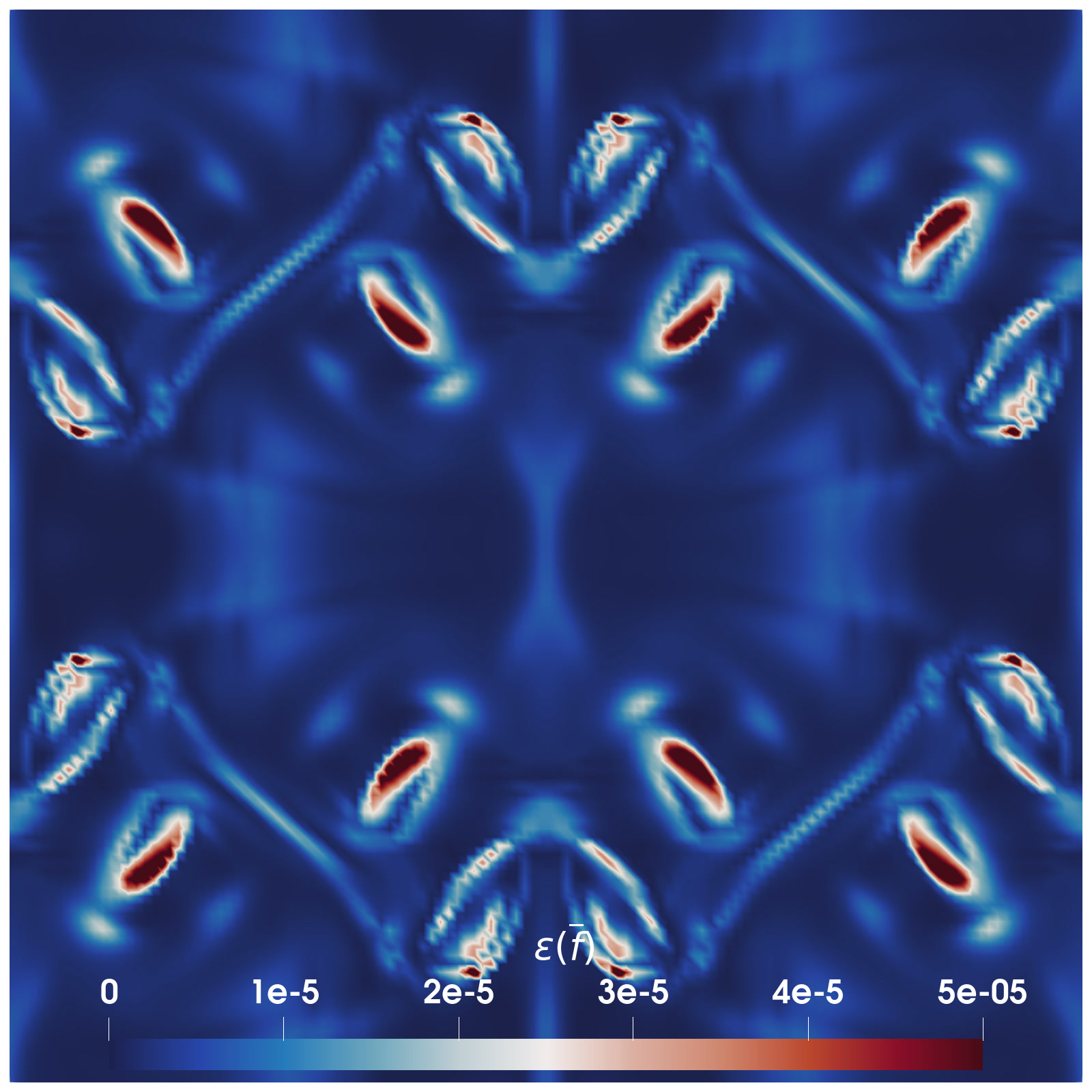}}}
        \newline
        \subfloat[Subgrid relative entropy]{
        \adjustbox{width=0.35\linewidth, valign=b}{\includegraphics[]{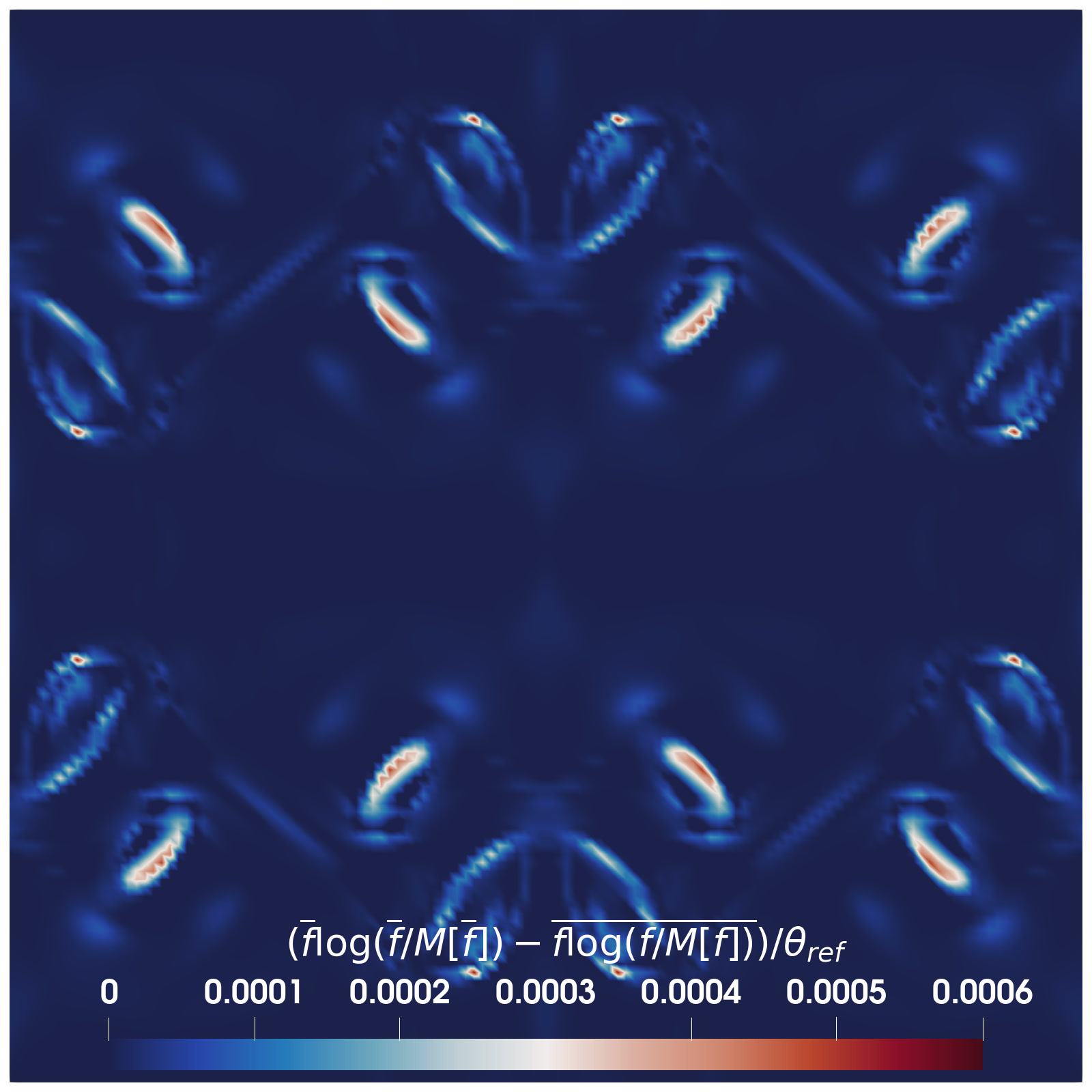}}}
        \hspace{2em}
        \subfloat[Subgrid dissipation]{
        \adjustbox{width=0.35\linewidth, valign=b}{\includegraphics[]{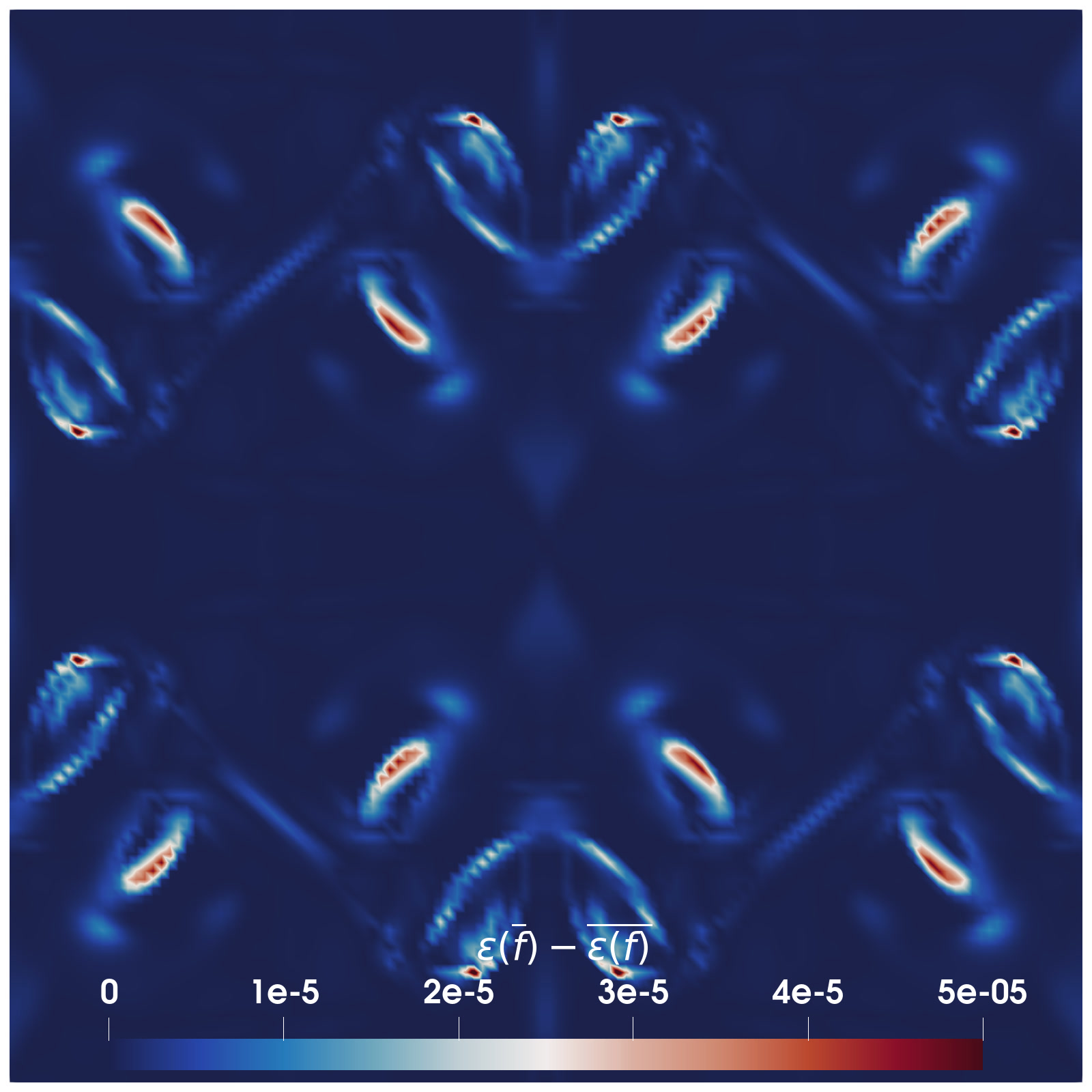}}}
        \newline
        \caption{
        Comparison of the filtered relative entropy/dissipation (top row), relative entropy/dissipation of the filtered solution (middle row), and relative entropy/dissipation residual (bottom row)  for the $M=1.25$, $Re = 1600$ case at $t=10$ on the plane $x = \pi$.}
        \label{fig:entropyfilter_M1p25Re1600x}
    \end{figure}
    
In comparing the behavior of the filtered relative entropy and filtered dissipation -- along with the relative entropy and dissipation computed from the filtered solution -- a number of qualitative similarities could be seen. For the plane $z = \pi$, the filtered relative entropy and filtered dissipation exhibited noticeable qualitative agreement. However, on the plane $x = \pi$, this agreement noticeably detoriated,  with the filtered/resolved relative entropy and filtered/resolved dissipation displaying distinct spatial features and differing in magnitude and structure. However, interestingly, \emph{even in cases were the filtered/resolved relative entropy and dissipation differed, the subgrid relative entropy and subgrid dissipation were still quite similar}. This is particularly evident in \cref{fig:entropyfilter_M0p5Re1600z}, albeit less so in \cref{fig:entropyfilter_M0p5Re1600x}. 

This was also observed as the Mach number increases, as it is shown in \cref{fig:entropyfilter_M1p25Re1600z} on the plane $z = \pi$ and \cref{fig:entropyfilter_M1p25Re1600x} on the plane $x = \pi$ for the case of $M = 1.25$ with $Re=1600$. In this case, the filtered, resolved, and subgrid relative entropy seemed to qualitatively match the associated dissipation for $x  = \pi$, and even for $z = \pi$ where the filtered and resolved relative entropy/dissipation notably differed, the subgrid relative entropy and dissipation still showed strong agreement. These results indicate that while the local relative entropy may not necessarily be a great model for the local dissipation, the \emph{subgrid} relative entropy may potentially be a good surrogate for the \emph{subgrid} dissipation. Furthermore, it can be seen that the filtered dissipation was typically of lower magnitude than the dissipation of the spatially-filtered solution. This observation is consistent with the observations in the work of \citet{Dzanic2024POF}, which indicate that underresolved simulations of turbulent flows via the Boltzmann equation overpredict momentum transport and turbulent kinetic energy. As previously mentioned, this same effect is present in the relative entropy.

\begin{figure}
        \centering
        \subfloat[Filtered relative entropy]{
        \adjustbox{width=0.35\linewidth, valign=b}{\includegraphics[]{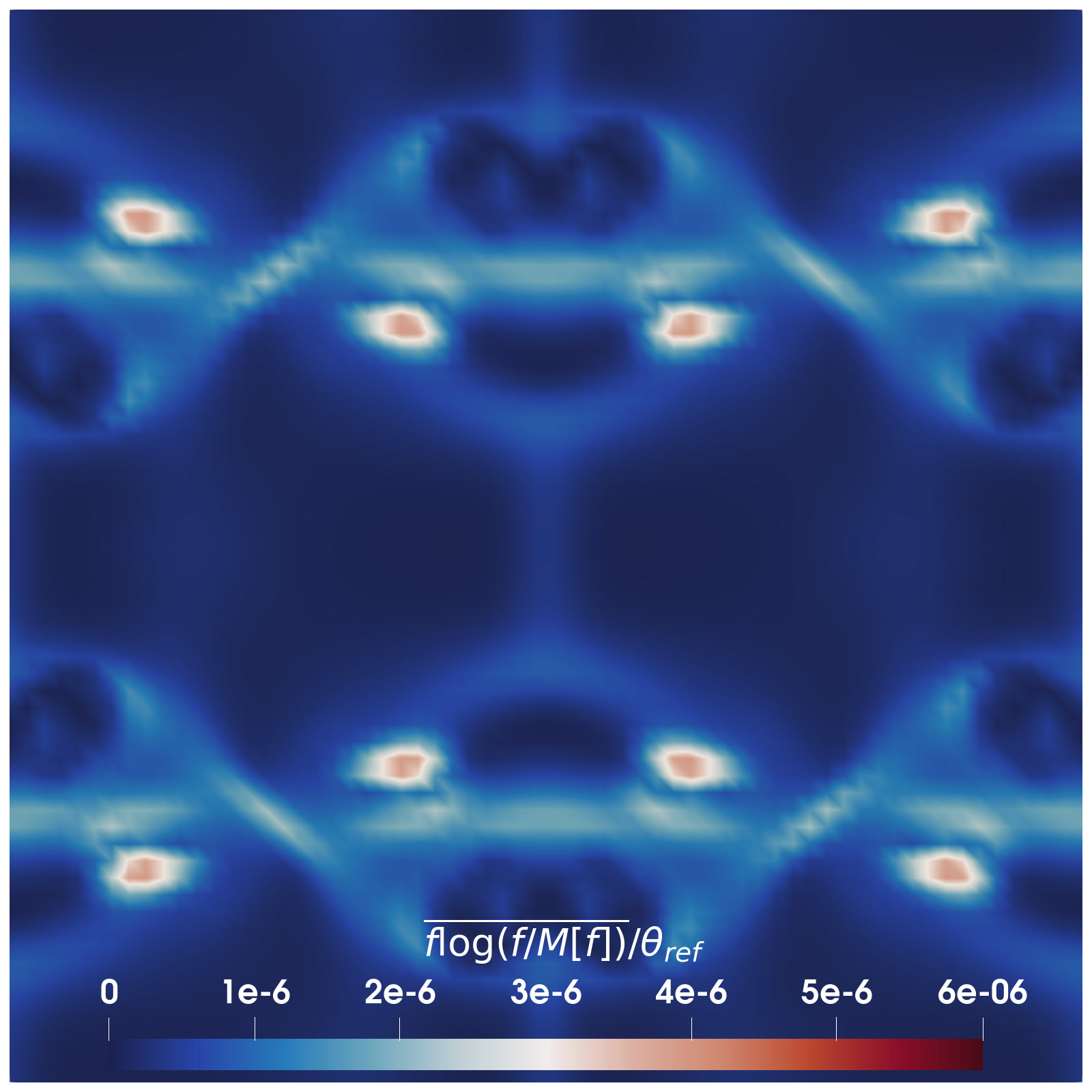}}}
        \hspace{2em}
        \subfloat[Filtered dissipation]{
        \adjustbox{width=0.35\linewidth, valign=b}{\includegraphics[]{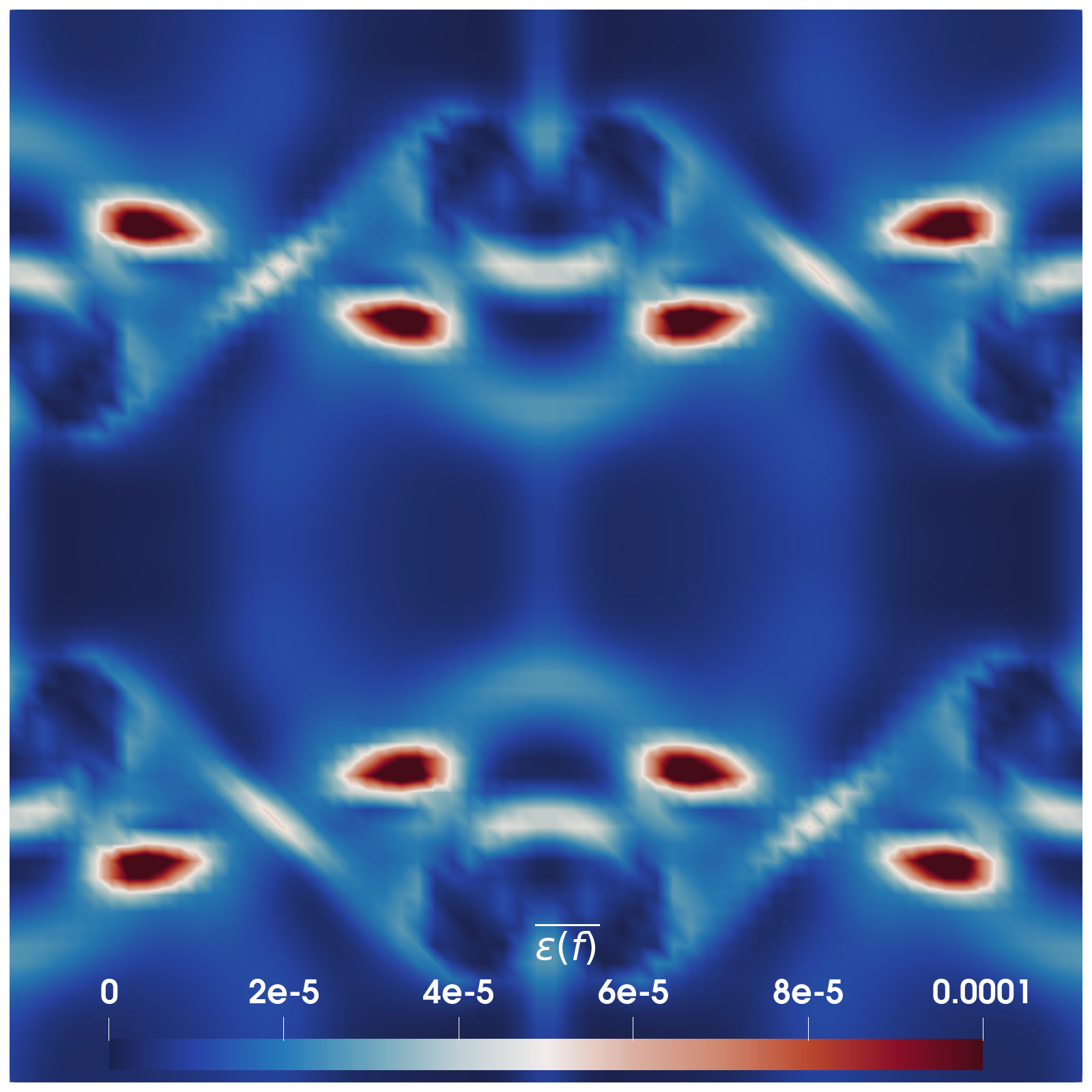}}}
        \newline
        \subfloat[Relative entropy of filtered solution]{
        \adjustbox{width=0.35\linewidth, valign=b}{\includegraphics[]{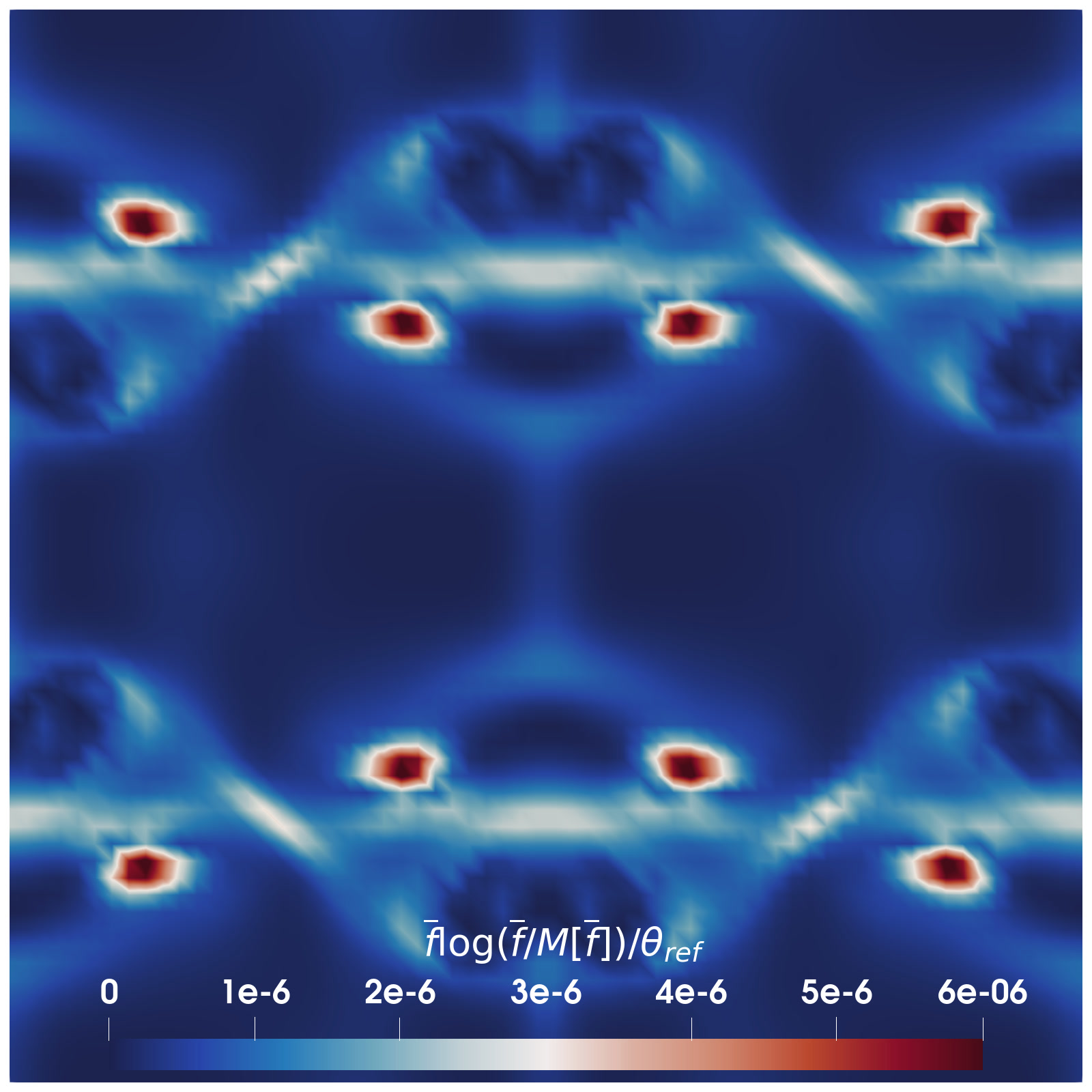}}}
        \hspace{2em}
        \subfloat[Dissipation of filtered solution]{
        \adjustbox{width=0.35\linewidth, valign=b}{\includegraphics[]{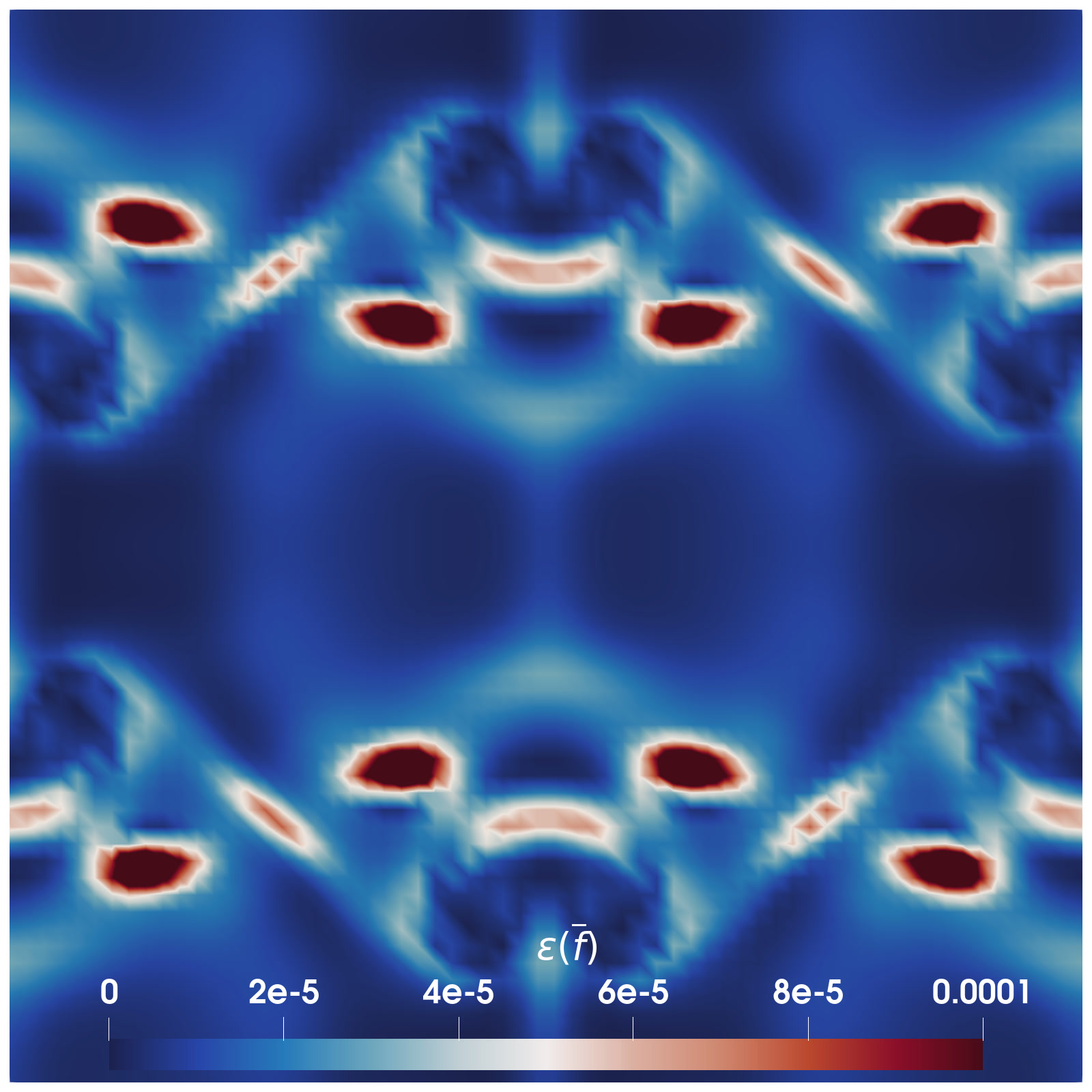}}}
        \newline
        \subfloat[Subgrid relative entropy]{
        \adjustbox{width=0.35\linewidth, valign=b}{\includegraphics[]{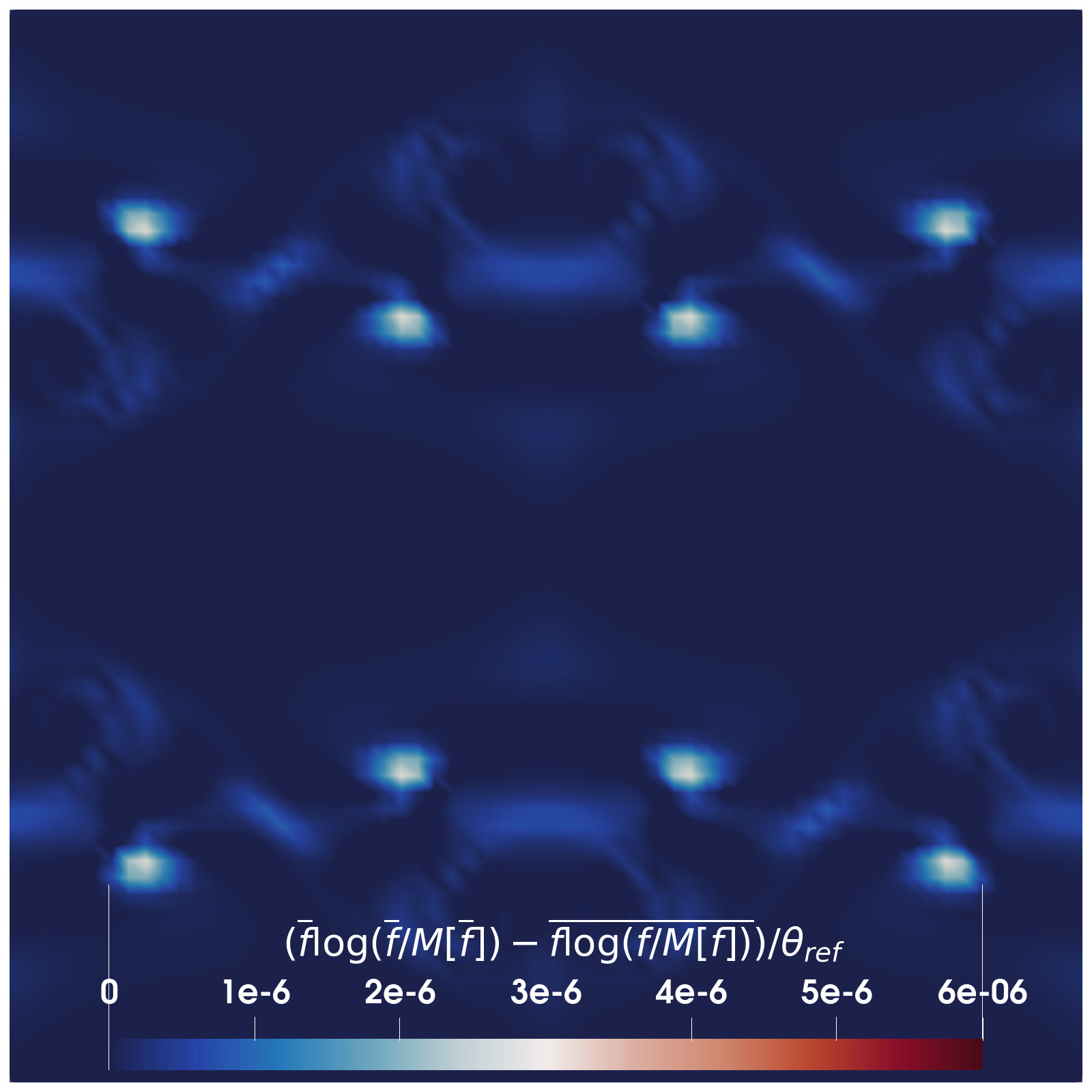}}}
        \hspace{2em}
        \subfloat[Subgrid dissipation]{
        \adjustbox{width=0.35\linewidth, valign=b}{\includegraphics[]{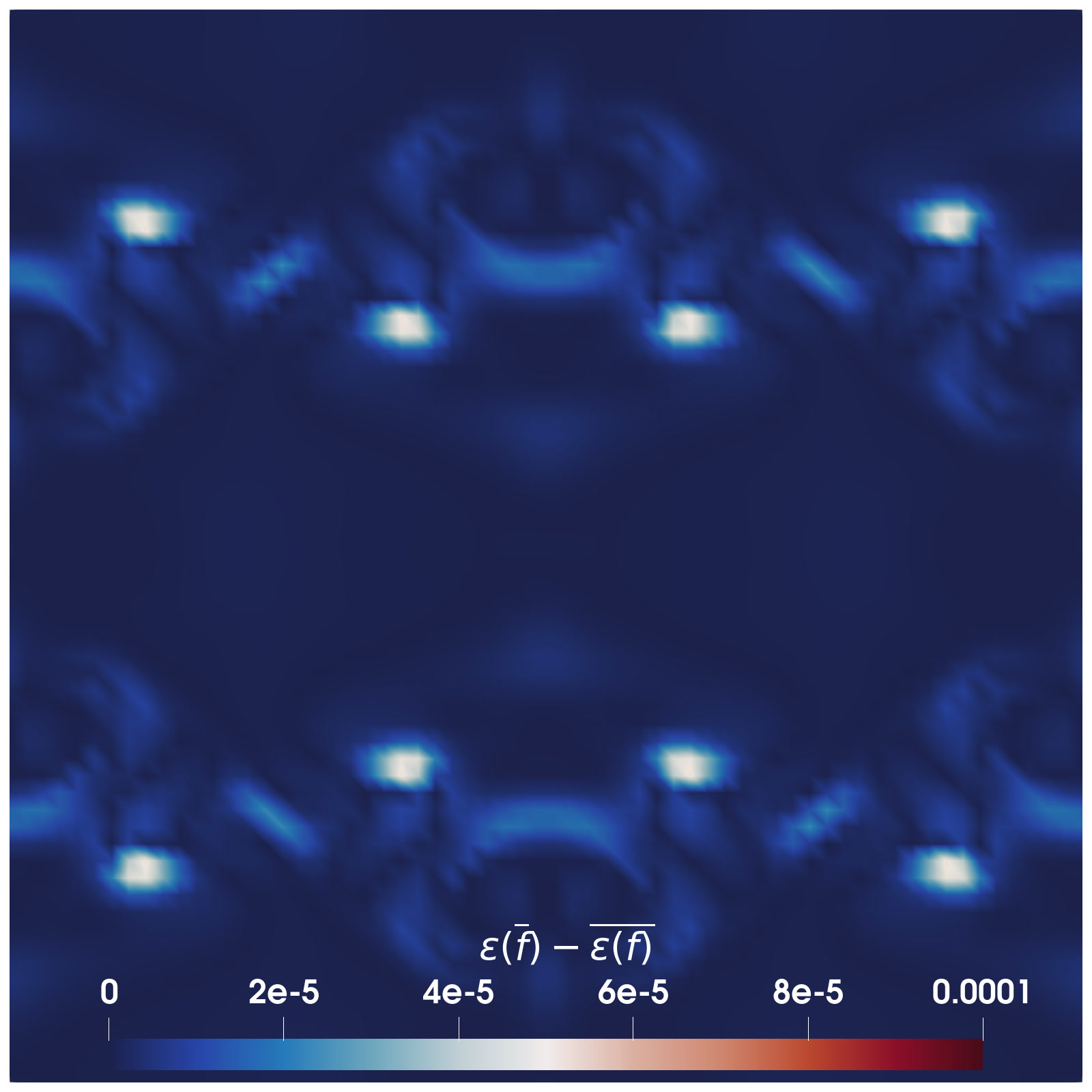}}}
        \newline
        \caption{
        Comparison of the filtered relative entropy/dissipation (top row), relative entropy/dissipation of the filtered solution (middle row), and relative entropy/dissipation residual (bottom row)  for the $M=0.5$, $Re = 400$ case at $t=10$ on the plane $x = \pi$.}
        \label{fig:entropyfilter_M0p5Re400x}
    \end{figure}

\begin{figure}
        \centering
        \subfloat[Filtered relative entropy]{
        \adjustbox{width=0.35\linewidth, valign=b}{\includegraphics[]{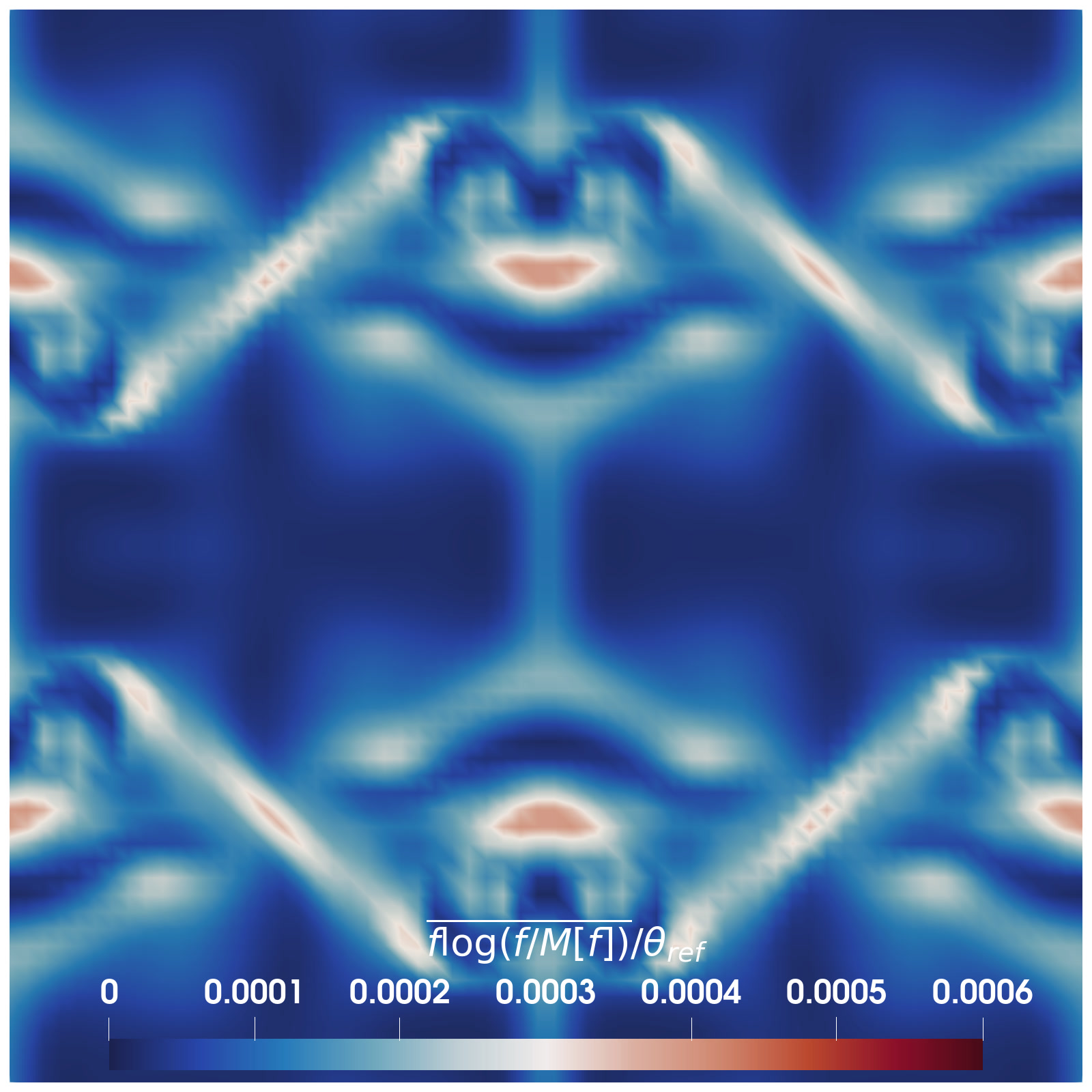}}}
        \hspace{2em}
        \subfloat[Filtered dissipation]{
        \adjustbox{width=0.35\linewidth, valign=b}{\includegraphics[]{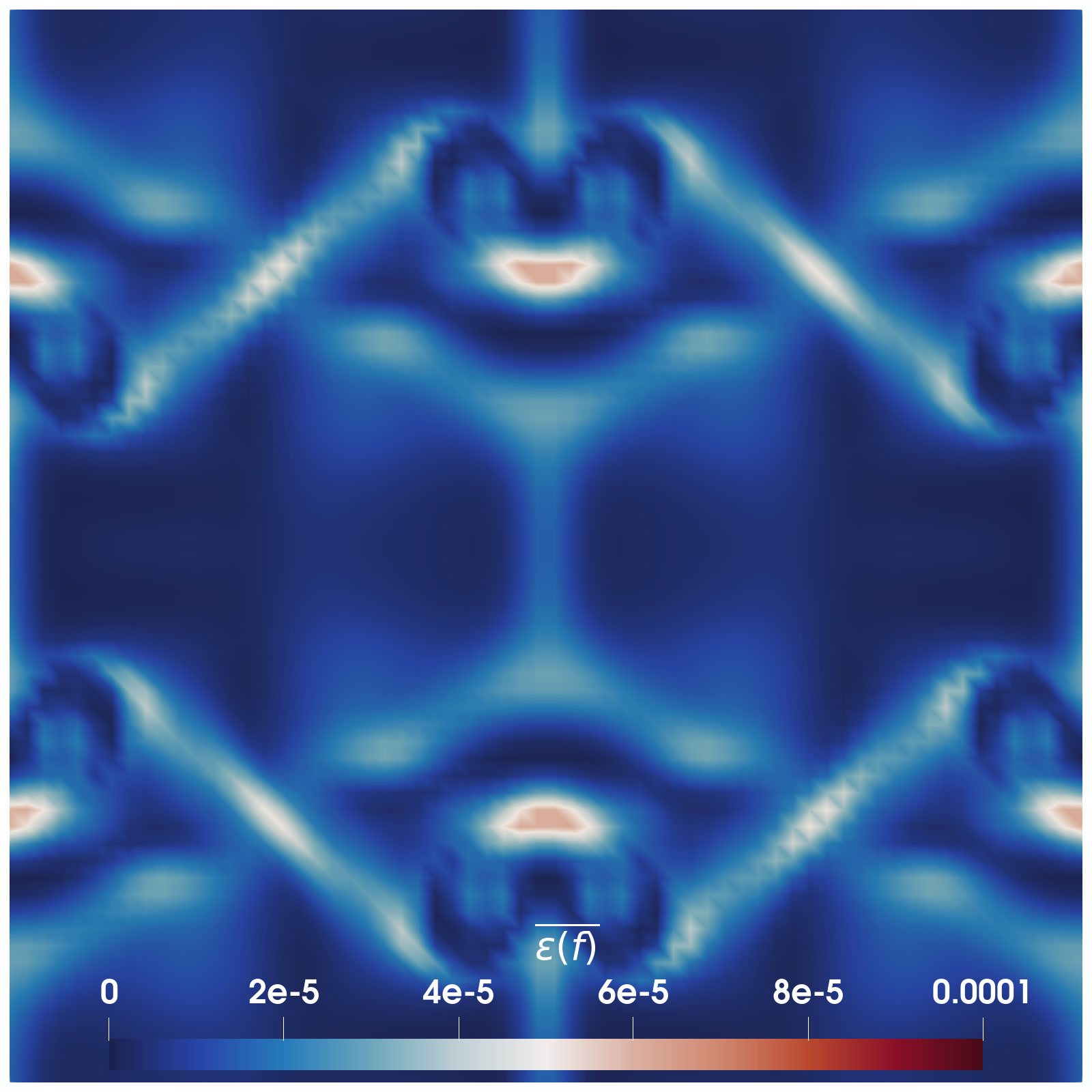}}}
        \newline
        \subfloat[Relative entropy of filtered solution]{
        \adjustbox{width=0.35\linewidth, valign=b}{\includegraphics[]{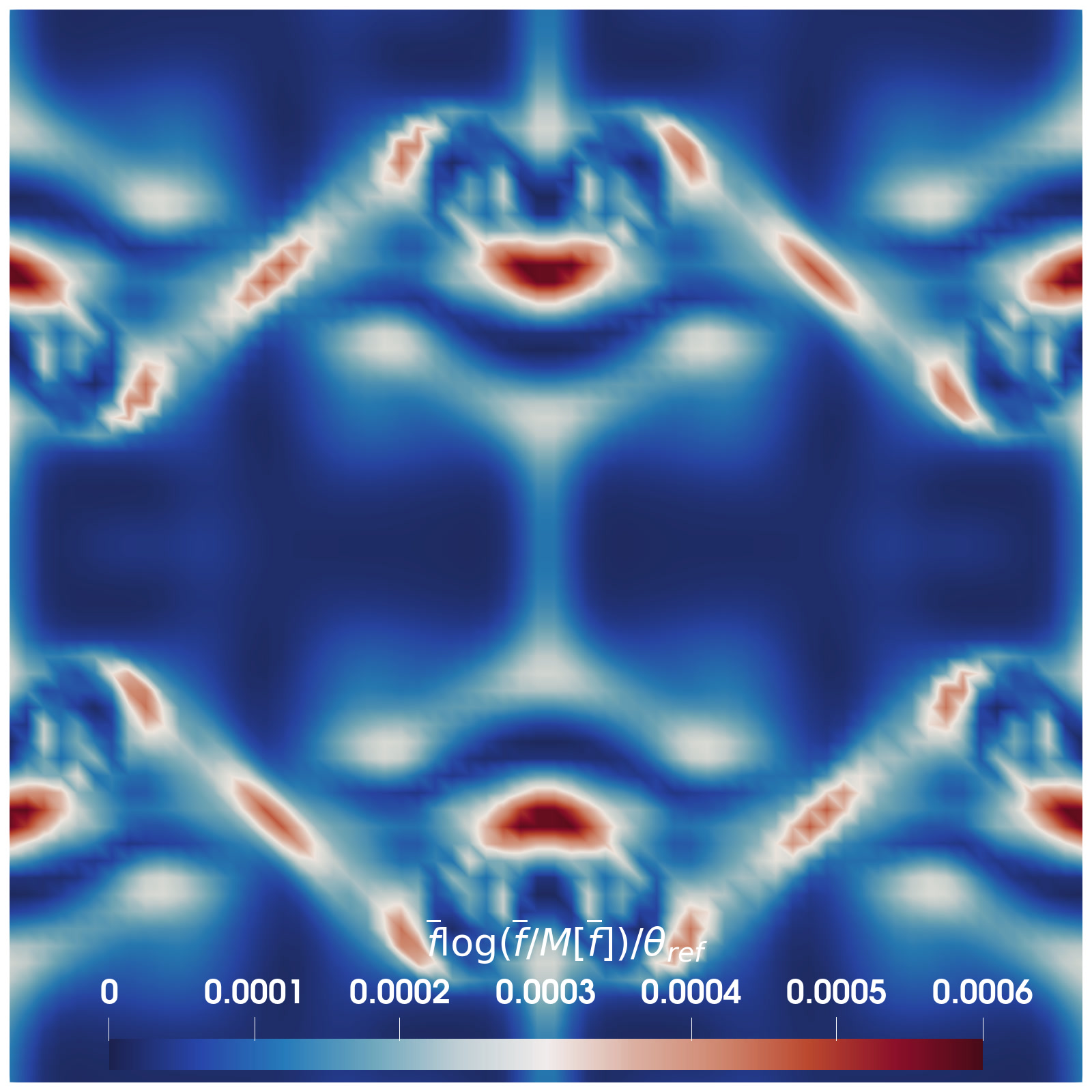}}}
        \hspace{2em}
        \subfloat[Dissipation of filtered solution]{
        \adjustbox{width=0.35\linewidth, valign=b}{\includegraphics[]{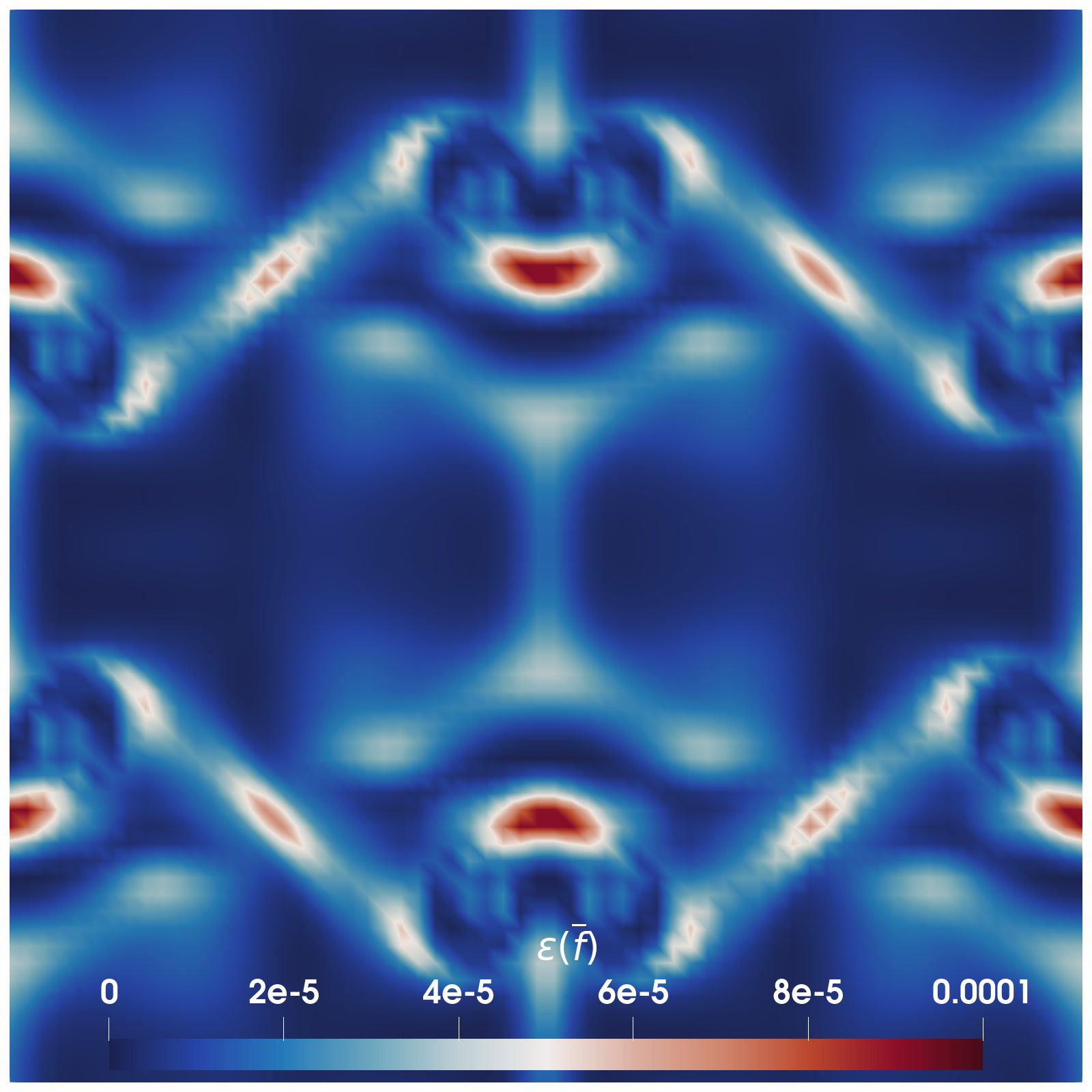}}}
        \newline
        \subfloat[Subgrid relative entropy]{
        \adjustbox{width=0.35\linewidth, valign=b}{\includegraphics[]{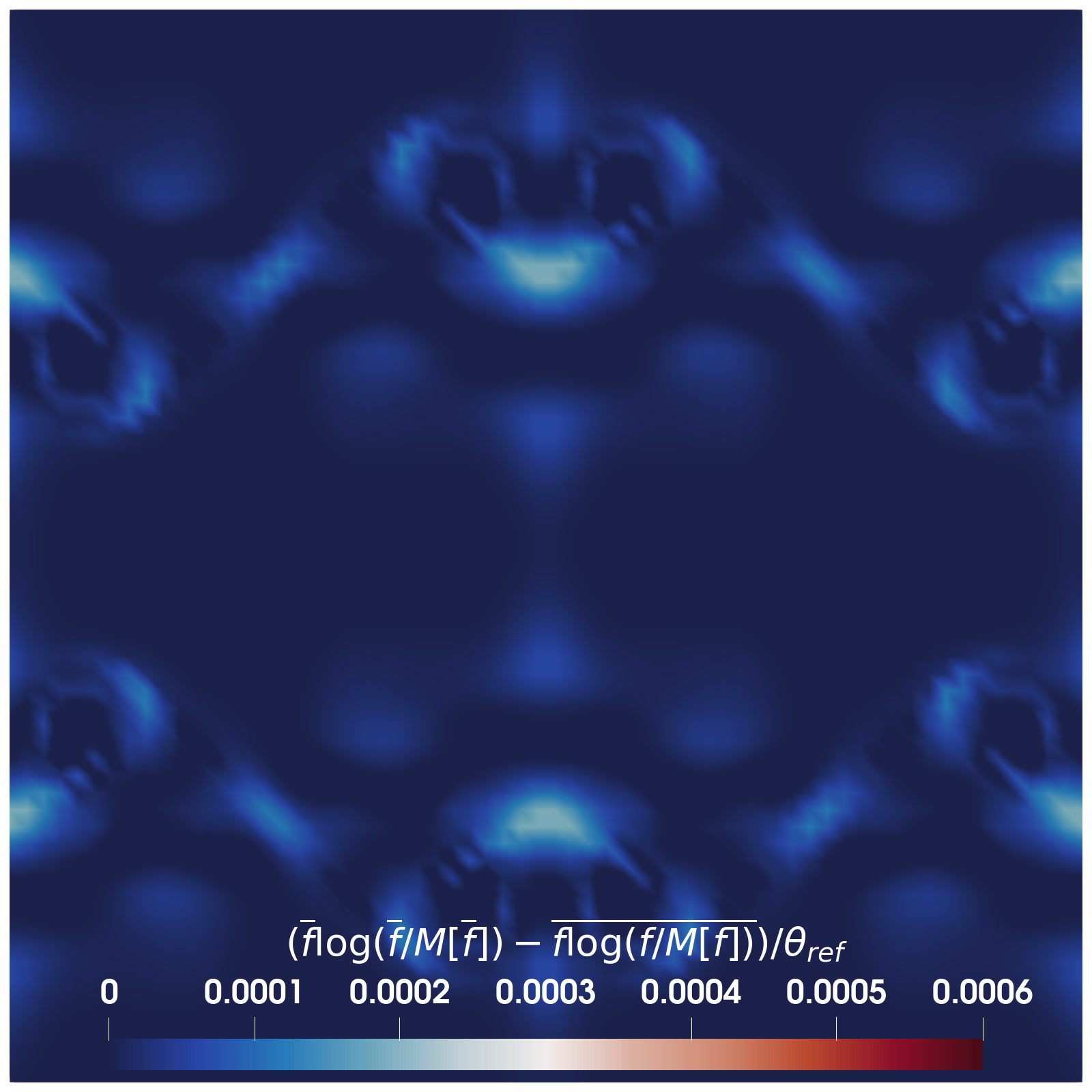}}}
        \hspace{2em}
        \subfloat[Subgrid dissipation]{
        \adjustbox{width=0.35\linewidth, valign=b}{\includegraphics[]{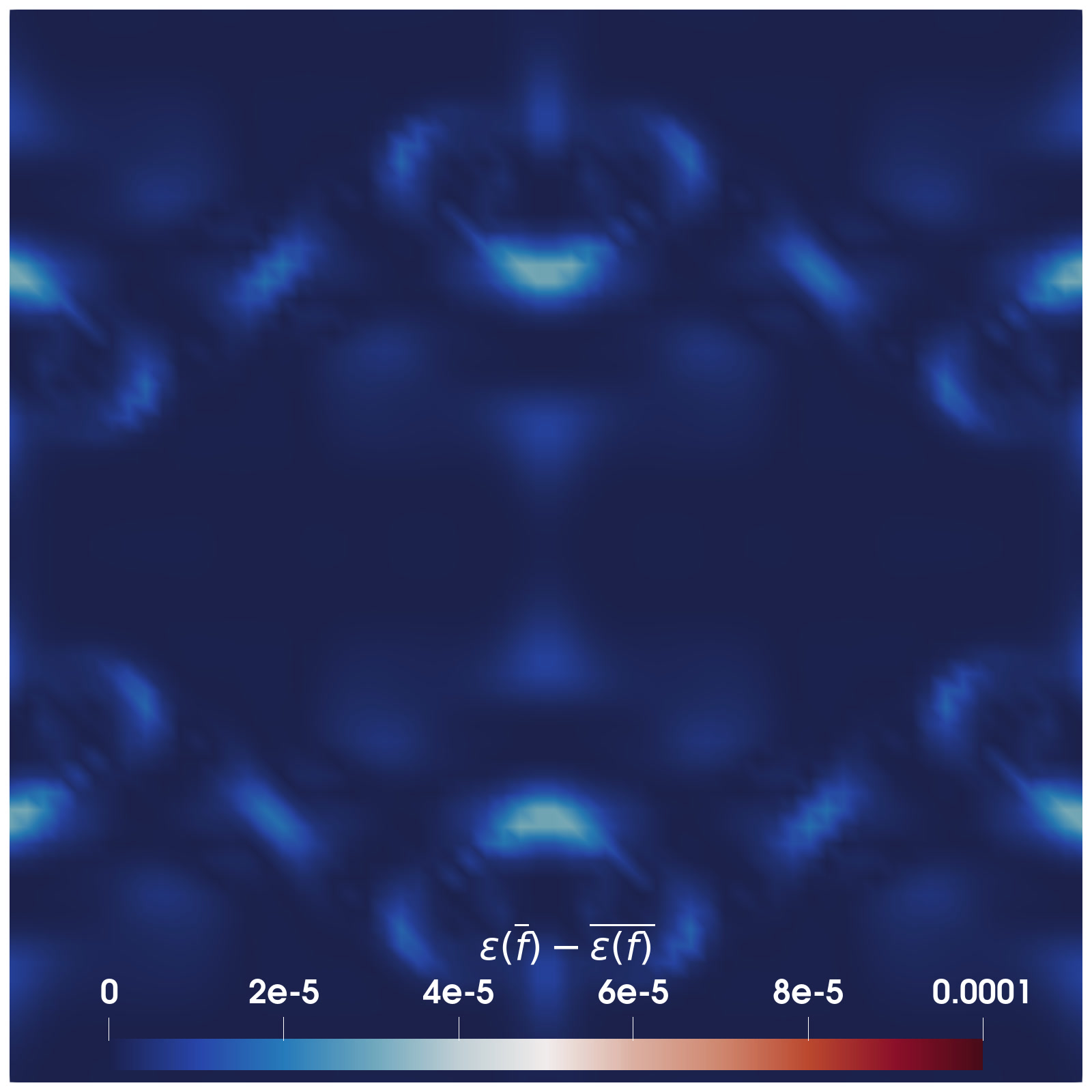}}}
        \newline
        \caption{
        Comparison of the filtered relative entropy/dissipation (top row), relative entropy/dissipation of the filtered solution (middle row), and relative entropy/dissipation residual (bottom row)  for the $M=1.25$, $Re = 400$ case at $t=10$ on the plane $x = \pi$.}
        \label{fig:entropyfilter_M1p25Re400x}
    \end{figure}

We also present this comparison briefly for the $Re = 400$ cases for $M = 0.5$ (\cref{fig:entropyfilter_M0p5Re400x}) and $M=1.25$ (\cref{fig:entropyfilter_M1p25Re400x}) on the plane $x = \pi$. The same observations were seen at $Re=400$, with strong agreement between the subgrid relative entropy and subgrid dissipation, showing consistency across both Reynolds and Mach numbers. Additionally, in these low Reynolds number cases, the filtered/resolved relative entropy and dissipation  did not show nearly as many differences as with the high Reynolds number cases. This is likely attributable to the reduced complexity and limited richness of flow structures at lower Reynolds numbers.


\section{Conclusion}\label{sec:conclusion}
This study investigates the dynamics of compressible turbulence through the perspective of kinetic theory by solving the Boltzmann equation for the three-dimensional compressible Taylor–Green vortex with the goal of obtaining insights into the interplay between molecular-scale phenomena and macroscopic turbulent flow behavior. The primary finding of this work is that the Kullback--Leibler divergence of the distribution function $f$ and its local equilibrium state $M[f]$ -- formally, the relative entropy functional introduced by \citet{Golse2014} -- is strongly correlated with macroscopic viscous dissipation rates. This correlation persists across a range of Mach and Reynolds numbers and suggests that relative entropy, a purely kinetic and locally computable (i.e., zeroth-order with respect to space/time) measure of non-equilibrium, may serve as a useful surrogate for turbulent kinetic energy dissipation. A secondary finding of this work is that under spatial averaging/filtering on the distribution function, the subgrid-scale relative entropy remains in strong qualitative agreement with the subgrid-scale dissipation, even in cases where the filtered or resolved relative entropy and dissipation diverge, such that it may act as a surrogate model for subgrid-scale dissipation. This, in conjunction with the primary finding, presents potential for alternate approaches to closure modeling of both macroscopic and kinetic subgrid-scale quantities of interest to compressible turbulence. 

\section*{Acknowledgements}
This work was partially performed under the auspices of the U.S. Department of Energy by Lawrence Livermore National Laboratory under Contract No. DE--AC52--07NA27344 and used resources of the Oak Ridge Leadership Computing Facility at the Oak Ridge National Laboratory, which is supported by the Office of Science of the U.S. Department of Energy under Contract No. DE--AC05--00OR22725.

\bibliographystyle{jfm}
\bibliography{references}

\end{document}